\documentclass[fleqn,usenatbib]{mnras}

\usepackage{latexsym,mathrsfs,amssymb,bm}
\usepackage[tbtags]{amsmath}
\usepackage[T1]{fontenc}
\usepackage{ae,aecompl,times}
\usepackage{epsfig}
\usepackage[dvipsnames]{xcolor}
\usepackage{multirow}
\usepackage{tikz,array,color,float}
\usepackage[utf8]{inputenc}
\usepackage{graphicx}
\usepackage{caption}
\usepackage{subcaption}
\usepackage{dsfont}

\renewcommand{\vec}[1]{ {\bmath #1} }

\title[Divergence-free MHD on a moving mesh]{Divergence-free magnetohydrodynamics on a moving mesh}

\author[V.~Springel, A.~C.~Mayer \& R.~Pakmor]{%
Volker Springel$^{1}$, Alexander C. Mayer$^{1}$, and R\"udiger Pakmor$^{1}$ 
\vspace*{0.1cm}\\%
$^{1}$Max-Planck-Institut f\"ur Astrophysik, Karl-Schwarzschild-Str. 1, D-85748, Garching, Germany\\%
}

\date{Accepted XXX. Received YYY; in original form ZZZ}
\pubyear{2025}

\begin{document}

\label{firstpage}

\pagerange{\pageref{firstpage}--\pageref{lastpage}}

\maketitle

\begin{abstract}
  Magnetic fields are ubiquitous in the Universe and play an important role in many astrophysical processes, such as star formation and the evolution of galaxies. Hydrodynamical simulation methods that account for ideal magnetohydrodynamics (MHD) are however challenged by the  need to achieve numerical stability and physical consistency. Both can be compromised by the occurrence of non-vanishing local divergence errors in the magnetic field. Different approaches to either eliminate or reduce these discretization errors  are in widespread use, including constrained transport or divergence cleaning techniques. However, thus far a divergence-free Galilean invariant formulation of MHD that can be readily applied in Lagrangian codes has been an elusive goal. Here we revisit the use of the vector potential for evolving the magnetic field on an unstructured moving-mesh and show how a particular gauge choice combined with a suitably discretized integration scheme provides a  stable and accurate formulation of MHD. Our method combines a finite-difference evolution of a cell-centred vector potential with  a finite volume representation of the other fluid quantities, and is realized in  the moving-mesh code {\small AREPO}.  The new approach is fully Galilean invariant and therefore shows vanishing errors for classic problems such as field loop advection. We demonstrate good accuracy and stability when our scheme is applied to various test problems and to practical applications in 3D, including when local timesteps and adaptive mesh refinement and derefinement are used.
\end{abstract}


\section{Introduction}
\label{sec:intro}

Much of the gas in the Universe is in the form of a highly conducting plasma that supports magnetic fields. These magnetic fields play a key role in a broad range of astrophysical systems, ranging from accretion disks, to stars, galaxies, or intergalactic filaments \citep[see][for a review]{Shukurov2021}, and in certain regimes reach sufficient strength to become dynamically important. Numerically simulating magnetohydrodynamics is thus a key task in computational astrophysics.

Superficially, extending ordinary non-viscous hydrodynamics to ideal magnetohydrodynamics seems relatively straightforward, at least in standard finite volume schemes. Here, the induction equation can be recast into an additional conservation law for the components of the volume averaged magnetic field. Together with additional terms in the energy and momentum fluxes, this extension can be readily accommodated in basic Godunov schemes. All that seems needed is a Riemann solver that is capable of dealing with an additional magnetic field in the states that meet at cell interfaces. The problem, however, is that this approach readily breaks down in practice. Unlike the continuum equations, discretized versions of the MHD equations are quite susceptible to various modes of numerical instability and do not generally maintain the magnetic field divergence constraint, $\nabla\cdot \vec{B}=0$, even if this is accurately the case in the initial conditions. And once numerical divergence errors appear, they tend to quickly grow and frequently lead to unphysical solutions \citep{Brackbill1980} or catastrophic code instability.

Stabilizing numerical MHD simulations hence requires special measures, something that can be broadly subsumed under the term divergence control. There is a large literature describing different strategies for dealing with this numerical problem. One class of methods works directly with the magnetic field, trying to limit the size of any appearing $\nabla\cdot \vec{B}$ error. Particular popular is the hyperbolic cleaning approach of \citet{Dedner2002}, where arising $\nabla\cdot \vec{B}$ errors are transported away and damped, thereby preventing them from strongly affecting solutions. Methods based on this approach are particularly popular in smoothed particle hydrodynamics (SPH) or mesh-free codes \citep{Price2005, Tricco2016a, Hopkins2016, Steinwandel2026}.

A related but somewhat different strategy is followed in the so-called Powell 8-wave formulation \citep{Powell1999}, which has successfully been implemented and used, for example, on moving meshes \citep{Pakmor2013} as well as in high-order Discontinuous Galerkin methods on adaptive Cartesian meshes \citep{Guillet2019}. Here all $\nabla\cdot \vec{B}$ terms are consistently retained in the derivation of the MHD equations and their physical effect is explicitly cancelled in the equations to the numerical discretisation level. While this leads to additional source terms in the momentum and energy equations, including them stabilizes the evolution equations against a catastrophic growth of non-zero divergences if they appear due to discretization errors, so that ideally they stay small enough to influence solutions only at a negligible level. The non-vanishing divergence errors can, however, cause incorrect jump conditions in certain strong magnetic shocks \citep{Toth2000}, and the source terms break the conservation of momentum and energy to machine precision that is otherwise guaranteed by construction in finite volume schemes.

Another, more drastic approach is to simply project out emerging divergence errors by regularly carrying out a Helmholtz decomposition of the magnetic field \citep[first proposed by][]{Brackbill1980}. Since this requires the expensive solve of an elliptic problem and can be quite diffusive, it is however rarely done in common codes \citep[but see][for a recent implementation]{Tsukamoto2026}. A method that is unfortunately likewise not very  helpful in practice  is to represent the magnetic field in terms of Euler potentials that are advected with the gas flow, as they cannot be combined consistently with the required numerical diffusion in non-linear flows \citep{Brandenburg2010}.

Instead, the family of so-called constrained transport methods \citep{Evans1988, Dai1998, Balsara1999, Toth2000, Gardiner2005, Gardiner2008} offer an elegant approach that can maintain  $\nabla\cdot \vec{B}=0$ to machine precision for a specific discrete version of the divergence operator. In classic constrained transport, the magnetic field is represented on a staggered mesh on the faces of cells, in terms of the perpendicular magnetic flux component through these faces. By computing electric fields at edges of the mesh, these magnetic fluxes can be updated in a way that guarantees a vanishing flux sum over the faces of all computational cells if this is the case in the initial conditions. Here divergence errors are thus prevented exactly, eliminating any need to cure these problems in the first place.

Closely related to constrained transport is the idea to evolve the magnetic vector potential instead of the magnetic field \citep{Price2010, Stasyszyn2015, Fragile2019, Tu2022}. The curl of a vector field is automatically divergence free, so by defining the $\vec{B}$-field in terms of the vector potential one can sidestep the divergence problem. Constrained transport methods can in fact be shown to be identical to particular versions of vector potential approaches \citep{Toth2000, Etienne2010, Mocz2017}, where the magnetic field is constructed from the vector potential stored at mesh corners through loop integrals that first compute magnetic fluxes through the faces from which then cell centred magnetic fields are recovered, rather than directly applying a finite difference version of the curl operator.

A question of considerable importance in the field is whether the results of cleaning methods can be trusted, or whether it is possible that small but finite divergence errors produce spurious results in some cases. This is often controversially discussed. Recently, for example, \citet{Tomida2026} argued that Dedner cleaning may cause a spurious amplification of magnetic fields in some situations. On the other hand, \citet{Pakmor2026} compared different realizations for constrained transport and Powell cleaning in simulations of driven subsonic turbulence, finding a close alignment of the amplification rates with differences that can be solely explained by the different numerical dissipation properties of the examined codes and methods \citep[see also][for a comparison of supersonic MHD turbulence]{Tricco2016b}. To further shed light on this, it is important to increase the variety of available methods, and in particular, to make progress on realizing stable constrained transport schemes in Lagrangian codes, which is elusive thus far, even though these methods are particularly powerful in problems with complex geometries and high bulk flows, such as cosmic structure formation.

In this paper, we propose to use the vector potential combined with a moving unstructured Voronoi mesh as realized in the  {\small AREPO} code, and derive a new method that is free of divergence errors in the magnetic field and avoids problems that troubled previous attempts along these lines. While our approach is inspired by an earlier vector potential implementation of ideal MHD  realized by \citet{Mocz2016}, it differs substantially in numerous aspects  that ultimately are critical for achieving numerical stability also in demanding flow situations. It likewise differs from an earlier realization of constrained transport in the simpler case of two-dimensional Voronoi meshes \citep{Mocz2014}, and from the constrained transport approach realized in the moving mesh-rings of the {\small DISCO} code \cite{Duffel2016}. Importantly, our new approach achieves Galilean invariance for a moving mesh despite the use of the vector potential. As we will see, this requires a special gauge which however invokes its own numerical challenges that need to be carefully addressed. To our knowledge, the present work represents the first successful divergence-free, Galilean-invariant, and quasi-Lagrangian formulation of numerical MHD that can be applied robustly in 3D. Furthermore, our approach also admits local timesteps and flexible local mesh refinement and derefinement operations.

This paper is structured as follows. In Section~\ref{secbasicMHD} we first recap the basic equations of MHD to set the stage. We then discuss the evolution equation for the vector potential on a moving mesh in Section~\ref{secA}, which also identifies the gauge we need for achieving Galilean invariance. In Section~\ref{SecNewApproach} we then describe the basic discretization strategy we adopt for our method, focusing on the time evolution of the vector potential in the induction equation and the approaches we adopt to realize this in a numerically stable fashion. A detailed outline of the final integration procedure is given in Section~\ref{SecDetails}. A variety of tests and their results are then presented and discussed in Section~\ref{sec:results}. We discuss how our method relates to other approaches in Section~\ref{sec:discussion}, and finally present our conclusions in Section~\ref{sec:conclusions}. For reference, we collect in  Appendix~\ref{SecCosmoEquations} the MHD equations in the form we use them in cosmological simulations.

\section{Basic magnetohydrodynamics} \label{secbasicMHD}

\subsection{Foundational equations}

In ideal MHD\footnote{We use Lorentz-Heaviside units throughout this paper.}, we have Faraday's law,
\begin{equation}
\frac{\partial \vec{B}}{\partial t} = - \nabla \times \vec{E}, 
\end{equation}
describing the change of the magnetic field $\vec{B}$ in terms of the electric field $\vec{E}$, and Ampere's law, giving the current density,
\begin{equation}
\vec{J} = \nabla \times \vec{B},
\end{equation}
in terms of the magnetic field. The equation of motion gets a Lorentz force term, $\vec{J}\times \vec{B}$, as follows:
\begin{equation}
\rho\left( \frac{\partial}{\partial t} + \vec{v}\cdot\nabla\right)
\vec{v} = \vec{J} \times\vec{B} - \nabla P,
\end{equation}
where $P = (\gamma -1)\rho u$ is the thermal pressure, $\rho$ is the gas density, $u$ denotes the thermal energy per unit mass, $\vec{v}$ the gas velocity, and $\gamma$ the adiabatic index. The Lorentz force term can alternatively be written as
\begin{equation}
 \vec{J}\times
\vec{B}  
= \vec{B}\cdot\nabla \vec{B}
- \nabla\left( \frac{\vec{B}^2}{2}\right),
\end{equation}
where the first term on the right hand side is a magnetic tension force, while the second term is an isotropic repulsive force due to the magnetic pressure. The tension term cancels out the parallel part of the magnetic pressure term. The combined effect is that the magnetic tension tries to straighten out field lines, while the magnetic pressure tends to create a force towards regions of lower magnetic energy density.

We also have Ohm's law in the form
\begin{equation}
\vec{E} +  \vec{v} \times \vec{B}  = \eta \vec{J},
\end{equation}
where $\eta$ is the Ohmic resistivity. For ideal MHD, the resistivity vanishes. We keep this non-ideal term for now, as it is useful for discussing how physical or numerical dissipation of the magnetic field enters.

Taking the curl of Ohm's law, and making use of Faraday's and Ampere's law yields the induction equation:
\begin{equation}
\frac{\partial \vec{B}}{\partial t} = \nabla \times (\vec{v} \times \vec{B})
- \nabla \times (\eta \vec{J}).
\end{equation}
Note that for spatially constant $\eta$ and vanishing $\vec{B}$-field divergence, the dissipative term can be written as $\nabla \times (\eta \vec{J}) =  - \eta \nabla^2 \vec{B}$, i.e.~here the Ohmic dissipation is simply a diffusion term in the magnetic field.

The induction equation is the central equation of MHD which augments the ordinary equation of ideal hydrodynamics. Assuming vanishing divergence of the magnetic field, $\nabla\cdot\vec{B} = 0$, they in fact retain their character as a set of conservation laws, although they are now only weakly hyperbolic. While the mass conservation law retains its usual form,
\begin{equation}
  \frac{\partial \rho}{\partial t}   + \nabla\cdot \left[  \rho \vec{v} \right]=0,
\end{equation}
and is not modified by the presence of a magnetic field, both the momentum and energy conservation laws get additional terms due to the magnetic field. They take the form
\begin{equation}
  \frac{\partial (\rho\vec{v})}{\partial t}   + \nabla\cdot \left[  \rho
  \vec{v}\vec{v}^T + p_{\rm tot} - \vec{B}\vec{B}^T \right] = 0,
\end{equation}  
and
\begin{equation}
  \frac{\partial  (\rho e)}{\partial t}   + \nabla\cdot \left[  \rho e
    \vec{v}
    + p_{\rm tot} \vec{v} - \vec{B} (\vec{v}\cdot\vec{B})  - \vec{B}
    \times (\eta \vec{J})\right] =0 ,
  \label{eqnegy}
\end{equation}  
where $p_{\rm tot} = p_{\rm therm} + \frac{1}{2} \vec{B}^2$ is the
total pressure, and $e = u + \frac{1}{2} \vec{v}^2 + \frac{1}{2\rho}
\vec{B}^2$ is the total energy per unit mass.
The induction equation itself can alternatively be written as 
\begin{equation}
  \frac{\partial \vec{B}}{\partial t} + \nabla\cdot \left[  \vec{B}\vec{v}^T
  - \vec{v}\vec{B}^T \right]  + \nabla \times (\eta \vec{J})= 0,
\end{equation}  
which in the limit of vanishing resistivity becomes a conservation law for the volume-averaged magnetic field.

As an aside, the origin of the term involving $\eta$ in the energy equation~(\ref{eqnegy}) can be understood as follows. The dissipation of the magnetic field creates a change of the energy density of the form
\begin{equation}
\left. \frac{\partial  (\rho e)}{\partial t} \right|_\eta  = \vec{B}\cdot
\left.\frac{\partial\vec{B}}{\partial t}\right|_{\eta} + \eta \vec{J}^2 ,
\end{equation}
where the first term on the right hand side describes the change of the magnetic energy density due to magnetic field changes induced by $\eta$, while the second term describes the Ohmic heating due to the current. Using vector identities, this term can be rewritten as
\begin{equation}
  \left. \frac{\partial  (\rho e)}{\partial t} \right|_\eta  =
  \nabla \cdot \left[  \vec{B} \times (\eta \vec{J})\right],
\end{equation}
explaining the origin of the corresponding term in equation~(\ref{eqnegy}).

Finally, we recall that there is a further conservation law in ideal hydrodynamics, for the entropy of the gas. We express it as
\begin{equation}
  \frac{\partial (\rho s)}{\partial t}   + \nabla\cdot \left[  \rho s \vec{v} \right]=  \eta \frac{\gamma -1} {\rho^{\gamma-1}}\vec{J}^2 ,
  \label{eqnentropy}
\end{equation}
where
\begin{equation}
s = \frac{P}{\rho^\gamma}
\end{equation}
is an entropic function that labels the entropy per unit mass of the gas. This conservation law is strictly valid only for smooth parts of the flow, while in shocks entropy production needs to happen (here the differential form of the ideal hydrodynamical equations breaks down). Since equation~(\ref{eqnentropy}) is in principle redundant if the total energy equation is used, and because use of the energy equation automatically yields the  entropy production in shocks and from averaging over the fluid state in mesh cells, the entropy equation  is typically not considered in most numerical MHD schemes. However, we report it here as it offers a potential `backup' in situations where solving for the temperature with the total energy equation becomes inaccurate (see the discussion in Section~\ref{SecLowBetaFlow}).
  
\subsection{Vector potential}

The magnetic field may alternatively be obtained  as 
\begin{equation}
\vec{B} = \nabla \times \vec{A}
\end{equation}
in terms of the vector potential $\vec{A}$. In this case, the induction equation can be expressed as 
\begin{equation}
\frac{\partial\vec{A}}{\partial t} = \vec{v} \times \vec{B} -
\nabla\psi -\eta \vec{J},
\label{eqninduction}
\end{equation}
where $\psi$ is an arbitrary scalar field, representing the gauge freedom in the choice of $\vec{A}$. Previous uses of the vector potential for magnetohydrodynamics have often employed the Weyl gauge, with $\psi =0$. As we will discuss in more detail below, we will not adopt this here but deliberately employ a  non-trivial gauge for the moving mesh case.

Use of the vector potential to follow MHD side-steps the problem of the appearance of magnetic monopoles, because the divergence of the curl of a field is always zero, at least in the continuum. In practice, a discretized version of the curl operator needs to be used, which may still incur some localized divergence error, unless special structure-preserving mimetic operators are used \citep[e.g.][]{Adler2021}.   Taking a numerical derivative is also often associated with an amplification of noise, which is a potential disadvantage. Furthermore, the induction equation for the vector potential cannot be expressed as a conservation law, simply because the volume averaged vector potential is not conserved. Another minor downside is that periodic boundaries for $\vec{A}$ imply a vanishing mean magnetic field, seemingly preventing problems with a non-vanishing mean background field. However, this can be easily cured, as we discuss later on.

An advantage of the vector potential is that it can be easily used to set initial conditions for a divergence-free $\vec{B}$-field. Likewise, even when $\vec{A}$ is perturbed locally, e.g., by local mesh refinement or de-refinement operations, the derived $\vec{B}$-field will always remain divergence-free, suggesting in principle a path to a fairly robust representation of magnetic fields.

\subsection{Gauge freedom}

As we will discuss in more detail below, an important consideration for our moving mesh formulation is the ability to solve the Riemann problems at cell interfaces in the frames of the moving faces themselves. The resulting fluxes however need to be transformed to a common frame (taken as the lab frame), so that they can be co-added and used for a consistent update of the conserved variables. This is done in the standard {\small AREPO} code for momentum and energy fluxes, whereas for mass fluxes no such conversion is needed as they are invariant under Galilei transformations. For MHD, magnetic fluxes likewise need no change as the magnetic field is invariant under (non-relativistic) Galilei transformations.

But what about the vector potential, and the electric field? Clearly, the electric field governing the evolution of $\vec{A}$ is not invariant under a Galilei transformation, as it directly depends on the velocity. If we want to evolve the vector potential on a moving mesh, we need to understand how $\vec{A}$ transforms between frames that move relative to each other. To this end, consider a frame that moves with constant velocity $\vec{v}_{\rm boost}$ relative to the lab frame. Without loss of generality, assume that the vector potential in the moving frame, $\vec{A}'(\vec{x}', t)$ coincides with the vector potential $\vec{A}(\vec{x}, t)$ of the lab frame at $t=0$, i.e.~we  have $\vec{A}'(\vec{x}', 0) = \vec{A}(\vec{x}, 0)$ at this moment. At some later time, the coordinates are related by $\vec{x}' = \vec{x} - \vec{v}_{\rm boost} t$, but in general, the evolution created by the induction equation~(\ref{eqninduction}) in the two frames with a choice of $\psi = 0$ will not necessarily maintain $\vec{A}'(\vec{x}', t) = \vec{A}(\vec{x}, t)$ at later times (this invariance is always given, however, for the magnetic field). 

This property can be recovered, however, with the gauge choice
\begin{equation}
\psi = \vec{v}\cdot \vec{A},
\end{equation}
which is called the helical or advective gauge \citep{Brandenburg2010,  Candelaresi2011}. This recovers an invariant vector potential in the above sense. While this is not evident from the induction equation in the standard form,
\begin{equation}
\frac{\partial\vec{A}}{\partial t} = \vec{v} \times \vec{B} - \nabla (\vec{v} \cdot \vec{A})  -\eta \vec{J},
\label{eqAnewton}
\end{equation}
and its convective derivative variant,
\begin{equation}
\left.\frac{{\rm d}\vec{A}}{{\rm d}t}\right|_{\vec{v}} = \frac{\partial\vec{A}}{\partial t}  + (\vec{v}\cdot\nabla) \vec{A},
\end{equation}
the latter can be can be reformulated  with the help of vector calculus identities \footnote{
Such as
$\nabla(\vec{v}\cdot\vec{A}) = \vec{v}\times(\nabla\times\vec{A}) + \vec{A}\times(\nabla\times\vec{v}) + (\vec{v}\cdot\nabla)\vec{A} + (\vec{A}\cdot\nabla)\vec{v}$.}
 as
\begin{equation}
\left. \frac{{\rm d}\vec{A}}{{\rm d}t}\right|_{\vec{v}} = - \vec{A}\cdot \nabla\vec{v}   -\eta \vec{J},
\label{eqnDADt}
\end{equation}
where 
\begin{equation}
(\nabla \vec{v})_{ij} = \frac{\partial {v}_i}{\partial {r}_j}
\end{equation}
is the Jacobian of the velocity field\footnote{$\vec{A}\cdot \nabla\vec{v}$ is here a vector with components $(\vec{A}\cdot \nabla\vec{v})_j = \sum_i A_i \frac{\partial {v}_i}{\partial {r}_j}$.}, 
and ${\rm d}/{{\rm d}t}|_{\vec{v}}$ denotes the convective derivative based on the local gas velocity. Interestingly, in the form (\ref{eqnDADt}) of the induction equation, which \citet{Brandenburg2010} called the `A-method', the spatial derivatives appear only in the velocities. A constant velocity boost will not change these derivatives, making it explicit that the evolution of the vector potential under this choice is Galilean-invariant. 

Using the formulation (\ref{eqnDADt}) directly for computing the evolution of $\vec{A}$ seems attractive at first sight, it is however fraught with numerical problems in our experience, and this matches various failure reports in the literature \citep[e.g.][]{Price2010, Tricco2023}. In particular, as the physical velocity field does not need to be continuous, the velocity derivatives are numerically more problematic than the alternative source term  $\vec{v}\times \vec{B}$, where discontinuities are treated via the solution of a Riemann problem at cell surfaces, in addition to providing a way to include proper upwinding into the numerical treatment. 

It may thus seem more promising to use equation~(\ref{eqAnewton}) directly for evolving the helical vector potential. But by taking the curl to get the magnetic field, we see that the gauge term contribution to the magnetic field must vanish, because a curl of a gradient is always zero. By its very nature the gauge term should not be responsible for driving a physical evolution of the magnetic field. When we directly discretize equation~(\ref{eqAnewton}) to evolve the vector potential values associated with stationary cells, we cannot safely avoid this. This is because the contribution created by the gauge term in the evolution of the discretized field will at the numerical level not be guaranteed to have vanishing curl. The situation is analogous to the problems encountered when one directly evolves a discretized version of $\vec{B}$.  Even though it is supposed to have vanishing divergence as it is the curl of a vector field, this property is not manifestly preserved at the discretized level.  

Numerical stability in the advective gauge can be recovered by computing the evolution of the scalar gauge potential itself (as opposed to its derivative), just as the vector potential comes to rescue for the magnetic field. In fact, if we follow the vector potential $\vec{A^{\rm W}}$ in the Weyl gauge with
\begin{equation}
\frac{\partial\vec{A^{\rm W}}}{\partial t} = \vec{v} \times \vec{B}  -\eta \vec{J},
\end{equation}
we can at the same time evolve a passive gauge field $\Lambda$ as
\begin{equation}
\left.\frac{{\rm d} \Lambda}{{\rm d}t}\right|_{\vec{v}} = - \vec{v}\cdot \vec{A^{\rm W}}
\label{eqnlambda}
\end{equation}
Using its time integral, $-\Lambda \int^t_0  \vec{v}\cdot \vec{A^{\rm W}} {\rm d}t$, one can then show \citep{Candelaresi2011} that the helical potential and the Weyl potential are related by a gauge transformation of the form
\begin{equation}
\vec{A} = \vec{A^{\rm W}} - \nabla \Lambda .
\label{eqngaugetransform}
\end{equation}
One numerical stable path to obtain the vector potential in the advective gauge for a stationary mesh is therefore to simply solve for $\vec{A}^{{\rm W}}$ in the Weyl gauge, but in addition to passively integrate equation~(\ref{eqnlambda}) in time. At the end, one may then use equation~(\ref{eqngaugetransform}) to obtain the helical vector potential from the Weyl vector potential via a gauge transformation. But of course, if we are not specifically interested in helicity (see below), the Weyl potential suffices and we may refrain from computing the gauge term in our numerical implementation to begin with. After all, for a fixed mesh we can always stay in the lab-frame, making it unnecessary to have a representation of the vector potential that can be readily expressed in a boosted frame. 

The situation changes again, however, when one  considers the moving-mesh case, where we would like to do our calculations in local rest frames of cells to achieve Galilean invariance. This requires some form of the advective gauge without spoiling numerical stability. We will describe our suggestion to solve this in the next section.

Before coming to this, note that for a purely two-dimensional problem, for example in the $xy$-plane with translational invariance in the  $z$-direction and $v_z = 0$, only the $A_z$ component of the vector potential is non-zero. Then the induction equation (assuming $\eta = 0$) actually becomes 
\begin{equation}
\left. \frac{{\rm d} A_z}{{\rm d}t}\right|_{\vec{v}} = 0,
\end{equation}
highlighting that the entire magnetic field dynamics can be understood in this case by a transport of $A_z$ with the local fluid elements,  which is reminiscent of the Euler potentials. But unlike for them, for the vector potential we can still easily prescribe dissipation. Furthermore, in this 2D case the gauge field $\psi = \vec{v}\cdot\vec{A}$ always vanishes, underlining the remarkable simplicity of the 2D case compared to the situation in 3D.

Another aspect worth mentioning in passing concerns the magnetic helicity density
\begin{equation}
h = \vec{A}\cdot \vec{B} ,
\end{equation}
which in general depends on the gauge choice for $\vec{A}$. For the advective gauge choice, the helicity is however Galilean-invariant, which underlines the special character of this gauge.  Furthermore, one can derive \citep{Candelaresi2011} the evolution equation 
\begin{equation}
\frac{\partial h}{\partial t} + \nabla\cdot \left[ h\vec{v} \right] = -\nabla \cdot [\eta \vec{J}\times \vec{A}] - 2\eta \vec{J}\cdot\vec{B}
\end{equation}
for the helicity in this gauge. This shows that for ideal MHD with $\eta = 0$, the helicity obeys a conservation law, and total helicity
\begin{equation}
{\cal H} = \int h\, {\rm d} V
\end{equation}
is a conserved quantity. Magnetic helicity is a crucial theoretical concept for understanding turbulent dynamos. Having direct access to this quantity in a numerical MHD formulation should therefore be advantageous for making full use of it in the analysis and testing of numerical simulations.

\begin{figure}
\resizebox{8.6cm}{!}{\includegraphics{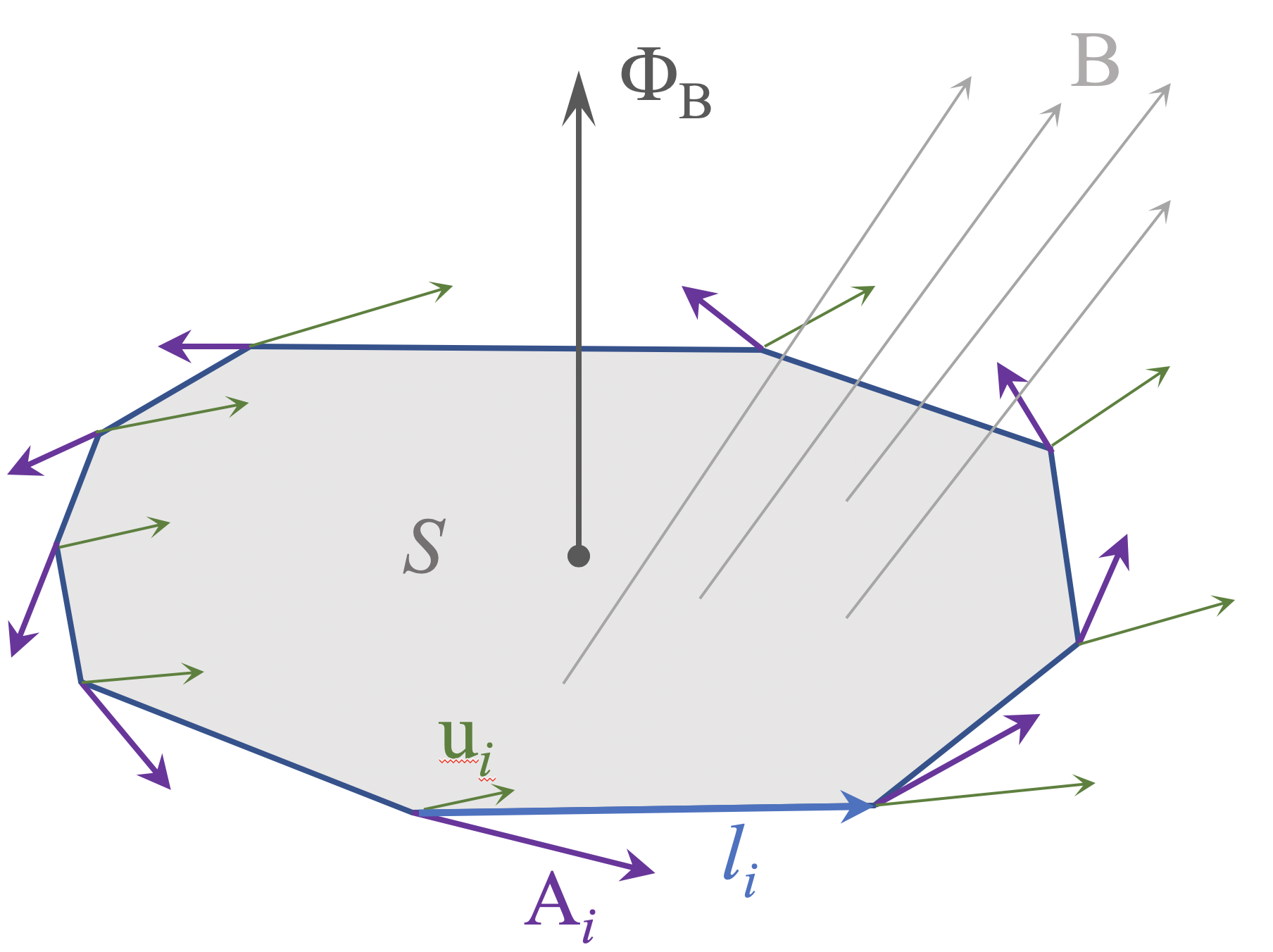}}%
\caption{Sketch of the magnetic flux through a face $S$ (shaded area), that is here approximated
as a polygon with edge vectors $\vec{l}_i$, each pointing from a corner $i$ to the next corner $i+1$. We assume that the vector potential is specified at the boundary of the face, with known values $\vec{A}_i$ at the corners of the polygon. These corners may move with velocities $\vec{u}_i$. The magnetic flux $\Phi_{\rm B}$ through the face can be obtained as a contour integral over $\vec{A}$, and its total time derivative is given by equation~(\ref{eqnfluxchange}) in the continuum limit.
\label{FigBflux}}
\end{figure} 

\section{The vector potential on a moving mesh} \label{secA}

To better motivate the particular gauge we are going to adopt in this work, let us consider the magnetic flux through a surface $S$ in a general unstructured moving mesh, as sketched in Figure~\ref{FigBflux}. We assume that the vector potential is defined at the boundary of the face, and that this boundary moves with a velocity field $\vec{u}$. The magnetic flux trough the surface is given by
\begin{equation}
  \Phi_{\rm B} = \iint_S \vec{B}\cdot {\rm d}\vec{S} =
  \oint_{\partial S} \vec{A}\cdot {{\rm d}\vec{l}} , \label{eqnperimeter}
  \end{equation}
where the second form expresses the flux as a line integral of the vector potential along the boundary of $S$. We may also compute the total time derivative of the magnetic flux \citep{Mocz2014}. This is given by 
\begin{equation}
\frac{{\rm d}\Phi_{\rm B}}{{\rm d}t} =\oint_{\partial S} [(\vec{v} - \vec{u}) \times \vec{B} ]\cdot {{\rm d}\vec{l}} ,\label{eqnfluxchange}
\end{equation}
where the ordinary gas velocity is denoted with $\vec{v}$ whereas $\vec{u}$ accounts for the boundary's motion.

For definiteness, we can approximate the geometry of the face through a polyhedron as shown in the sketch, with a set of discrete vector potential values $\vec{A}_i$ and edge vectors $\vec{l}_i$.  Let us now express the magnetic flux in terms of this polyhedral approximation, treating each line segment with a constant vector potential. This yields a sum of the form
\begin{equation}
   \Phi_{\rm B} \simeq \sum_i \vec{A}_i \cdot {\vec{l}_i} ,
\end{equation}
which is appropriate if the spatial variation of $\vec{A}_i$ is negligibly small over the scale of each segment\footnote{For simplicity of notation, we have refrained from using the edge-averaged vector potential in the discretisation instead of one-sided values. This makes no difference for the argument.}. Suppose now that the spatial locations at which we measure the $\vec{A}_i$ on the contour are moving according to a velocity field $\vec{u}$, and that this velocity field is smooth and differentiable, and varies only linearly over the scale of the face. Then the rate of change of the edge vectors is
\begin{equation}
   \frac{{\rm d} \vec{l}_i}{{\rm d}t} = \nabla\vec{u} \cdot  \vec{l}_i,
\end{equation}
where $\nabla\vec{u}$ is the Jacobian\footnote{Note that this Jacobian can be decomposed into an antisymmetric and symmetric part. The antisymmetric part can in turn be interpreted as a solid body rotation, while the symmetric part can be seen to create stretching/compression along three orthogonal directions.} of the smooth velocity field $\vec{u}$ with which the geometry of our surface being moved and distorted. The time derivative of the flux is then 
\begin{equation}
  \frac{{\rm d}\Phi_{\rm B}}{{\rm d}t} =
  \sum_i  \left(    \frac{{\rm d}\vec{A}_i}{{\rm d}t}    \cdot {\vec{l}_i} +  \vec{A}_i \cdot \frac{{\rm d} \, \vec{l}_i}{{\rm d}t}\right) .\label{eqnfacemov1}
\end{equation}
On the other hand, we may also discretize equation~(\ref{eqnfluxchange}) directly, yielding
\begin{equation}
  \frac{{\rm d}\Phi_{\rm B}}{{\rm d}t} \simeq \sum_i [(\vec{v}_i - \vec{u}_i) \times \vec{B}_i ]\cdot {\vec{l}_i}
  \label{eqnfacemov2} .
\end{equation}
Consistency between equation~(\ref{eqnfacemov1}) and (\ref{eqnfacemov2}) for arbitrary geometry of the polyhedron requires that we evolve the vector potential according to
\begin{equation}
  \left. \frac{{\rm d}\vec{A}}{{\rm d}t}\right|_{\vec{u}} = (\vec{v} - \vec{u}) \times \vec{B} -   \vec{A} \cdot \nabla\vec{u},
  \label{eqnbasicdadt}
\end{equation}
where we omitted the index $i$ to emphasize the generality of this result. The subscript marker on the total time derivate on the left hand side is meant as a reminder that this convective derivative is taken with respect to the velocity $\vec{u}$.

This evolution equation can just as well be derived from equation~(\ref{eqninduction}) by adopting a gauge equal to
\begin{equation}
  \psi= \vec{u}\cdot \vec{A},
\end{equation}
and by invoking the definition of the convective derivative
\begin{equation}
  \left. \frac{{\rm d}\vec{A}}{{\rm d}t}\right|_{\vec{u}} = \frac{\partial\vec{A}} {\partial t} + 
  (\vec{u}\cdot\nabla)\vec{A}
\end{equation}
for the velocity $\vec{u}$. Note that if one sets $\vec{u}$ equal to the gas velocity $\vec{v}$ itself, one recovers the earlier result of equation~(\ref{eqnDADt}). However, the above derivation underlines once more why this is not a good idea in practice. As we already pointed out earlier,  the gas velocity may contain physical discontinuities and is in general not sufficiently smooth to make the above work for discretized mesh loops of finite size. In a general polyhedral mesh where the vector potential is discretized in terms of values at mesh vertices, multiple edges (in particular more than three linearly independent directions) furthermore emanate from any given vertex. It is thus critical that the local mesh distortion can be fully described by the Jacobian $\nabla \vec{u}$ {\em alone}, otherwise the flux changes for all adjacent faces determined by equation~(\ref{eqnfacemov2}) could not be mapped back to a unique change of the  vector potential value at the vertex.

But how should the field $\vec{u}$ then be chosen? The primary requirement we need is that it is a smooth velocity field with a well defined Jacobian. In order to carry out the time integration accurately and stably (see below), we also require that this Jacobian is approximately constant over the scale of a cell. In order to obtain Galilean invariance, we furthermore need to tie the definition of $\vec{u}$ to the regular gas velocity field $\vec{v}$, such that the difference $\vec{v} - \vec{u}$ is invariant under a constant velocity boost. This can be straightforwardly achieved if $\vec{u}$ is defined as some smoothed version of the gas velocity, for example filtered on the scale of a few mesh cells. We will later discuss our practical choices for this. 

In practice, it will furthermore be useful if the motion of individual mesh vertices can deviate from $\vec{u}$. To allow for this, we introduce a further velocity that describes the actual motion of the points where the vector potential is stored\footnote{These points could be chosen as the mesh generating points of our Voronoi tessellation, or be identified with the geometric centers of cells. We will opt for the latter.}. This velocity may not necessarily coincide with the smooth velocity field -- in fact, in {\small AREPO}, the velocity with which the mesh-generating points move will typically be chosen to be equal to the gas velocity $\vec{v}$ of the {\em local cell}, modulo an optional mesh correction velocity that can change from cell to cell. The velocity $\vec{w}$ with which the geometric centers of cells move will in general slightly deviate from the velocity with which the mesh generators move, and thus in turn also both from $\vec{v}$ and $\vec{u}$. To account for this, the evolution equation~(\ref{eqnbasicdadt}) for the vector potential acquires an additional purely advective term. It then takes on the  form
\begin{equation}
\left.  \frac{{\rm d}\vec{A}} {{\rm d}t} \right|_{\vec{w}}
  =  
  (\vec{v} - \vec{u}) \times {\vec{B}}  - \vec{A}\cdot \nabla \vec{u}  +  
  [(\vec{w}-\vec{u})\cdot\nabla]\vec{A}  -\eta \vec{J}
  \label{eqncentral},
\end{equation}
where we have also reintroduced a dissipative term if there is a non-vanishing resistivity $\eta$. This form of the induction equation is the central evolution equation for the vector potential that we propose for numerical discretization in this paper.

The form of the equation makes it explicit that the evolution of $\vec{A}$ is  Galilean invariant, because any velocity boost will not change the velocity differences appearing in the first and third terms on the right hand side, and neither will it affect the velocity gradients appearing in the second term on the right-hand side. Importantly, gradients directly acting on the velocities appear only for the smoothed version of the velocity field, which we specifically introduced to allow for this without severe numerical problems. 

Note that all four terms on the right hand side of  equation~(\ref{eqncentral}) have a straightforward physical interpretation. The first term  can be understood as accounting for  non-trivial inductive changes of the magnetic field, while the second term takes care of purely geometric modifications due to a local stretching, compression or rotation of the field representation. The third term is just accounting for additional shifts of the reference points where we store the values of the vector potential. Finally, the fourth term is an optional Ohmic loss term in case one wants to include such a deviation from ideal MHD. 

It is clear that each of these terms requires a careful choice of time integration to ensure numerical stability. But even if this is done, it turns out that the time integration is still tricky and prone to develop instabilities, particularly after some simulation time has passed. The origin of these instabilities is not readily apparent from the form of equation~(\ref{eqncentral}), but casting it into the  alternative form 
\begin{equation}
  \frac{\partial\vec{A}} {\partial t} 
  =  
  \vec{v}  \times {\vec{B}}  - \nabla (\vec{u}\cdot\vec{A})   -\eta \vec{J}
  \label{eqnaltform}
\end{equation}
makes again explicit that we are using a gauge field $\psi =\vec{u}\cdot\vec{A}$. While the adoption of this gauge does not change the magnetic field itself, it does modify the evolution of the vector potential. In particular, it injects divergence into $\vec{A}$, which can be seen by taking the divergence of equation~(\ref{eqnaltform}). Adopting $\eta = 0$ for simplicity, we see that $\nabla \cdot\vec{A}$ evolves as 
\begin{equation}
\frac{\partial}{\partial t}\nabla\cdot\vec{A} =  \nabla\cdot  (\vec{v}  \times {\vec{B}})    -\nabla^2(\vec{u}\cdot\vec{A}) \simeq -\nabla^2 \psi.  \label{eqndivAevolv}
\end{equation}
The term with the gauge field $\psi$ is expected to strongly dominate as we have $\nabla\cdot  (\vec{v}  \times {\vec{B}}) = -\nabla \cdot \vec{E} \simeq 0$ in ideal MHD without an accumulation of a macroscopic charge density. We thus see that the Laplacian of the gauge scalar is a source of divergence in $\vec{A}$.

Numerically, this injection of irrotational modes into the vector potential can make it hard, however, for the curl-operator to reliably project them out, particularly if the divergence in $\vec{A}$ becomes very large, which may well happen according to equation~(\ref{eqndivAevolv}). For numerical stability, it is therefore highly desirable to prevent such a situation and make sure that $\nabla\cdot \vec{A}$ stays small. In other words: We would like to have a divergence control for $\vec{A}$. 

This sounds eerily familiar to the problem of divergence control for the $\vec{B}$-field in schemes that directly evolve the magnetic field. Have we thus not gained anything by moving to the vector potential? Note, however, that there is an important difference. As long as the divergence in the vector potential is sufficiently small\footnote{A finite value for $\nabla\cdot\vec{A}$ is perfectly permissible, but if this becomes large against $\nabla \times \vec{A}$, there is a danger that numerical finite difference estimates for $\nabla \times \vec{A}$ become polluted by $\nabla\cdot\vec{A}$, causing numerical errors or instability.}, there is no error in the induction equation and the magnetic field at all  -- unlike when the $\vec{B}$-field is directly evolved.  We think this makes the use of the vector potential combined with a cleaning method clearly superior compared to corresponding techniques applied to the magnetic field. The latter aim to limit the impact of errors that are already present in the magnetic field, whereas in the vector potential case we aim to prevent errors that could arise from limitations of a numerical discretization operator.  

For ensuring a small divergence of $\vec{A}$ one can use similar techniques as developed for the $\vec{B}$-field. For example, one could regularly project out the divergence in $\vec{A}$ and establish and maintain a Coulomb gauge in this way. Because second partial derivatives are not guaranteed to permute in our numerical discretization, this however risks to introduce unwanted perturbations into the magnetic field. Alternatively, one can use a variant of the hyperbolic cleaning technique of Dedner, which we adopt here. To this end, we introduce a further gauge scalar with convective evolution equation
\begin{equation}
\left.\frac{{\rm d}\phi}{{\rm d}t}\right|_{\vec{w}} = \alpha c_s^2\, \nabla \cdot \vec{A} - \frac{\beta}{\tau}\phi.
\label{eqncleaningscalar}
\end{equation}
Here $c_s$ is the local soundspeed, and
$\tau$ is a local decay timescale, which in practice we parametrize as $\tau = {r_{\rm cell}}/{c_s}$, where $r_{\rm cell}$ is the local cell radius. The dimensionless coefficients $\alpha$ and $\beta$ regulate how fast $\phi$ responds to a non-zero divergence in $\vec{A}$, and how fast it decays again. Our default values are $\alpha=1$ and $\beta=0.15$.  We furthermore couple $\phi$ to the induction equation in the form
\begin{align}
\left.  \frac{{\rm d}\vec{A}} {{\rm d}t} \right|_{\vec{w}}
  =  
  (\vec{v} - \vec{u}) \times {\vec{B}}  - \vec{A}\cdot \nabla \vec{u}  +  
  [(\vec{w}-\vec{u})\cdot\nabla]\vec{A}  -\eta \vec{J} + \nabla \phi ,
  \label{eqncentral2}
\end{align}
without changing the $\vec{B}$-field itself. The coupled dynamics of the equations~(\ref{eqncleaningscalar}) and (\ref{eqncentral2}) can then limit the size of the term $\nabla\cdot\vec{A}$, so the equations we actually integrate for the evolution of $\vec{A}$ is thus equation (\ref{eqncentral2}) combined with equation~(\ref{eqncleaningscalar}).

If we neglect the convective terms for the moment (and assume $\eta=0$), we can combine the time-derivative of equation~(\ref{eqncleaningscalar}) and the curl of equation~(\ref{eqncentral2}) to obtain
\begin{align}
\frac{\partial^2\phi}{\partial t^2} &= \alpha c_s^2 ( \nabla^2 \phi - \nabla^2\psi) - \beta \frac{c_s}{r_{\rm cell}} \frac{\partial \phi}{\partial t} \nonumber \\
& +\left( \frac{\partial \phi}{\partial t} + \beta \frac{c_s}{r_{\rm cell}}\right) \frac{\partial \ln c_s^2}{\partial t} -\beta \phi \frac{\partial}{\partial t} \frac{c_s}{r_{\rm cell}} . \label{eqncleaning}
\end{align}
In the second line we have kept terms that arise from temporal variations of the so-called transport speeds if these are not kept constant spatially. \citet{Tomida2026} have recently argued that these terms can compromise the accuracy of the Dedner cleaning approach, especially when the transport coefficient have an explicit timestep dependence as used in \citet{Tomida2026}. Note, however, that we do not have such a dependence. It is therefore not clear whether spatial/temporal variations of the transport coefficients spoil the $\nabla\cdot\vec{A}$ cleaning approach we use here. So far we have not observed any evidence for this in our simulations. In any case, if we neglect the corresponding terms in the second row of equation~(\ref{eqncleaning}), we obtain a modified telegraph equation
\begin{align}
\frac{r_{\rm cell}}{\beta c_s^2}\frac{\partial^2\phi}{\partial t^2} 
 + \frac{\partial \phi}{\partial t} = \frac{\alpha c_s r_{\rm cell}}{\beta}   \nabla^2 (\phi - \psi)
\end{align}
with diffusivity 
\begin{equation}
D = \frac{\alpha c_s r_{\rm cell}}{\beta}.
\end{equation}
Note that unlike in the normal case of Dedner cleaning for the magnetic field, we here have an explicit source term $\psi= \vec{u}\cdot\vec{A}$ for the injection of divergence in $\vec{A}$. But $\psi$ itself is influenced by $\phi$, because the latter can modify the vector potential. Overall, we simultaneously expect that the combined dynamics will evolve $\phi$ and $\psi$ towards an attractor solution where
\begin{equation}
\nabla^2 \phi \simeq \nabla^2\psi,
\label{eqnlabplacephilaplacepsi}
\end{equation}
and furthermore, we expect that $\phi$ will adjust to a value of order
\begin{equation}
\phi \sim  D \, \nabla\cdot\vec{A}
\label{eqnphinablaA}
\end{equation}
based on equation~(\ref{eqncleaningscalar}). These two conditions in turn should limit the expected residual magnitude of the divergence in $\vec{A}$ to something of order
\begin{equation}
\nabla\cdot \vec{A} \sim \frac{1}{D} \psi, 
\label{eqndivAbeta}
\end{equation}
where we assumed that the velocity field gauge scalar $\psi = \vec{u}\cdot\vec{A}$ is not modified significantly by $\phi$ but rather stays largely fixed independent of the cleaning. This suggests that the diffusivity $D$ should ideally be as large as possible within the boundaries of an explicit time integration scheme, whereas the hyperbolic transport speed ($c_s$) itself is of lesser importance. We shall later test explicitly to what extent these expectations are borne out. 

We note that one could augment our representation of the smoothed large-scale velocity field $\vec{u}$ to also determine smoothly varying fields for $c_s'$ and $r_{\rm cell}'$, and to compute the transport coefficients with these, in order to guard against sharp local variations of the transport speeds, as recommended by \citet{Tomida2026}. We thus far have not found it necessary to adopt this, however. To make sure that our local timestepping always fulfils the stability constraints for hyperbolic and diffusive transport carried out with the corresponding speeds, we impose the following additional timestep criteria:
\begin{equation}
\Delta t < \frac{r_{\rm cell}}{\alpha c_s'}
\end{equation}
and
\begin{equation}
\Delta t < \frac{r_{\rm cell}^2}{D'},
\end{equation}
where $r_{\rm cell}$ is the actual size of a local cell. We note that these criteria are normally fulfilled already by the Courant timestep for our default  parameter choices for the cleaning.

\begin{figure}
\centering \resizebox{8.0cm}{!}{\includegraphics{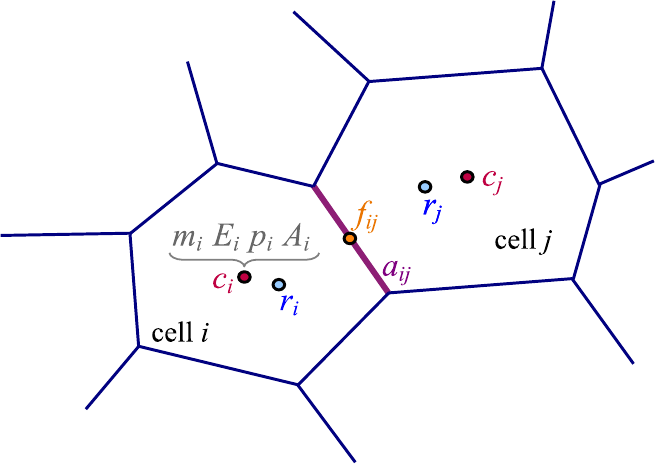}}
\caption{Local environment of cell $i$, illustrating the locations of various points that are relevant for our discretization approach on a dynamic Voronoi grid. The geometric
cell centres $\vec{c}_i$ (red circles) are in general offset from the locations $\vec{r}_i$ of the mesh-generating points themselves (blue circles). Between two cells $i$ and $j$, there is a facet of area $a_{ij}$, width midpoint $\vec{f}_{ij}$ (orange circle). The conserved fluid quantities mass $m_i$, energy $E_i$, and momentum $\vec{p}_i$ are used to reconstruct the primitive quantities density $\rho_i$, velocity $\vec{v}_i$ and pressure $P_i$, all expanded around the geometric cell centre. The vector potential value $\vec{A}_i$ of a cell is likewise associated with the coordinate of the geometric cell centre. \label{FigSketchCell}}
\end{figure}

\section{Discretization strategies}
\label{SecNewApproach}

We now describe the central practical aspects of our new formulation of MHD on an unstructured and moving Voronoi mesh.  We first discuss how we derive the magnetic field from $\vec{A}$ and then how we treat the time evolution equation of the vector potential. After addressing both of these points, we discuss how we cope with a non-vanishing average magnetic field in the presence of periodic boundary conditions. Later we shall also discuss how the scheme can be refined to add  dissipation if desired, to ensure entropy stability in low-$\beta$ flow, and how to generalize it to cope with local timesteps that can be different for neighbouring cells. In Section~\ref{SecDetails}, we furthermore give a detailed, step-by-step summary of our method.

\subsection{Basic mesh geometry}

The sketch in Figure~\ref{FigSketchCell}, for simplicity drawn in 2D, illustrates the basic cell geometry we are concerned with. For every Voronoi cell $i$, we store the coordinate $\vec{r}_i$ of its mesh-generating point, and the conserved quantities given by mass $m_i$, momentum $p_i$, and energy $E_i$ of the cell. Upon construction of the corresponding Voronoi mesh (based just on the $\vec{r}_i$ coordinates), the volumes and geometric centers $\vec{c}_i$ of all cells become known. The vector potential value $\vec{A}_i$ is associated with the location $\vec{c}_i$. The three conserved quantities then give rise to cell-averaged densities and velocities, and, once the cell-averaged magnetic field is known, we can also obtain the average temperature from the thermal energy, with the latter being computed by subtracting the kinetic and magnetic energies from the total energy. 

Evidently, for our vector potential formulation we need to explicitly specify how we obtain the magnetic field from the vector potential at the discretized level. We shall discuss this important question next.

\subsection{Estimation of the magnetic field and its derivatives}
\label{subsecbderivative}

One conceptual disadvantage of using the vector potential is that the actual physical quantity of interest, the magnetic field, only arises by computing a numerical derivative. Furthermore, for spatial reconstructions within cells and for computing the current $\vec{J}$, one needs spatial derivatives of the magnetic field, which  in turn correspond to second derivatives of the vector potential. A linear reconstruction of the current density within cells (if needed) involves already third derivatives of the vector potential. Fortunately, the vector potential is continuous and comparatively smooth, which somewhat mitigates the numerical challenges associated with these derivatives.

For obtaining numerical derivatives on our unstructured Voronoi mesh, several different strategies can be used. One is to estimate the first and second derivatives of the vector potential simultaneously, through a higher-order least square reconstruction. In this case, one can also guarantee that second-order partial derivatives with respect to different dimensions commute at the discretized numerical level. This approach requires, however, a comparatively large computational stencil that can manifest in a loss of resolving power at sharp features in the flow. A considerably simpler approach that avoids this disadvantage is to use a least square gradient estimator on a smaller stencil for the first derivative only, and to  compute second derivatives simply by applying this operator twice successively. This approach is also computationally more efficient, but the resulting second derivatives can be  noisier.

For the special case of the magnetic field, it is also possible to carry out an integral construction to directly obtain face-averaged normal components of the $\vec{B}$-field, and then to extend these to volume averaged magnetic fields, as pursued by many constrained transport methods and by \citet{Mocz2016} for a Voronoi mesh. We shall discuss these different approaches in turn, and motivate the choices we have made in this study.

\subsubsection{Least square gradient estimate of the $\vec{B}$-field} \label{secstdlsq}

In the {\small AREPO} code, \citet{Pakmor2016} replaced the formerly used Green-Gauss gradient estimates \citep{Springel2010} with linear least-square reconstructions of the target quantities around a cell and its immediate Voronoi neighbours, as this proved to yield more accurate gradients for general mesh geometries. The corresponding fits can be devised as interpolating polynomials that are forced to go through the values of the central point \citep[as in][]{Pakmor2016}, or instead be constructed as a fit that only depends on the neighbouring values and does not necessarily go through the values of the central cell \citep[see also][]{Maron2012}.

For reasonably regular cell geometries, this makes hardly any difference in practice. Still, we here opt for the latter, on the grounds that this approach is slightly more resilient to noise in the considered field. We thus use an ansatz of the form 
\begin{equation}
\vec{A}^{{\rm fit}}_i(\vec{x}) = \overline{\vec{A}}_i + \nabla\vec{A}_i \cdot(\vec{x} - \vec{c}_i),
\label{eqnAfit}
\end{equation}
where $\overline{\vec{A}}_i$ is the reconstructed value at the location of cell $i$'s center of mass, 
and
\begin{equation}
\nabla\vec{A}_i = \left(
\begin{array}{ccc}
\frac{\partial A_x}{\partial x} & \frac{\partial A_x}{\partial y}  & \frac{\partial A_x}{\partial z}\vspace*{0.15cm} \\
\frac{\partial A_y}{\partial x} & \frac{\partial A_y}{\partial y}  & \frac{\partial A_y}{\partial z} \vspace*{0.15cm}\\
\frac{\partial A_z}{\partial x} & \frac{\partial A_z}{\partial y}  & \frac{\partial A_z}{\partial z} \\
\end{array}
\right) 
\end{equation}
is the cell's Jacobian of the vector potential. We next define a quadratic error function
\begin{equation}
L_i = \sum_{{j\, \in\, {\rm neighbours\; of\;}i}} w_{ij} [\vec{A}_j - \vec{A}^{\rm fit}_i (\vec{c}_j) ]^2
\end{equation}
and treat the coefficients of the Jacobian and $\overline{\vec{A}}_i$ as unknowns that we determine through minimization of $L_i$. For the weights $w_{ij}$, we use the area of the facet between cells $i$ and $j$.

The least-square problem separates into three independent and equal optimization problems, one for each dimension. Defining a coefficient vector 
\begin{equation}
\vec{X}_j  =  [1,\; 
(\vec{c}_j-\vec{c}_i)_x,\; (\vec{c}_j-\vec{c}_i)_y,\; (\vec{c}_j-\vec{c}_i)_z]
\end{equation}
for each neighbouring point $j$  based on the cell centres, we can compute a square matrix $\vec{M}_i$ encoding the local mesh geometry of cell $i$ as 
\begin{equation}
\vec{M}_i = \sum_j w_{ij} \vec{X}_j \vec{X}_j^T.
\end{equation}
The fit is then obtained as a solution of a linear system of equations. For example, for the $x$-coordinate, we have
\begin{equation}
\vec{M}_i \vec{\chi}_x = \vec{b}_x,
\label{eqnleastssqr}
\end{equation}
where the vector of unknowns is
\begin{equation}
\vec{\chi}_x = \biggr[ \overline{{A}}_{x}, \frac{\partial A_x}{\partial x} , \frac{\partial A_x}{\partial y}  , \frac{\partial A_x}{\partial z}  \biggr],
\end{equation}
and the right-hand side is
\begin{equation}
\vec{b}_x = \sum_j w_{ij} A_{x,j} \vec{X}_j.
\end{equation}
One may apply LU-decomposition to solve equation~(\ref{eqnleastssqr}) or invert the symmetric $4\times 4$ matrix $\vec{M}_i$ once. If the latter is done, we can also define
\begin{equation}
\vec{q}_{ij} = w_{ij} \vec{M}_i^{-1} \vec{X}_j,
\label{eqnvectorweight}
\end{equation}
and then readily apply this vector weight to compute 
any of the three  gradients, for example as
\begin{equation}
\vec{\chi}_{x} = \sum_j \vec{q}_{ij} A_{j,x}.
\end{equation}
The gradient of any scalar can thus simply be computed as a weighted sum over the values of the direct neighbours. These weights can be stored, allowing the  efficient computation of many gradients, if desired.

With the estimate of $\nabla\vec{A}$ in hand, we can then  set  
\begin{equation}
\vec{B}_i =
\left(
\begin{array}{c}
\frac{\partial A_z}{\partial y} - \frac{\partial A_y}{\partial z} \vspace*{0.15cm} \\
\frac{\partial A_x}{\partial z} - \frac{\partial A_z}{\partial x} \vspace*{0.15cm} \\
\frac{\partial A_y}{\partial x} - \frac{\partial A_x}{\partial y} 
\end{array}
\right)
\label{eqnBset}
\end{equation}
to define the cell-centred $\vec{B}$-field based on the definition of the curl. Note that we do not apply slope limiting to $\nabla\vec{A}$ when obtaining magnetic field values. We do, however, apply slope limiting to $\nabla\vec{A}$ when we use the gradients to extrapolate the vector potential to cell faces for the purpose of computing the advection term in equation~(\ref{eqAcentral}), as we discuss below.

For obtaining the Jacobian of $\vec{B}$ we can can use the standard approach in {\small AREPO} for estimating gradients of cell-centred quantities and apply it to $\vec{B}$, which effectively realizes a consecutive application of the linear derivative operator to obtain second derivatives of $\vec{A}$. For estimating the cell-centred current density, $\vec{J} = \nabla \times \vec{B}$, we again refrain from slope-limiting the  numerical derivates of $\vec{B}$. However, the estimated gradients of the magnetic field are slope-limited when used to extrapolate the magnetic field to cell surfaces, in order to avoid the introduction of new extrema. Note that this way of estimating the spatial derivatives of the magnetic field at cell centres does not guarantee  $(\nabla \cdot \vec{B})_i = 0$ to machine precision (unlike the method discussed in the next subsection), but this is inconsequential for the numerical stability of the method (note that it is also not fulfilled in constrained transport methods).

\begin{figure}
\centering
\resizebox{8.5cm}{!}{\includegraphics{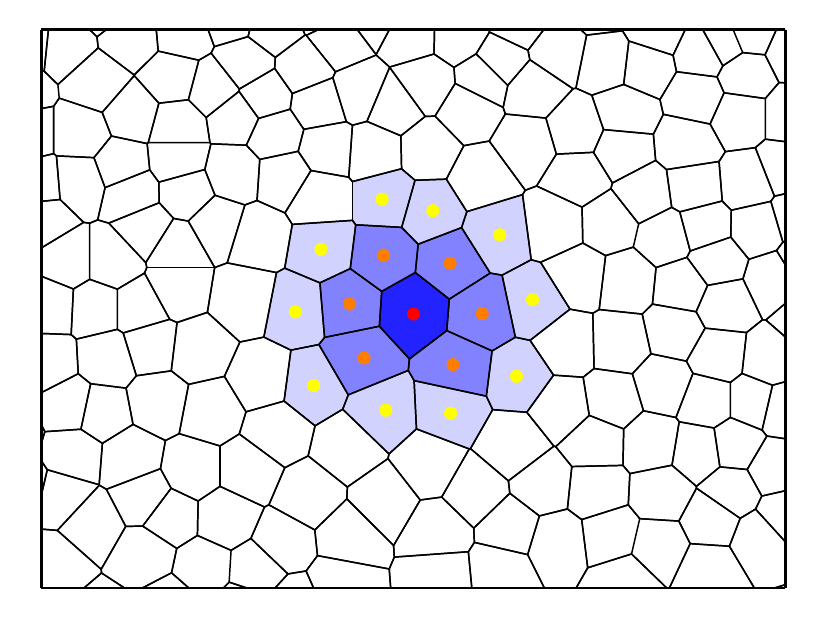}}%
\caption{Sketch of a `double stencil', which can optionally be used to estimate gradients and second-order derivatives of the vector potential in a single least square fit. For the central cell (shaded in dark blue), we consider  its immediate neighbours (cells marked with orange points) as the primary stencil (which is used in the normal gradient estimate). If we then add the neighbours of the neighbours of cell $i$ (marked by yellow points), we obtain an extended stencil that 
can be used for a least-square quadratic reconstruction of the vector potential that simultaneously estimates the Jacobian and Hessians of the vector potential. 
 \label{FigSketchDoubleStencil}}
\end{figure}

\subsubsection{Concurrent estimate of first and second derivatives}

As an alternative, we may also estimate first and second derivatives of the vector potential in one go,  with an ansatz of the form
\begin{equation}
\vec{A}^{{\rm fit}}_i(\vec{x}) = \overline{\vec{A}}_i + \nabla\vec{A}_i\cdot (\vec{x} - \vec{c}_i)
+\frac{1}{2} (\vec{x} - \vec{c}_i)^T  \cdot\nabla^2\vec{A}_i \cdot (\vec{x} - \vec{c}_i).
\label{eqnAfitDouble}
\end{equation}
Again, we do not force the fit to go precisely through the intrinsic value $\vec{A}_i$ stored for the target cell. Compared to the approach above,
we have added a  third-order tensor
$\nabla^2\vec{A} = (\nabla^2\vec{A}_x, \nabla^2\vec{A}_y, \nabla^2\vec{A}_z ) $   collecting the Hessians of the three dimensions, given for example for the $x$-coordinate by
\begin{equation}
\nabla^2\vec{A_x} = \left(
\begin{array}{ccc}
\frac{\partial^2 A_x}{\partial x^2} & \frac{\partial^2 A_x}{\partial x \partial y}  & \frac{\partial^2 A_x}{\partial x \partial z}\vspace*{0.15cm} \\
\frac{\partial^2 A_x}{\partial y \partial x} & \frac{\partial^2 A_x}{\partial y^2}  & \frac{\partial^2 A_x}{\partial y \partial z}\vspace*{0.15cm} \\
\frac{\partial^2 A_x}{\partial z \partial x} & \frac{\partial^2 A_x}{\partial z \partial y}  & \frac{\partial^2 A_x}{\partial z^2} \\
\end{array}
\right).
\end{equation}
Of course, we expect the Hessians to be symmetric, i.e.~the second-order partial derivatives with respect to different coordinate dimensions commute, something that we can build into our fit by adopting a symmetric form for the Hessians, with correspondingly fewer degrees of freedom. In order to robustly estimate the full set of coefficients of the Jacobian, the Hessians, and $\overline{\vec{A}}_i$, we need  a sufficiently large stencil. Here we choose the points of a `double-stencil' that includes neighbours of neighbours of $i$, as illustrated in Figure~\ref{FigSketchDoubleStencil}. We do this to ensure that our fit is always well-defined and in fact slightly overdetermined. 

The weights  $w_{ij}$  for each included point $j$ are in principle arbitrary. We here use an inverse distance weighting for definiteness. Note that we can use the offset between $\vec{A}_{i}^{\rm fit}(\vec{c}_i)= \overline{\vec{A}}_i$ and $\vec{A}_i$ as an indicator for the local smoothness and whether local dissipation (see Section~\ref{SecDissipation}) may be needed.

We now generalize the coefficient vector for a neighbouring point to
\begin{multline}
\vec{X}_j  =  \biggr[1,\; 
(\vec{c}_j-\vec{c}_i)_x,\; (\vec{c}_j-\vec{c}_i)_y,\; (\vec{c}_j-\vec{c}_i)_z,  \\%
\hspace*{1cm}\frac{1}{2}(\vec{c}_j-\vec{c}_i)_x^2, \frac{1}{2}(\vec{c}_j-\vec{c}_i)_y^2, \frac{1}{2}(\vec{c}_j-\vec{c}_i)_z^2, \\%
(\vec{c}_j-\vec{c}_i)_x (\vec{c}_j-\vec{c}_i)_y,\; (\vec{c}_j-\vec{c}_i)_x (\vec{c}_j-\vec{c}_i)_z,\; (\vec{c}_j-\vec{c}_i)_y (\vec{c}_j-\vec{c}_i)_z \biggr], 
\end{multline}
and the vector of unknowns becomes
\begin{multline}
\vec{\chi}_x = \biggr[ {A}_{0,x}, \frac{\partial A_x}{\partial x} , \frac{\partial A_x}{\partial y}  , \frac{\partial A_x}{\partial z},  \frac{\partial^2 A_x}{\partial x^2} , \frac{\partial^2 A_x}{\partial y^2}  , \frac{\partial^2 A_x}{\partial z^2},\\
\frac{\partial^2 A_x}{\partial x\partial y} , \frac{\partial^2 A_x}{\partial x\partial z}  , \frac{\partial^2 A_x}{\partial y\partial z}  \biggr],
\end{multline}
otherwise everything proceeds as above, except that we now have to solve a symmetric $10\times 10$ matrix problem for $\vec{M}_i$.

With the solutions in hand, we can then directly define the gradients $\nabla \vec{B}_i$ of the magnetic field in cell $i$ as
\begin{equation}
\nabla \vec{B}_i = 
\left(
\begin{array}{ccc}
\frac{\partial^2 A_z}{\partial x\partial y} - \frac{\partial^2 A_y}{\partial x\partial z}  & 
\frac{\partial^2 A_z}{\partial y^2}         - \frac{\partial^2 A_y}{\partial y\partial z}  & 
\frac{\partial^2 A_z}{\partial y\partial z} - \frac{\partial^2 A_y}{\partial z^2}  \vspace*{0.15cm} \\

\frac{\partial^2 A_x}{\partial x\partial z} - \frac{\partial^2 A_z}{\partial x^2}  & 
\frac{\partial^2 A_x}{\partial y\partial z} - \frac{\partial^2 A_z}{\partial x\partial y}  & 
\frac{\partial^2 A_x}{\partial z^2}         - \frac{\partial^2 A_z}{\partial x\partial z}  \vspace*{0.15cm} \\

\frac{\partial^2 A_y}{\partial x^2}         - \frac{\partial^2 A_x}{\partial x\partial y}  & 
\frac{\partial^2 A_y}{\partial x\partial y} - \frac{\partial^2 A_x}{\partial y^2}  & 
\frac{\partial^2 A_y}{\partial x\partial z} - \frac{\partial^2 A_x}{\partial y\partial z} \\
\end{array}
\right).
\end{equation}
Note that the trace of this Jacobian vanishes, i.e.~we manifestly have  $(\nabla \cdot \vec{B})_i = 0$ to machine precision. We do not think, however, that this provides a particularly  meaningful advantage in this context, because this property is contingent on using a particular discretization operator. The subsequent evolution of the system is oblivious to this vanishing divergence, because the important feature of our method is that the magnetic field is always defined via the curl of the vector potential. This  guarantees a vanishing divergence of the magnetic field up to local discretization errors of the divergence operator -- it does not really get any better than this, since the form of the divergence operator is not unique. 

A case in point is that we are well advised to apply slope limiting to the estimated gradients $\nabla \vec{B}$ of the magnetic field before using them for extrapolation to cell faces. While we have found that this is not strictly necessary in many problems because the $\vec{A}$-field is comparatively smooth, it does reduce oscillations in the resulting $\vec{B}$-field, for example behind magnetic shocks. For slope limiting, we treat the three spatial gradients as independent and make sure that extrapolated $\vec{B}$-field values at the centroids of all facets of a cell $i$ do not over- or undershoot the maximum/minimum $\vec{B}$-field values of all neighbouring cells. But if slope limiting is applied component-wise, it  will in general  cause a violation of the condition $(\nabla \cdot \vec{B})_i = 0$ for the cell centred divergence operator. Again, this has no negative consequences,  highlighting that this is ultimately not critical for the stability of the numerical scheme, because a vanishing magnetic field divergence is robustly encoded through the definition of the $\vec{B}$-field in terms of the vector potential.

The second-order fit can of course also be used to define the cell-centred magnetic field values $\vec{B}_i$. However, since the double stencil involves a relatively large region, this procedure is comparatively diffusive, especially when there are magnetic shocks and discontinuities in $\vec{B}$ itself. 

We therefore usually prefer to only use -- if at all -- the double stencil estimate for obtaining the Jacobian of the $\vec{B}$-field, and for defining  the current density of cell $i$ as
\begin{equation}
\vec{J}_i = 
\left(
\begin{array}{c}
\frac{\partial^2 A_y}{\partial x\partial y} + \frac{\partial^2 A_z}{\partial x\partial z} 
- \frac{\partial^2 A_x}{\partial y^2}   - \frac{\partial^2 A_x}{\partial z^2}  \vspace*{0.15cm} \\
\frac{\partial^2 A_z}{\partial y\partial z} + \frac{\partial^2 A_x}{\partial x\partial y} 
- \frac{\partial^2 A_y}{\partial z^2}   - \frac{\partial^2 A_y}{\partial x^2}  \vspace*{0.15cm} \\
\frac{\partial^2 A_x}{\partial x\partial z} + \frac{\partial^2 A_y}{\partial y\partial z} 
- \frac{\partial^2 A_z}{\partial x^2}   - \frac{\partial^2 A_z}{\partial y^2} 
\end{array}
\right).
\label{eqncurrentdensity}
\end{equation}
But note that  the current density is actually not explicitly needed as long as we restrict ourselves to ideal MHD.

\subsubsection{Obtaining the B-field through an integral reconstruction}

In \citet{Mocz2016}, a vector potential method for Voronoi meshes was proposed in which the vector potential values are stored at the mesh generating points. These are the vertices of the topologically dual Delaunay triangulation. \citet{Mocz2016} proposed to compute the mean magnetic flux for every triangular facet of all Delaunay tetrahedra, based on contour integrals of the vector potential along the Delaunay edges. In this case, the sum of the magnetic fluxes over the surfaces of all tetrahedra vanishes by construction, which thus expresses a manifestly divergence-free magnetic field, in direct analogy to standard constrained transport methods.

Furthermore, for each tetrahedron, a spatially uniform magnetic field in its interior can be determined such that the surface fluxes are reproduced. This thus can be used to define a piece-wise constant magnetic field tessellation of space that is divergence free and is consistent with the prescribed vector potential values. A volume-averaged magnetic field for the Voronoi cells can then be defined by computing the volume overlap with the Delaunay cells and weight their contributions accordingly.

While this construction is elegant and avoids the need for a spatial derivative operator to obtain the magnetic field, it also has a number of downsides. One is that the Delaunay tessellation can discontinuously change due to an infinitesimal shift of one of the mesh-generating points. As a result, the reconstructed magnetic field can likewise be sensitive to small shifts of the mesh generating points. It is also not clear whether the divergence-free field obtained for the Delaunay tetrahedra is ultimately ideal  for an integration of MHD on the Voronoi-based control volumes. Since  the mean field of the Voronoi cells is obtained from averaging over parts of  several Delaunay tetrahedra, followed by an independent gradient estimate needed for second order extrapolation to cell faces, the flux sum over all faces of Voronoi cells will in general not vanish. One could instead interpolate the vector potential values from Voronoi cell centres to vertices and average them there, followed by a magnetic flux reconstruction. We have tried this approach but ultimately abandoned it, both because it is quite cumbersome, but also because we doubt that there is ultimately a fundamental advantage to guarantee a flux sum for a computational cell that vanishes to machine precision. The key aspect is that the magnetic field is defined in terms of the vector potential. We for this reason prefer the simpler direct differential operator in terms of our least square estimates for an unstructured mesh,  due to its weak sensitivity to the geometric details of the Voronoi mesh.

\begin{figure*}
\centering
\resizebox{18.0cm}{!}{\includegraphics{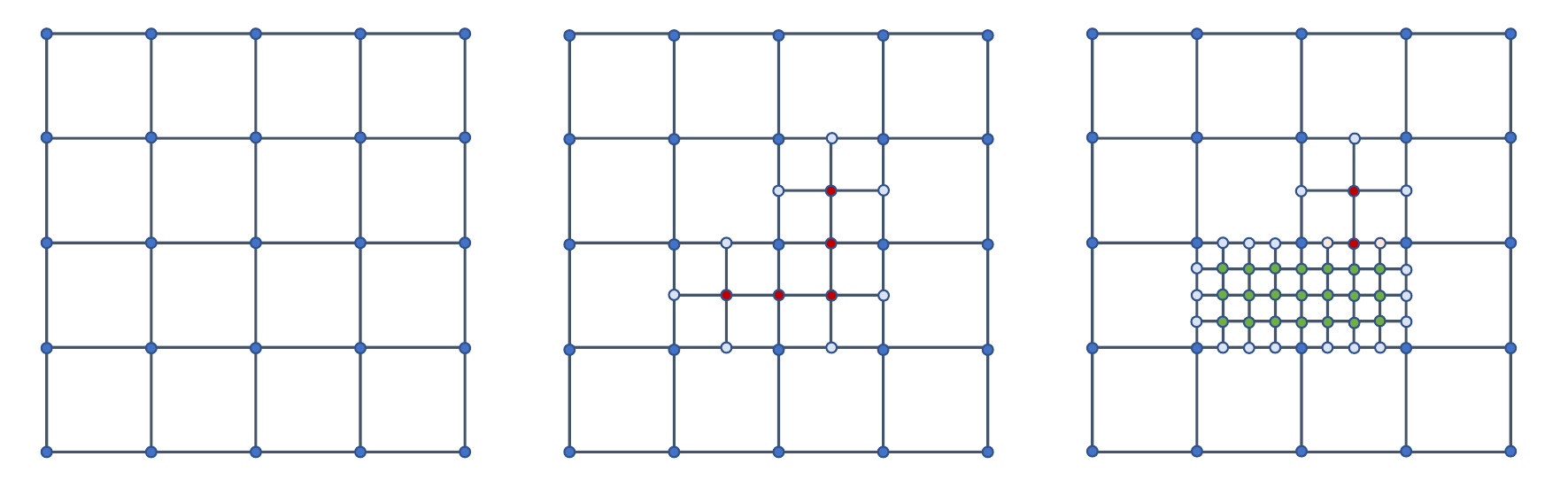}}%
\caption{Adaptive tricubic interpolation mesh used to represent the large-scale velocity field $\vec{u}$ in 3D, here illustrated in 2D for graphical clarity. The panel on the left shows a regular Cartesian grid covering the full simulation volume. Input values for the smoothed velocity field are given at the corners of the grid, marked with blue points. These values combined with numerical derivates computed with a stencil just involving neighbouring corners give rise to a tricubic interpolation polynomial within each cubical cell. These interpolations and their first derivatives are continuous across the cubes. In the middle panel, we show how a refinement level can be added to this grid. In the example, three cubes are refined by replacing them with cubes of half the size. This gives rise to new corners, some of the may coincide with corners of the parent mesh. Among them, we identify interior corners (marked in red) that have a full set of neighboring corners at the refined grid resolution. These interior points can be assigned new independent smoothed values for $\vec{u}$, and new numerical derivatives can be estimated for them on the refined grid, whereas all other corner points (light and dark blue) get their values and derivatives assigned by evaluating the interpolation polynomial of the corresponding parent cell. With this construction, the adaptive polynomial interpolation remains ${\rm C}^1$ across all cube boundaries. The panel on the right shows a subsequent refinement step of the new cubes obtained in the middle panel. Here the dark green points are interior points that allow the addition of new information about the described field. \label{FigTricubic}}
\end{figure*}

\subsection{Defining a smooth large-scale velocity field}

In order to treat the vector potential's time evolution in a Galilean invariant formulation, a suitable definition of the velocity field $\vec{u}$ is required. Recall that this should be a smoothed version of the actual velocity field, so smooth in fact that it can be described well as a linear field over the scale of single mesh cells. Furthermore, we need accurate spatial derivatives of this field. To avoid that the field $\vec{u}$ introduces errors or noise in the solution, the velocity derivates of $\vec{u}$ should be as accurate as possible, and preferably exact.

There are many, non-unique ways to realize such a smoothed field. One possibility that we have tried during the development of our new code is to realize the field $\vec{u}$ via real-space averaging with a spherical kernel of adaptive size $r^{\rm smth}_i$ around a given cell's geometric centre $\vec{c}_i$. In this scheme, we considered all cells whose mesh-generating point falls within this radius, and computed a local fit to the velocity field $\vec{u}$ and its Jacobian $\nabla\vec{u}$, using our least-square gradient estimate methodology described in subsection~\ref{secstdlsq} with the volumes of the individual cells as weights. We chose the radius $r^{\rm smth}_i = \chi\,  \overline{r_i^{\rm cell}}$ as a fixed multiple $\chi \simeq 5.0 - 7.0$ of the average local cell radius, so that we ended up reconstructing a smooth linear velocity field by averaging over $\sim 125-275$ cells in 3D. Each of these fits then yields both the velocity $\vec{u}_i$ and the value of the Jacobian $\nabla\vec{u}_i$ at the location of the cell centre $\vec{c}_i$. While this produces reasonably smooth realizations of  $\vec{u}$, there is invariably a fair amount of noise introduced by fluctuations between neighbouring cells in exactly which flow features are included in the averaging kernel. Furthermore, the spatial derivatives of $\vec{u}$ will only be approximate and likewise be a source of noise in the evolution of the vector potential. These effects can be reduced by enlarging the size of the averaging kernel, but this makes the adaptive real-space smoothing increasingly more expensive  without fundamentally  curing the problem.

We have therefore abandoned this approach in favour of representing $\vec{u}$ though an auxiliary (adaptively refined) Cartesian grid, combined with a tricubic interpolation on each cubical mesh element. We base our method on the 3D interpolation approach described and analyzed by \citet{Lekien2005} in which  a scalar field $f$ is expanded as a polynomial 
\begin{equation}
f(x, y, z) = \sum_{i,j,k = 0}^3 a_{ijk} \,x^i y^j z^k
\end{equation}
inside a cubical element. The 64 coefficients $a_{ijk}$ of this expansion are uniquely determined by specifying $f$, its first derivatives $\partial f/\partial x$, $\partial f/\partial y$, $\partial f/\partial z$, and the mixed partial derivatives $\partial^2 f/\partial x\partial y$, $\partial^2 f/\partial x\partial z$, $\partial^2 f/\partial y\partial z$, and $\partial^3 f/\partial x\partial y\partial z$ on each of the 8 corners of the cube. Inside the cube, the representation of $f$ is evidently perfectly smooth. What \citet{Lekien2005} show is that if a regular grid of cells with such expansions is considered (note that the corners of adjacent cubes and thus the corresponding input values for $f$ and the  partial derivatives above coincide), then the globally reconstructed field $f$ is continuous over all faces, and the first derivatives are continuous as well. In fact, also the above mixed partial derivatives are continuous on all faces. This high degree of smoothness and continuity of the piece-wise representation, and the ability to evaluate exact first derivatives of the field $f$ at any spatial point make this ideal for our needs with respect to the velocity field $\vec{u}$\footnote{Each velocity field component is simply treated as a separate scalar field.}. Also note that the partial derivatives of $f$ needed as input can easily be computed by finite differencing $f$ if it is specified on a regular grid. A stencil just based on neighbouring points is sufficient for this purpose, even for the higher order derivatives because they remain first order in every dimension.

However, the original method of \citet{Lekien2005} applies only to a regular grid with uniform resolution, which would be a serious limitation for applications with a high dynamic range in the employed local Voronoi mesh resolution. We have therefore generalized the scheme to an adaptive mesh. The basic idea is illustrated in Figure~\ref{FigTricubic}. The left panel shows a piece of a regular Cartesian grid where we assume that input values of $\vec{u}$ are given at the corners of each cubical element. We then consider a hierarchical refinement algorithm where all nodes at the finest level of the current mesh are optionally replaced with a set of nodes of half the linear size. We then distinguish two groups among all the corners of these new cells. {\em Interior corners} are those which have a full set of 26 neighbouring corners at the new finest resolution of the mesh hierarchy, whereas {\em boundary corners} are simply those for which this is not the case. For the latter group, we define the values of $f$ and the needed partial derivatives simply by evaluating the polynomial expansion of the parent node at the corresponding spatial location of the corner. For the interior points, we can specify new (in principal arbitrary) input values for $f$, and then compute the needed derivatives at these locations via finite differences (by construction, the corners needed for this do exist). With this in hand, the polynomial expansions for the newly inserted cubes can be fully defined. Furthermore, the global solution remains continuous, and has continuous first (mixed) derivatives. This approach can be applied recursively, yielding a fully adaptive piece-wise tricubic interpolation technique for the field $f$.

In our application, we define the input velocity values needed at each of the corners of the interpolation mesh always as the volume-weighted spatial average of all cells that fall within a cubical region of equal size to the grid spacing around the corner. The averaging is thus staggered by half a grid spacing relative to the interpolation elements. Our refinement criterion is chosen such that we split an interpolation element only if the biggest Voronoi cell falling into each of the 8 daughter cells is smaller than a given maximum fraction $\epsilon_{\rm vol}$ of the daughter element's volume. This ensures that each interpolation element contains at least of order $\sim 1/\epsilon_{\rm vol}$ Voronoi cells. Our default choice is $\epsilon_{\rm vol} = 0.002 - 0.01$, meaning that we at least smooth over $\sim 100 - 500$ input cells. In order to avoid that a given interpolation cube may potentially have input values for $\vec{u}$ at its 8 corners that stem from vastly different averaging volumes (this could happen if highly different mesh resolutions are directly adjacent to each other), we typically impose the restriction that cubes may only be refined at a given level if all their 8 corners are internal points at that level. This is however not strictly needed for the scheme to be viable.

Note that the resolution of this adaptive interpolation mesh is much coarser than the actual Voronoi mesh of the underlying simulation. With at least a factor 100 fewer cells, the total storage cost of this auxiliary mesh is negligible compared to that of the main simulation itself. Similarly, the construction cost is small compared to the cost of a full timestep. However, in a local timestep scheme with a deep timestep hierarchy it would become expensive to renew the interpolation mesh on every single (potentially thinly populated) timestep, since this operation relies on information from all cells. We therefore at present only renew the large-scale velocity field on timesteps when all cells are synchronized (i.e.~on full timesteps), which should be sufficient in almost all practical applications. Recall that the field $\vec{u}$ is merely an auxiliary computational object that at the continuum level does not affect any of the results. This is why there is great flexibility in how precisely $\vec{u}$ is chosen, and results do not depend sensitively on its update frequency. Finally, note that the retrieval costs for evaluating values for $\vec{u}$ and $\nabla\vec{u}$ are acceptably small as well. This is because evaluating the 64 polynomial terms involves only simple local computations that are quickly executed by modern processors, and this is in particular much faster than finding and operating on a large number of neighbouring cells as required by a kernel-based local smoothing.

\subsection{Time integration of the vector potential equation}

A critical aspect of our new scheme is the way in which we discretize the time evolution of the induction equation for the vector potential. This means we need to specify how we integrate each of the different terms of equation~(\ref{eqncentral2}). Note that we do not apply a standard finite-volume strategy to this equation because $\vec{A}$ is not a conserved quantity. Rather we treat the individual terms by finite difference methods, augmented with upwinding.

Specifically, for the first term in equation~(\ref{eqncentral2}), which involves the electric fields, we use the discretized form
\begin{align}
 \left. \frac{{\rm d}\vec{A}_i} {{\rm d}t}\right|_{\vec{w}}^{\rm induction}
 & = \frac{1}{S_i}\sum_j a_{ij}  ({\vec{v}}_{ij} - {\vec{u}}_{i}) \times {\vec{B}}_{ij}.
  \label{eqAcentral}
\end{align}
Here the sum over $j$ adds contributions from all neighbouring cells that share facets of area $a_{ij}$ with cell $i$, and $S_j= \sum_j a_{ij}$ is the surface area of the cell. The quantities ${\vec{v}}_{ij}$ and ${\vec{B}}_{ij}$ refer to the lab-frame states\footnote{We actually solve the Riemann problem itself in the frame of the moving facet, but the solution can then be transformed to the lab-frame.} returned by the Riemann solver for the corresponding facet. The computation of this term is for this reason inlined with the usual hydrodynamical flux computation. Note that we here effectively estimate the rate of change of the vector potential, i.e.~the electric field, at the location of the geometric centre of a cell by averaging the electric fields at facet centres. No further upwinding beyond what the Riemann solver does for each interface is applied.

Note that our averaging is here done by using the surface areas as weights. This weighting is not unique, but it is plausible and corresponds  to what is done in  various Cartesian constrained transport schemes \citep[e.g.][]{Toth2000} to estimate the cell centred electric field based on estimates obtained at face centres from the solutions of Riemann problems. In some constrained transport codes, the electric fields of the facets is similarly averaged at mesh corners. Note that this averaging can also be viewed as a means to introduce necessary dissipation into the method. Instead of carrying out volume averages of conserved fluid quantities like in the base hydrodynamic scheme, we here average over the point-wise update rates of the vector potential, sampled from several locations placed around the target point.

For the second term, which describes the change in the values of the vector potential due to relative motions of the points where it is stored, we directly discretize the term as
\begin{equation}
 \left. \frac{{\rm d}\vec{A}_i} {{\rm d}t}\right|_{\vec{w}}^{\rm distortion}
  = -\vec{A}_i \cdot \nabla\vec{u}_i
\end{equation}
in a cell-centered fashion. Here $\nabla\vec{u}_i$ is the Jacobian of our smooth velocity field evaluated at the location $\vec{c}_i$. This quantity is computed as part of our definition of the large-scale velocity field, as discussed in the previous subsection.

For the advective term, we compute a special upwinded derivative in the form
\begin{equation}
 \left. \frac{{\rm d}\vec{A}_i} {{\rm d}t}\right|_{\vec{w}}^{\rm advection}
  =   \sum_j  \vec{A}_{ij}[ (\vec{w}_i - \vec{u}_i) \cdot \vec{g}_{ij} ]
\end{equation}
where the $\vec{g}_{ij}$ are the weights of a least square derivative operator defined for the particular cell $i$ for values given at the face centers of the surrounding neighbours $j$, analogous to what we described in subsection~\ref{secstdlsq} for ordinary gradients. The value $\vec{A}_{ij}$ is taken as the upwind selection of the vector potential from both sides relative to the velocity $(\vec{w}_i - \vec{u}_i)$,~i.e. $\vec{A}_{ij} = \vec{A}_i$ for $[({\vec{w}}_{i} - {\vec{u}}_{i})\cdot {\vec{f}}_{ij}] < 0$ and $\vec{A}_{ij} = \vec{A}_j$ otherwise, where ${\vec{f}}_{ij}$ is the vector pointing from $\vec{c}_i$ to the center of the facet between $i$ and $j$. The vector potential quantities for the left and right states of each facet are linearly extrapolated from cell centers using slope-limited versions of the gradients of $\nabla\vec{A}$ estimated at cell centers. 

The term involving the cleaning scalar is discretized as
\begin{equation}
 \left. \frac{{\rm d}\vec{A}_i} {{\rm d}t}\right|_{\vec{w}}^{\rm cleaning}
  =   \sum_j  \phi_j \vec{q}_{ij} ,
\end{equation}
where the $\vec{q}_{ij}$ are the coefficients of our standard linear least square derivative operator for a cell when the reference values are given at centers of neighbouring cells, as defined in equation~(\ref{eqnvectorweight}).

Finally, in case we explicitly include Ohmic losses,
we discretize this as a pure diffusion term in $\vec{A}$, which is adequate provided the divergence of $\vec{A}$ is small, something we have arranged for through our explicit $\nabla\cdot\vec{A}$ control. Then we can write
\begin{equation}
 \left. \frac{{\rm d}\vec{A}_i} {{\rm d}t}\right|_{\vec{w}}^{\rm dissipation}
  =   \eta  \nabla^2 \vec{A}_i ,
\end{equation}
where $\nabla^2 \vec{A}_i$ stands for the cell-centred Laplacian for each of the components of $\vec{A}$. For the Ohmic loss term, second derivatives of $\vec{A}$ are explicitly needed, whereas they are otherwise only used indirectly when  gradients of the $\vec{B}$-field are computed for linear reconstructions within cells. Finally, the time evolution of the cleaning scalar $\phi$, described by equation~(\ref{eqncleaningscalar}), is done straightforwardly in a cell-centred fashion.

To obtain second-order accuracy in time for smooth flows we actually split each timestep in two computations of the above terms, one at the beginning of the timestep, and one at the end, with the latter carried out already for a new mesh geometry corresponding to the updated coordinates of the mesh-generating points at  the end of the  timestep. The corresponding details are described in Section~\ref{SecDetails}.

\subsection{Treatment of a non-vanishing mean magnetic field} \label{secBmean}

While our approach requires periodic boundary conditions and thus in principle a periodic vector potential, we would still like to be able to treat problems with  a periodic  $\vec{B}$-field but non-vanishing mean magnetic field. However, such a field cannot be represented with a purely periodic vector potential, as discussed by \citet{Mocz2016}. To work around this,   \citet{Mocz2016} expressed the vector potential as a sum of a periodic, time-dependent component, and a non-periodic component constant in time. The latter can be expressed, for example,
through
\begin{equation}
{\vec{A}}^{\rm const}(\vec{x}) =  \frac{1}{2}\vec{B}_{\rm mean} \times \vec{x}.
\label{eqnAperiodicity}
\end{equation}
This approach, however, requires special care at the boundaries of the box, in particular when cells are mapped to the other side of the box, in order to avoid that neighbouring cells see spurious jumps in ${\vec{A}}^{\rm const}(\vec{x})$.

To simplify this treatment, we here proceed differently than \citet{Mocz2016}.  We define the full vector potential $\vec{A}(\vec{x}, t)$ such that
\begin{equation}
\vec{B}(\vec{x}, t)  = \vec{B}_{\rm mean}  + \nabla \times\vec{A}(\vec{x}, t), \label{EqnBmean}
\end{equation}
which sidesteps the periodicity problem. In fact, the vector potential appearing in equation~(\ref{EqnBmean}) is periodic, as is $\vec{B}(\vec{x}, t)  - \vec{B}_{\rm mean}$. So effectively we are only evolving the difference of the magnetic field around a stationary mean field $\vec{B}_{\rm mean}$.

In practice, only a few small changes are needed in the simulation code to realize this. When we compute the magnetic field by applying a difference operator to the vector potential, we have to afterwards add in $\vec{B}_{\rm mean}$ to get the total magnetic field $\vec{B}(\vec{x}, t)$, following equation~(\ref{EqnBmean}). In addition, the evolution equation for the magnetic vector potential acquires an additional term of the form
\begin{equation}
\left.\frac{{\rm d}\vec{A}} {{\rm d}t}\right|_{\rm mean} = \vec{u} \times \vec{B}_{\rm mean}  \label{eqnAmean}
\end{equation}
that has to be added to the right hand side of equation~(\ref{eqncentral}). One way to see this is by revisiting our original derivation of equation~(\ref{eqncentral}), and by considering the variable part of the magnetic flux only, i.e.~the one created by $\vec{B}(\vec{x}, t) - \vec{B}_{\rm mean}$ instead of $\vec{B}(\vec{x}, t)$. Equation~(\ref{eqnperimeter}) then stays in the form of a line integral over the vector potential, whereas in equation~(\ref{eqnfluxchange}) we need to exclude magnetic flux changes arising from $\vec{B}_{\rm mean}$ and a change in area. The net result is then the modification~(\ref{eqnAmean}) above.

\subsection{Numerical dissipation}
\label{SecDissipation}

When the magnetic field is evolved directly with a conservative finite volume scheme\footnote{In practice this cannot be done in its pure form without some form of divergence cleaning as otherwise numerical instabilities will appear.}, one follows the reconstruct-evolve-average paradigm. This means that after the Riemann solutions at interfaces have been computed (which assume piece-wise constant input states that meet at the interface), they are effectively evolved over a finite time $\Delta t$ and are then averaged in space to obtain the new averaged fluid states in each cell. Conveniently, this spatial averaging procedure can be done implicitly on the basis of the fluxes alone. It in any case has the important effect of creating entropy where necessary, and providing numerical dissipation that stabilizes the scheme. For the magnetic field, it also means that dissipated magnetic field energy appears as heat, an effect that one can describe as numerical resistivity. 

When the magnetic field is instead evolved in terms of the vector potential as described above, this type of averaging is not really present. In fact, while there is a bit of averaging involved when the electric fields are interpolated to the cell centres and averaged there, this is not equivalent to a finite volume average of the magnetic field over individual cell volumes. Rather, the evolution of the vector potential is more akin to what is commonly done in  finite difference methods, and here explicit forms of numerical dissipation are commonly included. Only for the conserved variables of mass density, momentum density and energy density, sufficient dissipation should still be guaranteed by our finite
volume averaging. 

There could thus be a potential dearth of numerical dissipation in our scheme for the magnetic field. Note that too little dissipation can be an even worse problem as the opposite, as it can make a scheme prone to artifacts or serious instability in challenging flow situations. The corresponding problem could become more acute in the moving mesh approach because here dissipation due to fluid advection is essentially eliminated. For example, in a 2D problem such as the Orszag-Tang vortex, the initially assigned vector potential values for each cell are in principle simply advected along with the motion of each cell. For a differentially rotating fluid, gradients between adjacent cells could hence become extreme, a situation that cannot be expected to stay numerically robust and accurate.

We therefore consider it advantageous to at least foresee the possibility to add explicit numerical dissipation to our new MHD scheme in a suitably parameterized form, although we have not yet seen a clear need to actually do this in practice. In particular, the test problems we discuss in this study have all not needed this. In any case, having such optional dissipation can serve two purposes, (1) to provide the necessary magnetic dissipation to  handle magnetic reconnection events in situations where this cannot be avoided at the available  numerical  resolution, and (2) to eliminate numerical discretization noise in the $\vec{A}$-field, arising, for example, by truncation errors in numerical fluxes, large mesh-correction motions, or mesh refinement/derefinement operations.

In this context it is also helpful to recall that the vector potential is a continuous function, and that it needs to be quite smooth over the scale of the local stencil to correspond to a meaningfully discretized physical state. This is because an erratically  fluctuating vector potential on the scale of neighbouring cells could not be meaningfully differentiated twice, and thus would not provide a sound basis for time integration by the code.  

In order to ensure this smoothness, a straightforward possibility is to simply  invoke an artificial Ohmic resistivity, and to treat this correspondingly just like the physical resistivity in the MHD equations. We can thus parameterize extra numerical dissipation in the vector potential, if desired,  as  
\begin{equation}
\left.\frac{{\rm d}\vec{A}_i}{{\rm d}t}\right|_{\rm diss} =   \alpha_{\rm diss} {v_A}{r_{\rm cell}}
\nabla^2 \vec{A}_i,
\end{equation}
where  $\alpha_{\rm diss}$ is a dimensionless parameter we set typically to $0.1-0.3$, $v_A$ is the Alfvén speed of the cell, and $r_{\rm cell}$ is the cell radius based on the cell volume.   We also add a corresponding increase in the gas entropy, at a rate
\begin{equation}
\left. \frac{{\rm d} (\rho s)}{{\rm d} t}\right|_{\rm diss} = - \frac{\gamma - 1}{\rho^{\gamma-1}} \vec{J}\cdot \left.\frac{{\rm d}\vec{A}_i}{{\rm d}t}\right|_{\rm diss}
\end{equation}
in case the entropy is (optionally) integrated along the other fluid quantities. This expression assumes that the dissipation can be described as an effective Ohmic resistance, which appears to work well in practice. We make sure, however, that the scalar product between $\vec{J}$ and the change in $\vec{A}$ is negative, otherwise we do not modify the entropy. Note that no corresponding change in the energy needs to be done, as use of the total energy equation will automatically ensure that any dissipated magnetic energy appears as heat.

\subsection{Magnetically dominated flows}
\label{SecLowBetaFlow}

In a conservative finite volume scheme, the total energy equation is normally used to obtain the evolution of the thermal energy by subtracting the kinetic energy from the total energy. If a magnetic field is included, one additionally subtracts the magnetic energy from the total energy. While this guarantees exact conservation of the total energy (at least in the absence of self-gravity), it is well known that significant challenges arise when the kinetic energy is much larger than the thermal energy, i.e.~in highly supersonic flows. In this case, the thermal energy arises as a small difference of two large terms, and its evolution may become dominated by truncation errors. To avoid this, one can resort to evolving the thermal energy equation instead, or alternatively the entropy equation, in order to obtain more accurate values for the temperature and thus the pressure, even though this means to give up manifest conservation of total energy.

A similar problem arises when the magnetic energy is larger than the thermal pressure, i.e.~in low $\beta$-flows, where
\begin{equation}
\beta \equiv \frac{P_{\rm therm}}{P_{\rm mag}}
\end{equation}
characterizes the ratio of thermal to magnetic pressure. Again, this tends to make the temperature evolution inaccurate, because it is then  prone to suffer relatively strongly from truncation errors in the evolution of the total energy and the magnetic energy.

To address this issue, our code optionally allows the use of the entropy equation to set the thermal energy of a cell. If this option is activated, we always evolve the entropy equation in addition to the total energy equation. The default is to give precedence to the total energy equation, unless special criteria are fulfilled. We only consider the use of the entropy for closing off a timestep if the plasma-$\beta$ of a cell is lower than a prescribed threshold $\beta_{\rm low}$ both at the beginning of a timestep, and at the end of a timestep based on a provisional calculation of the end state with the total energy equation. In addition, the largest Mach number detected by any of the Riemann problems that have been solved for the cell during the step  needs to be smaller than a prescribed threshold ${\cal M}_{\rm low}$. If both conditions are met, we give the entropy precedence over the total energy to close off the timestep by using it to compute the new thermal energy. We then also reinitialize the total energy of the cell based on the this updated thermal energy. We furthermore log the corresponding adjustment of the total energy, as this is exactly the incurred error in total energy conservation. This error can simply be monitored to make sure that it stays sufficiently small. Alternatively, one may also add this energy back to the system in a smoothed fashion to restore formal total energy conservation.

Otherwise, we proceed with the default behaviour, i.e.~the thermal energy at the end of the timestep is computed using the total energy equation. We then also re-initialize the entropy of the cell with the corresponding result. The decision of whether or not one should use the entropy instead of the total energy question can thus be made individually on a cell-by-cell basis, and in principle differently for every timestep.

\section{Detailed implementation}
\label{SecDetails}

For definiteness, we now give a step-by-step description of the discretization of our MHD algorithm. We will first describe the approach for a global timestep equal for all cells, and then later discuss what needs to be changed for local timesteps. 

\subsection{Step by step procedure for a single timestep}

A timestep starts out with a set of mesh-generating points at coordinates $\vec{r}_i$, and values of the conserved fluid quantities mass $m_i$, momentum $\vec{p}_i$, total energy $E_i$, and entropy $S_i$ of the associated cell, as well as the vector potential $\vec{A}_i$ of the cell and the value $\phi_i$ of its cleaning scalar. We then carry out the following steps:

\begin{enumerate}

\item Voronoi mesh construction. This yields cell volumes $V_i$, the geometric-cell centres $\vec{c}_i$, and for every cell $i$ a list of the neighbouring cells $j$ that share a facet with area $a_{ij}$ and centroid coordinate $\vec{f}_{ij}$ with cell $i$. We note that this step is only really needed for the very first timestep, otherwise the mesh constructed at the end of the previous timestep, see step \ref{stepmeshconstruction} below, can simply be reused.

\item \label{nextstep} Construction of the $\vec{B}$-field. We compute a least-square estimate of the Jacobian $\nabla \vec{A}_i$ for each cell $i$, taking into account the direct neighbours, using geometric cell centres as reference coordinates and the face areas as weights. This yields cell-centred values of the magnetic field, $\vec{B}_i$, from the  components of $\nabla \vec{A}_i$. 

\item Optionally, only if the current density $\vec{J}_i$ and/or the Laplacian of  $\vec{A}_i$ is desired, we determine the Hessians $\nabla^2\vec{A}_i$, either using a higher order reconstruction of the magnetic field with an enlarged `double stencil' that includes the neighbours of neighbours of $i$, or by applying our standard difference operator to the results obtained for $\nabla \vec{A}_i$. From the components of the Hessians of the vector potential we extract the current density $\vec{J}_i$. 

\item We now slope limit the gradients $\nabla \vec{A}_i$ such that they can be used for piece-wise linear reconstruction of $\vec{A}_i$ within the cell, which is needed for the advection term in the induction equation.

\item Computation of new primitive variables. This is done for density $\rho_i$, velocity $\vec{v}_i$, and entropic function $s_i$ of the cell. We also determine the specific thermal energy $u_i$. This is normally computed from the total energy of the cell by subtracting the kinetic energy and magnetic energy (using the $\vec{B}_i$-value determined in the previous step), resulting in  the thermal energy. As this can become inaccurate for very low $\beta$-plasma, we have also implemented an optional alternative way to determine the thermal energy based on the entropy. When this is enabled, our default approach is still to use the total energy equation, but to replace it with entropy when $\beta < \beta_{\rm low}$. In the latter case, total energy conservation is not manifestly guaranteed any longer, but we can accumulate for every cell the difference in energy implied by use of the entropy. We can then use this difference to monitor the error in total energy conservation and use it as a means of quality control, or optionally inject the energy difference back into the system in a suitably smoothed way to restore exact energy conservation. In any case, with $u_i$ in hand, we determine the thermal gas pressure $P_i = (\gamma-1)\rho_i u_i$.

\item Computation of cell-centred gradients for the primitive variables. This is done for $\rho$, $\vec{v}$, $P$, $\vec{B}$, and the entropic function $s$ (if optionally enabled), using  our default linear least square approach. For the slope estimates, we are usually not forcing the  reconstruction to go through the central value of the cell, although this can also be optionally required.

\item Slope-limiting of the estimated gradients. We require that the spatially extrapolated values at the face centres of a cell may not overshoot the maximum (or undershoot the minimum) of the cell-averaged values among all neighbouring cells.  
  
\item Setting of new timestep sizes $\Delta t_i$. This is based on the Courant condition and other timestep criteria if needed. We also employ the tree-based timestep scheme introduced in \citet{Springel2010} to anticipate the arrival time of hydrodynamical signals from distant cells, and to reduce the timestep appropriately before they arrive.
  
\item Determination of the motion of the mesh-generating points. The velocities $\tilde{\vec{w}}_i$ of the mesh-generating points can in principle be chosen freely. For the moving-mesh case, the default choice  is to make $\tilde{\vec{w}}_i$  equal to the cell's gas velocity $\vec{v}_i$. For some of the cells, additional small `mesh-correction' motions may be added with the goal to improve mesh regularity. For a stationary mesh, the velocities of the mesh-generating points are simply set to zero. Note that there can be a small difference between $\tilde{\vec{w}}_i$ and the velocity $\vec{w}_i$ with which the geometric centres of a cell actually moves, as the latter also depends on the $\tilde{\vec{w}}_j$ values of all neighbours.

\item Determination of the large-scale smoothed velocity field $\vec{u}_i$ at the location of the cell centre $\vec{c}_i$, as well as the Jacobian $\nabla\vec{u}_i$ of this velocity field at these locations.

\item First half-step flux computation. To  compute the first half of the hydrodynamical fluxes across cell faces (at the beginning of the timestep), we spatially extrapolate $\rho$, $\vec{v}$, $P$, $\vec{B}$ and the entropic function $s$ from the geometric cell centres to the mid-points of cell facets, yielding the left and right states at each interface. In the case of a moving mesh, we solve the Riemann problem in the rest-frame of the moving face.  We therefore transform the left and right states into the frame of the facet by subtracting the facet's velocity $\vec{w}_{ij}$ from the left and right states' velocities. We next use the HLLD Riemann solver of \citet{Miyoshi2005} to determine the hydrodynamical fluxes. The entropy flux is computed as a purely upwind flux, based on the extrapolated state of the side that donates mass according to the HLLD solver. The computed fluxes are multiplied with $\Delta t/2$ and transformed back to the lab frame. They are then used to update the conserved states of both involved cells, i.e.~their mass, momentum, total energy, and entropy content are updated accordingly.

\item Vector potential change from the first half of the step. Using the states returned by the Riemann solver for the facets between cells $i$ and $j$ (namely $\vec{B}_{ij}$ and lab-frame ${\vec{v}}_{ij}$) we estimate the electric field at the cell centre in a frame moving with velocity $\vec{u}_i$, allowing us to compute the genuine induction part of the evolution equation for $\vec{A}_i$.  We also determine up-wind solutions $\vec{A}_{ij}$ for the vector potential as described earlier. This yields a  rate of change of the vector potential equal to  
\begin{align}
\dot{\vec{{A}}}_i^{(n)}  = & \frac{1}{S_i}\sum_j a_{ij}  ({\vec{v}}_{ij} - {\vec{u}}_{i}) \times {\vec{B}}_{ij}   +  \sum_j  \phi_j \vec{q}_{ij}   \label{eqndAdtA} \\
  & -  \vec{A}_i \cdot \nabla\vec{u}_i 
   + \sum_j  \vec{A}_{ij}[ (\tilde{\vec{w}}_i - \vec{u}_i) \cdot \vec{g}_{ij}  + \eta  \nabla^2 \vec{A}_i \nonumber   
  \end{align}
at the beginning of the timestep $(n)$. Note that we here use $\tilde{\vec{w}}_i$ in lieu of  $\vec{w}_i$, simply because we do not know the precise value of $\vec{w}_i$ yet at the beginning of a timestep\footnote{In our moving-mesh approach, only the velocities of the mesh-generating points of the cells have a prescribed velocity that is known already at the beginning a timestep, normally set equal to the gas velocity of the cell modulo small correction velocities in some cases, but in principle the choice of these velocities is unrestricted. A side-effect of the definition of Voronoi cells is that an accurate computation of the resulting velocities of the geometric cell centres based on given motions of the generators is computationally very involved. However, we can, at the end of a timestep, based on the new position of the cell-centre, compute the velocity $\vec{w}$ for the given step. Our strategy in dealing with the advection term will therefore be that for the calculation of the first half step, we assume an approximate velocity for $\vec{w}$, which we then correct in the second half step, when the new mesh geometry is known, to the correct average velocity $\vec{w}$.}.  The difference between the two will be corrected in the computation of the second half of the timestep. We also compute the time derivative of the cleaning scalar, as
\begin{equation}
\dot{\phi}_i^{(n)} = 
 \alpha c_{s,i}^2 \nabla \cdot \vec{A}_i - \beta \frac{c_{s, i}}{r_{{\rm cell}, i}}\phi_i.
\end{equation}
\item The mesh generating points are now moved with their velocities $\tilde{\vec{w}}_i$ to their positions at the end of the timestep.
\begin{equation}
\vec{r}^{n+1}_i = \vec{r}^{n}_i  + \tilde{\vec{w}}_i \Delta t.
\end{equation}
  
\item \label{stepmeshconstruction} New voronoi mesh construction. This yields new cell volumes $V_i$, new geometric-cell centres $\vec{c}_i$, and for every cell a list of new neighbouring cells $j$ that share facets of area $a_{ij}$ with cell $i$.

\item Second hydrodynamical flux computation at the end of the timestep. We use Heun’s method for time integration, a 2-nd order accurate Runge-Kutta type method \citep{Pakmor2016}. To this end, we average the fluxes at the beginning and end of the steps, with the latter being computed with a system state extrapolated in time to the end of the step. We thus need to both spatially and temporarily extrapolate the primitive variables from the geometric cell centres to the new mid-points of cell facets. The time derivatives of the primitive variables are obtained from the hydrodynamical equations in their primitive variable form, based on the spatial gradients computed above \citep[see][]{Springel2010}. Note that in the MHD case, the equations for the partial time derivate of density and of pressure are not affected by the magnetic field, only the equation for the partial time derivate of the velocity gets an additional contribution from the Lorentz force,
\begin{equation}
\left.\frac{\partial \vec{v}}{\partial t}\right|_{\rm MHD}
= \left.\frac{\partial \vec{v}}{\partial t}\right|_{\rm hydro}
+ \frac{1}{\rho} \vec{J} \times \vec{B}.
\end{equation}
We express the Lorentz force when needed for the time extrapolation through the (slope-limited) Jacobian $\nabla \vec{B}$ of the magnetic field, rather than through the optionally computed (and not slope-limited) form of the current density $\vec{J}$ obtained through equation~(\ref{eqncurrentdensity}). The partial time derivative for $\vec{B}$ is simply given by the induction equation. We can then estimate, for example, the needed ``left'' input density for the Riemann solver as
\begin{equation}
\rho^{\rm L}_{ij} = \rho_i + (\nabla\rho)_i (\vec{f}^{n+1}_{ij} - \vec{c}^{n}_i) + \left.\frac{\partial \rho}{\partial t}\right|_{\vec{c}_i} \Delta t ,
\label{eqnrhoextrastationary}
\end{equation}
while the corresponding ``right'' value $\rho^{\rm R}_{ij}$ would come from cell $j$. Here
the coordinate $\vec{f}^{n+1}_{ij}$ is the new centre of the facet, while $\vec{c}^{n}_i$ still refers to the old centre of the cell at the beginning of the timestep, because this is where the gradients were computed, where $\partial \rho / \partial t$ can be inferred, and where $\rho_i$ is known. In the stationary mesh case, we use the above form for the extrapolation, while in the moving mesh case, we express the spatial extrapolation relative to a coordinate that moves with the gas velocity so that the temporal extrapolation can be expressed through the convective derivative, as 
\begin{equation}
\rho^{\rm L}_{ij} = \rho_i + (\nabla\rho)_i (\vec{f}^{n+1}_{ij} - \vec{c}^{n}_i - \vec{v}_i \Delta t) + \frac{{\rm d} \rho}{{\rm d} t} \Delta t. 
\label{eqnrhoextramoving}
\end{equation}
As our mesh motion follows the fluid, we roughly expect $\vec{f}^{n+1}_{ij} \approx \vec{f}^{n}_{ij} + \vec{v}_i \Delta t$, so that this Galilean-invariant form limits the magnitude of the spatial extrapolation vector to be of order the cell radius, irrespective of the speed of the mesh motion. Otherwise, everything else proceeds like in the first hydrodynamical flux computation.

\item Vector potential change from the second half of the step. As for the first half of the step, we now compute a rate of change of the vector potential for cell $i$ at the end of the timestep, using the new cell geometry and the Riemann solver solutions obtained during the second flux computation:
\begin{align}
\dot{\vec{{A}}}_i^{(n+1)}  = & \frac{1}{S_i'}\sum_j a'_{ij}  ({\vec{v}}_{ij} - {\vec{u}}'_{i}) \times {\vec{B}}_{ij}   +  \sum_j  \phi'_j \vec{q}'_{ij}  \label{eqndAdtB} \\
  & -  \vec{A}'_i\cdot \nabla\vec{u}'_i 
   + \sum_j  \vec{A}_{ij}[ (\hat{\vec{w}}_i - \vec{u}'_i) \cdot \vec{g}'_{ij}  + \eta  \nabla^2 \vec{A}_i .\nonumber  
\end{align}
Here the apostrophes indicate that the corresponding quantities are computed for the new mesh geometry at the end of the timestep. Furthermore, we use predicted values
\begin{equation}
\vec{A}_i' = \vec{A}_i + \dot{\vec{{A}}}_i^{(n)} \Delta t
\label{eqnAextrapolate}
\end{equation}
and
\begin{equation}
\phi'_i = \phi_i + \dot\phi_i^{(n)} \Delta t
\label{eqnPhiextrapolate}
\end{equation}
to realize second-order accuracy in time for the corresponding terms. For the advection velocity we here use
\begin{equation}
\hat{\vec{w}}_i = 2 \vec{w}_i - \tilde{\vec{w}}_i ,
\end{equation}
with $\vec{w}_i = (\vec{c}_i^{n+1} - \vec{c}_i^n)/\Delta t$, so that the average advection speed from the first and second half-step computations recovers $\vec{w}_i$. We also compute an updated rate of the cleaning scalar at the end of the step, as
\begin{equation}
\dot{\phi}_i^{(n+1)} = 
 \alpha c_{s,i}^2 \nabla \cdot \vec{A}_i - \beta \frac{c_{s, i}}{r'_{{\rm cell}, i}}\phi'_i.
\end{equation}
Here the values of the soundspeed and the divergence of $\vec{A}_i$ are retained from the beginning of the timestep.
\item Update of the cell-centred vector potential. We finally close the timestep by updating the vector potential stored for the cell to the new value 
\begin{equation}
\vec{A}^{n+1}_i  =  \vec{A}^{n}_i  + \frac{1}{2} \left( \dot{\vec{{A}}}_i^{(n)} + \dot{\vec{{A}}}_i^{(n+1)}\right) \Delta t ,
\end{equation} 
which  incorporates all the changes accumulated during the timestep. Likewise, we update the cleaning scalar as
\begin{equation}
\phi^{n+1} = \phi^{n}_i  + \frac{1}{2} \left( \dot{\phi}_i^{(n)} + \dot{\phi}_i^{(n+1)}\right) \Delta t .
\end{equation} 
As mentioned earlier, the next timestep can then proceed by starting at step \ref{nextstep}, reusing the already constructed mesh geometry.

\end{enumerate}

\subsection{Local timesteps}

Many problems of interest feature a high-dynamic range in the timestep sizes required by local cells, especially when the cell-size itself is adaptive and varies spatially by large factors. For example, when {\small AREPO} is applied to problems in galaxy formation using a quasi-Lagrangian tracking of the gas mass, both the cell sizes and the required local timestep sizes vary by orders of magnitude. It then becomes numerically wasteful to determine the minimum timestep required by any of the cells and to use this globally for all cells to advance the system. Instead, it can be computationally much more efficient if cells are evolved on their individual, locally required timesteps. 

{\small AREPO} uses a power-of-two hierarchy of permissible timestep sizes, and all cells are sorted into this hierarchy. Local timestepping is then carried out with an algorithm where fluxes between cells that share a common face are always computed with the smaller timestep of the two neighbouring cells. This readily preserves  conservation of total mass, momentum and energy to machine precision. In practice, this primarily requires a specification of how passive cells, i.e.~cells that have a larger timestep than the active side and end their timestep later than the current time, are handled in such flux computations. We address this by computing new primitive variables from accumulated updates of the conserved variables only when a timestep completes, and otherwise temporal and spatial extrapolation are used -- as described in the previous subsection -- to compute the input Riemann solver states for such passive cells. Note that this does not require any special treatment in our formalism because it is analogous to the procedure for estimating the state at the end of a regular timestep for our second half-step flux computation, except that for a passive cell, such a prediction is also required in the middle of its timestep (if there is an active partner cell that has a timestep half as large), or at additional intermediate times (if there are neighbouring cells with timestep sizes 4 times smaller, or more).

In the case of MHD, we additionally need to define how the vector potential is treated, which is not a conservative field\footnote{But note that the volume-averaged magnetic field is automatically conserved when using the vector potential combined with periodic boundary conditions.} and thus takes on a special role. We address this by computing the time derivative of the vector potential of any given cell only at the beginning and end of the timestep, i.e.~there is no need for an accumulation of intermediate changes if the cell happens to be a passive partner cell in a flux computation of an active neighbour at an intermediate time. However, when computing ${\rm d}\vec{A}/{\rm d}t$ with equations~(\ref{eqndAdtA}) and (\ref{eqndAdtB}) it may happen that we encounter a neighbouring cell that is passive in one of the sums over neighbours. These are treated with spatial and temporal extrapolations just as in the ordinary flux computations. The only additional point is that we also need the time-extrapolated vector potential and cleaning scalar of the neighbouring cell, which are then calculated with equations~(\ref{eqnAextrapolate}) and (\ref{eqnPhiextrapolate}) as needed.

The $\vec{B}$-field of a cell is reconstructed from the vector potential always at the beginning of a cell's timestep. If this involves neighbouring cells that are passive at this time, their vector potential values are likewise predicted forward in time. The coordinates of mesh-generating points of passive cells at such intermediate times (needed also for the mesh construction) are obtained by linear motion with the prescribed velocities of the mesh-generating points. If a cell is passive, its centre-of-mass coordinate may be required in a neighbouring active cell's gradient estimate. In this case, the centre-of-mass position of the passive cell is estimated by advecting it with the velocity of its mesh-generating point as well.

A final technical difficulty concerns the optional use of a double stencil. Using this stencil for an active cell would normally not be possible in case a neighbouring cell is passive, because such mesh cells are not reconstructed in our default algorithm. We address this by modifying our mesh construction algorithm if the double-stencil option is enabled such that not only the mesh cells of all active cells are constructed in any given mesh construction, but also all the neighbours of these cells. Once this is ensured, we can then iterate over neighbours of neighbours of all active cells, and thus evaluate double stencil quantities for them.

\subsection{Mesh refinements and derefinements}

Classic constrained transport methods that work with a staggered mesh of stored face-centred magnetic flux values require extra care in applying adaptive mesh refinement approaches on-the-fly, otherwise the detailed zero flux-sum condition for every cell can be lost. Working with the vector potential directly provides a far simpler alternative. Note that a geometric integral construction (i.e.~doing contour integrals over the vector potential along mesh edges) can at any time be used to recompute the magnetic fluxes from the vector potential. They will then automatically obey the zero flux-sum condition (simply because every edge is traversed twice in opposite directions for the faces making up a cell). But as we argue in this paper, it appears equally accurate to simply apply a finite-difference operator to compute the magnetic field based on the curl of the vector potential. Maintaining the divergence-free property of an MHD code in the presence of mesh refinement and de-refinement operations then boils down to the problem of consistently remapping vector potential values to new storage locations.

In the new `next-generation' version of the {\small AREPO-2} code utilized in this paper, new mesh-generating points can be created at arbitrary locations, not just by splitting mesh generating points into close pairs as originally described in \citet{Springel2010}. The conservative fluid quantities of new cells are obtained based on volume fraction overlaps between the old and the new Voronoi cells. For the vector potential, we instead initialize its value at the location of a new cell centre according to a best-fit linear reconstruction based on neighbours of the new point identified in the old mesh, using essentially our basic gradient estimation approach. Here we thus effectively take the value $\overline{\vec{A}}_i$ in equation~(\ref{eqnAfit}) to initialize the new vector potential value.

Similarly, when a cell is removed, the ordinary conserved fluid quantities (mass, momentum, energy) are distributed to neighbouring cells according to the fraction of the volume they claim from the removed Voronoi cell. The vector potential value of the removed cell is dropped. However, the vector potential values of the neighbouring cells are corrected for the induced spatial offset their cell centres experience. Because the geometry of these cells changes (and their volume increases), the positions of their geometric cell centres, which are the storage locations associated with the vector potential, shift slightly. We use the most recently determined Jacobian of the vector potential for each of these cells to adjust for this shift. Note that for a spatially constant magnetic field, the mesh refinement and derefinement operations then do not introduce any perturbation in the magnetic field.

In more detail, our mesh refinement and derefinement algorithm works as follows. For every cell,  arbitrary, user-controlled criteria for whether or not a cell should either be removed (derefined) or split (refined) can be defined. These are evaluated at the beginning of a timestep for all active cells, first with a view to carry out possible derefinements, then for refinements. In the code, this is done right after a new mesh for all active cells has been constructed, and this mesh is then subsequently modified as needed to implement the desired refinement and derefinement operations.

In practice, we impose the condition that a cell that is derefined should not have a neighbouring cell that is also derefined and should not have a neighbouring cell that is passive, and likewise for refinement. This condition simplifies the technical implementation, and it in general selects a subset among all cells flagged by the (de)refinement criteria for actual (de)refinement. In particular, if two neighbours both want to be (de)refined at the same time, we decide randomly which one can go ahead in the present timestep. If large coherent mesh patches of a simulation want to be refined or derefined concurrently, our procedure slows this down to be executed over the coarse of a couple of timesteps. This is not restricting the flexibility of the approach in any practically relevant way. Note that the refinement and derefinement procedures operate fully at a per-cell basis, and they can be seamlessly combined with local timestepping.

For refinement, an important new approach introduced in this paper is that we do not split cells by creating a new mesh-generating point a small distance $\epsilon$ in a random direction around the old mesh generating point, as done in previous versions of  {\small AREPO}. An advantage of this former approach is that it only affects the geometry of the targeted cell by splitting it along a (randomly oriented) plane. However, the two new cells that result correspond to a  bad mesh quality that subsequent mesh-correction motions need to fix. For this reason, we now rather select the position of the most distant vertex of the target cell relative to its mesh-generating point as the position for the new mesh-generating point. This location, which  in some sense is the ``farthest corner'' of the cell, is in fact the circumsphere centre of a Delaunay tetrahedron. The other corners of the tetrahedron (or triangle in 2D) are formed by mesh-generating points of neighbouring cells that are equally far away from the new point, and in this sense, this new point is a natural location ``in the middle'' of the already present points. In fact, this type of placement tends to improve the local mesh quality in terms of the regularity of the Voronoi cells, which is beneficial for the overall accuracy of the simulation code. Note, however, that this type of refinement operation will affect more than just the target cell. In fact, its impact can also go beyond the immediate neighbours of the target cell. Our algorithms for carrying out the mesh refinement are however general enough to treat this properly. In subsection~\ref{secTestrefinement}, we will discuss an illustrative example for testing our mesh refinement and de-refinement procedures.

\subsection{Computing an initial vector potential}

If the initial conditions only specify the magnetic field but not the vector potential, it is helpful to have a method that can automatically derive the vector potential from a given magnetic field. We have integrated such a scheme in our code as a convenient start-up feature.

Suppose we are given an initial magnetic field $\vec{B}^\star$, not necessarily divergence free. This can always be written as
\begin{equation}
\vec{B^\star} = \nabla \times \vec{A} + \nabla \phi,
\end{equation}
where $\phi$ is a scalar field. Taking the curl of this equation and adopting the Coulomb gauge with $\nabla\cdot\vec{A} = 0$, we obtain
\begin{equation}
\nabla^2 \vec{A} = - \vec{J} ,
\label{eqnJpoisson}
\end{equation}
where we have defined the initial current as 
\begin{equation}
\vec{J}= \nabla \times \vec{B}^\star .
\label{eqnJsource}
\end{equation}
We can thus obtain the vector potential by solving a Poisson equation for each spatial component, using the initial current as a source. Here it is important to exactly use the same finite difference operators that we use to evaluate the gradients on our unstructured grid, and in particular, the Laplace operator needs to be computed by subsequently applying our discrete gradient and divergence operators. Only then we can obtain an accurate reproduction of the input magnetic field from the constructed vector potential $\vec{A}$, irrespective of the resolution and precise geometry of the mesh.

Note that equation~(\ref{eqnJpoisson}) corresponds to a sparse linear system, which is only approximately symmetric for our unstructured grid. However, this system can be efficiently and robustly solved in an iterative fashion with the stabilized bi-conjugate gradient algorithm (BiCGSTAB) proposed by \citet{vanderVorst1992}, which we have implemented for this task in {\small AREPO}. An alternative would be to try to use the gravity solver of {\small AREPO} for this purpose, but its effective gradient operator is not exactly commensurate with the gradient estimates in our MHD solver at the level of individual cells, so that the local mesh geometry is accounted less accurately. We thus here prefer the iterative bi-conjugate gradient solver.


\section{Test results}
\label{sec:results}

\subsection{Circularly polarized Alfvén wave}

To verify the accuracy and expected convergence rate of our code for a non-trivial smooth MHD problem, we use the circularly polarized Alfvén wave test originally proposed by \citet{Toth2000}. It represents an analytic, non-linear solution of the MHD equations and thus is particularly suitable as a test. \citet{Gardiner2008} generalized this test to 3D at an oblique angle relative to the principle coordinate axes of the mesh, and we here follow their setup closely \citep[see also][]{Stone2008}. In a coordinate system $x'y'z'$, the planar wave front can be described by
\begin{align}
B_{x'}  = &\; 1 , \\
B_{y'}  = &\; v_{y'} = 0.1 \sin(2\pi\, x'),\\
B_{z'}  = &\; v_{z'} = 0.1 \cos(2\pi\, x'),
\end{align}
with a wavelength $\lambda = 1$. We adopt density $\rho=1$, pressure $P=0.1$ and $\gamma =5/3$. For $v_{x'}= 0$, the wave is travelling along the negative $x'$-direction, returning to its original configuration after time $t=1$, while for $v_{x'} = 1$, a stranding wave is obtained. The corresponding vector potential for the $\vec{B}$-field can be written as
\begin{align}
A_{x'} = &\;0,\\
A_{y'} = &\;\frac{0.1}{2\pi} \cos(2\pi\, x')  + \frac{y'}{2},\\
A_{z'} = &\;\frac{0.1}{2\pi} \sin(2\pi\, x') - \frac{z'}{2}.
\end{align}
Note that the two terms linear in $y'$ and $z'$, respectively, generate $B_{x'}=1$ and are not compatible with periodic boundary conditions for the full vector potential, but they can instead be straightforwardly treated with the technique described in subsection~\ref{secBmean}. Following \citet{Gardiner2008},  we let the wave travel at an oblique angle relative to the lab-frame, into the direction $\vec{k} = 2\pi \lambda (1/3, 2/3, 2/3)$. This can be realized by rotating the wave via 
\begin{equation}
\left(
\begin{array}{c}
x\\ y \\ z   
\end{array}
\right) = 
\left(
\begin{array}{ccc}
\cos\alpha \cos\beta & -\sin\beta & -\sin\alpha \cos\beta \\
\cos\alpha \sin\beta & \cos\beta  & - \sin\alpha \sin\beta \\
\sin\alpha           &   0        &  \cos\alpha
\end{array}
\right)
\left(
\begin{array}{c}
x'\\ y' \\ z'   
\end{array}
\right)
\end{equation}
from the $x'y'z'$-system into the lab frame, with $\sin \alpha = 2/3$ and $\sin\beta = 2/\sqrt{5}$. By choosing a period box size with dimensions $3 \times 3/2 \times 3/2$, we can make sure that all periodic displacements in the lab-coordinates $x$, $y$, and $z$ map to multiples of one wavelength in $x'$, so that the inclined wave should propagate without disturbances by the box boundaries.

We consider this test both for a Cartesian grid with mesh resolution $2N \times N \times N$, and for a body-centred cubic lattice with $2 \times (2N \times N \times N)$ cells in which all Voronoi cells are truncated octahedra with 14 faces. The latter configuration corresponds much more closely to a typical mesh structure encountered in general 3D-flow by {\small AREPO} than a Cartesian grid. We carry out the test for a moving-mesh and a stationary mesh, in each case both for the standing and the moving wave.

\begin{figure}
\resizebox{8.5cm}{!}{\includegraphics{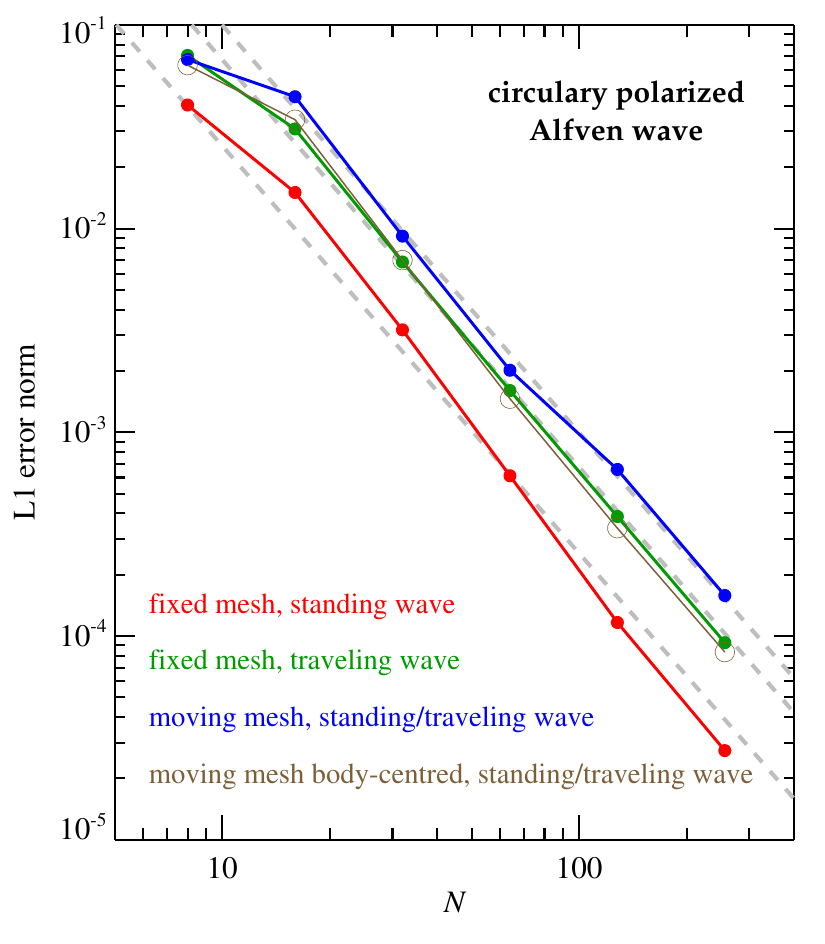}}%
\caption{Convergence rate of an oblique circularly polarized Alfvén wave when computed with different configurations of our code as a function of resolution $N$. We compare results for a Cartesian grid with mesh resolution $2N \times N \times N$, and for a body-centred cubic lattice with $2 \times (2N \times N \times N)$. Both grid configurations are tested for a moving- and a stationary mesh, and each case both for the standing and the moving wave, as labelled. In all cases, the convergence rate accurately follows the $L_1 \propto N^{-2}$ scaling expected for a second-order accurate code. The lowest errors are obtained for the stationary wave in the fixed mesh case. Note, however, that for the travelling wave the errors are substantially larger in the stationary mesh case, reflecting the higher advection errors in this case. In contrast, the errors in the moving-mesh case are {\em exactly identical} for the travelling and the stationary wave, as expected for a Galilean invariant code. \label{L1AlfvenWave}}
\end{figure} 

\begin{figure*}
\resizebox{18cm}{!}{\includegraphics{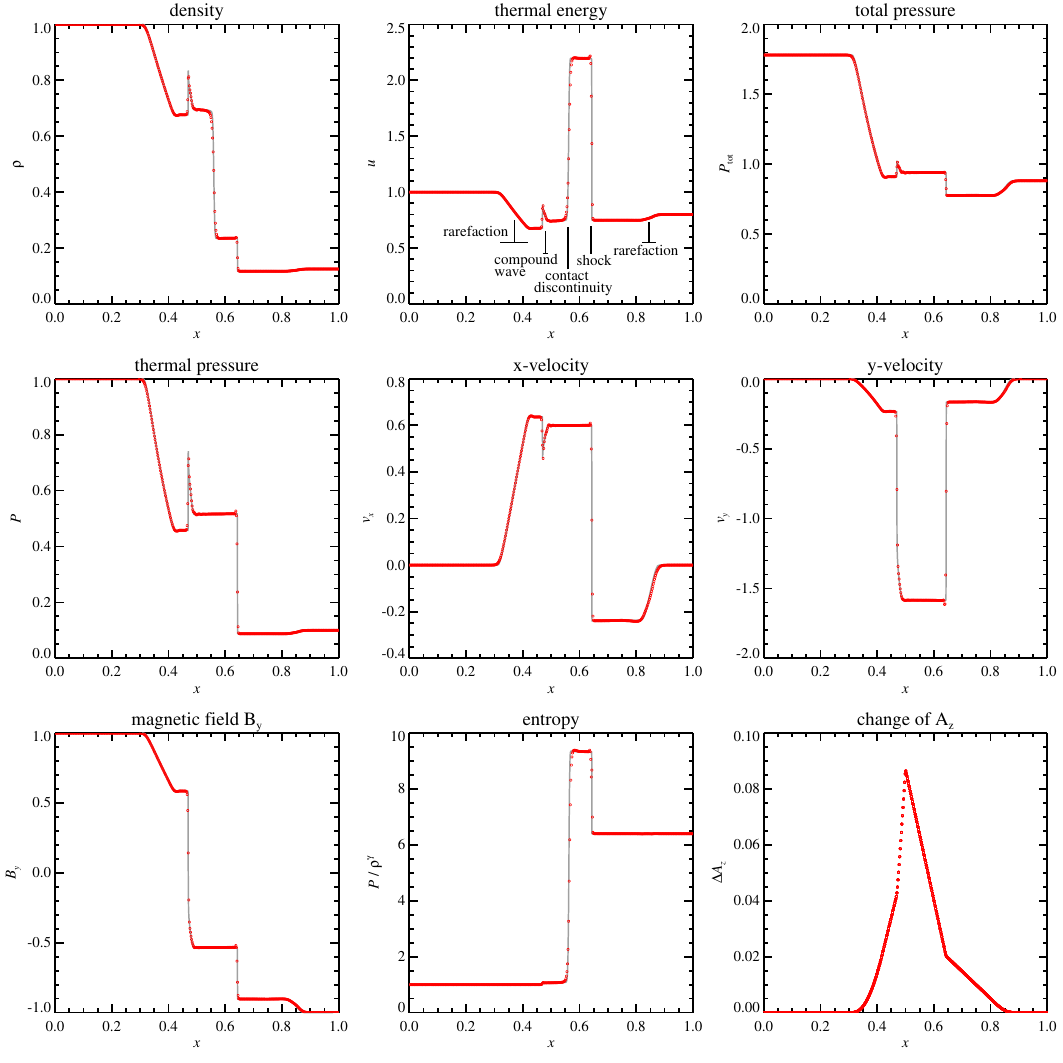}}%
\caption{The MHD shock tube test of Brio \& Wu, analysed at time $t=0.1$ for a simulation with 512 cells in the range $[0,1]$  (red dots). The panels show, from top left to bottom right, the values of individual cells for density, thermal energy, total pressure, thermal pressure, velocity components in the $x$- and $y$-directions, magnetic field in the $y$-direction, and change of the $A_z$-component of the vector potential relative to its value in the initial conditions. The grey background lines show a calculation carried out with the independent {\small ATHENA++} code \citep{Stone2008} at much higher resolution, using 8192 cells in the $x$-dimension within $[0,1]$, and piece-wise constant reconstruction to guarantee a result without oscillations at discontinuities. We take this as a proxy for the exact result. We also note that our code reproduces the expected constant value of $B_x = 0.75$ everywhere  to machine procession (not shown explicitly) -- this is something where cleaning schemes like Powell and Dedner show substantial errors that persist independent of resolution,  because $\nabla \cdot \vec{B}$ terms and/or errors modify the jump conditions of the Riemann problem \citep{Toth2000}. Finally, also note that rightwards of $x \simeq 0.65$ the plasma beta is low, whereas elsewhere the thermal pressure dominates. 
\label{FigBrioWu}}
\end{figure*}

In Figure~\ref{L1AlfvenWave}, we show the L1 error norm in the transverse magnetic field component as a function of resolution for this problem when evolved until $t=1$. We compare results obtained for a fixed Cartesian mesh both for a standing and a travelling wave, with corresponding results for a moving mesh, again for a travelling and a standing wave. In all cases, the convergence rate accurately follows the $L_1 \propto N^{-2}$ scaling expected for a second-order accurate code, except for the lowest resolution runs which deviate slightly on the side of lower errors, probably due to discreteness effects from the very small number of points sampling the wave in this case. The lowest errors are obtained for the stationary wave in the fixed mesh case, which is unsurprising, as this should provide the ideal setup for a quasi-stationary setting. Note, however, that for the travelling wave the errors are substantially larger in the stationary mesh case, reflecting the higher advection errors in this case. Yet higher travel velocities of the wave will lead to still larger errors in the fixed mesh case, reflecting the lack of numerical Galilean invariance in a Eulerian treatment. In contrast, the errors in the moving-mesh case are {\em exactly identical} for the travelling and the stationary wave, as expected for a Galilean invariant code. The absolute error is slightly higher than for the travelling wave fixed-mesh case, and is here generated by the circular motion of the mesh cells. When a body-centred Voronoi mesh is used, which is close to the typical mesh configuration encountered in free 3D flow, the errors are almost identical to the fixed mesh case, but the total number of cells is also larger by a factor of 2.  The $L_1$ errors we find for the fixed mesh case are close to but slightly larger than those reported in \citet{Gardiner2008}. We suspect this is due to the superior resolving power of a staggered-mesh representation of the magnetic field compared to the larger stencil we need to compute the magnetic field by finite-differencing the vector potential.

\subsection{Brio Wu shock tube}

The magnetic shock-tube problem introduced by \citet{Brio1988} is often used \citep[e.g.][]{Stone1992} to verify the basic capability of an MHD code to reproduce the array of shock waves, rarefaction fans and MHD compound waves that are possible in MHD flows, which give rise to a considerably richer set of waves than plain hydrodynamics. In the corresponding test, we consider a magnetic shock tube problem where the gas is initially at rest. Its initial state on the left is given by $\rho_{\rm L} =1$, $P_{\rm L} = 1$, $\vec{B}_{\rm L} = (0.75, 1.0)$, while the state on the right is given by $\rho_{\rm R} =0.125$, $P_{\rm R} = 0.1$, $\vec{B}_{\rm R} = (0.75, -1.0)$. A usual, we prefer to use the vector potential to define the initial magnetic field.  The adiabatic index is given by $\gamma =2$. Notice that the $x$-component of the magnetic field is continuous and non-vanishing. As we simulate the problem in a large periodic box in 2D, the mean $\vec{B}$-field is thus not vanishing.

In Figure~\ref{FigBrioWu} we show results computed for a stationary mesh with grid spacing $\Delta x = 0.01$, i.e.~100 points over the nominal $[0,1]$ extension of the simulation domain in the $x$-dimension. In the $y$-dimension, the problem is translationally invariant, so we actually simulate it in a box that is narrower in the $y$-direction for computational efficiency. All expected MHD waves and jump conditions are reproduced well. The accuracy of the results matches that seen in other grid codes, but is much better than reached by SPH \citep[e.g.][]{Price2004, Price2010} and MFM/MFV \citep{Hopkins2016}.

\begin{figure}
\begin{center}
\resizebox{7.9cm}{!}{\includegraphics{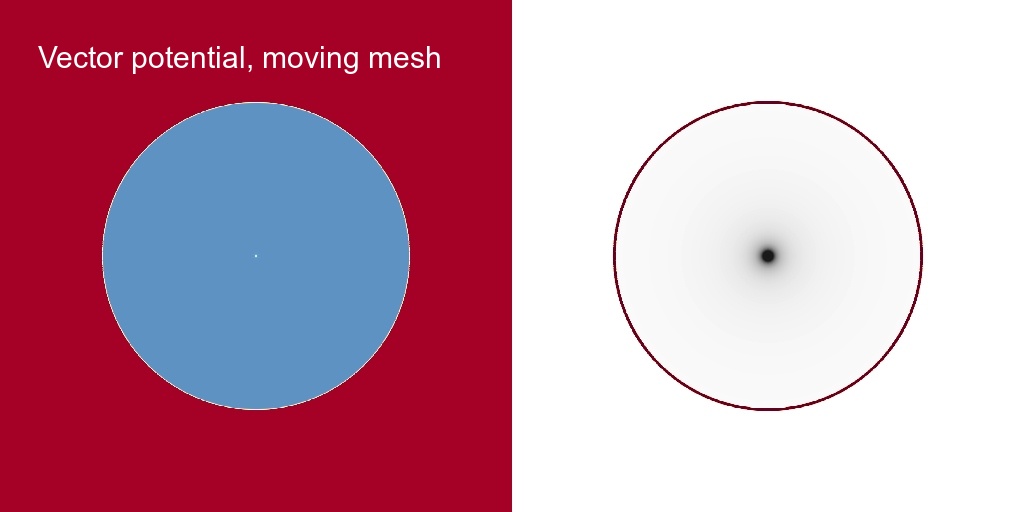}}\\
\resizebox{7.9cm}{!}{\includegraphics{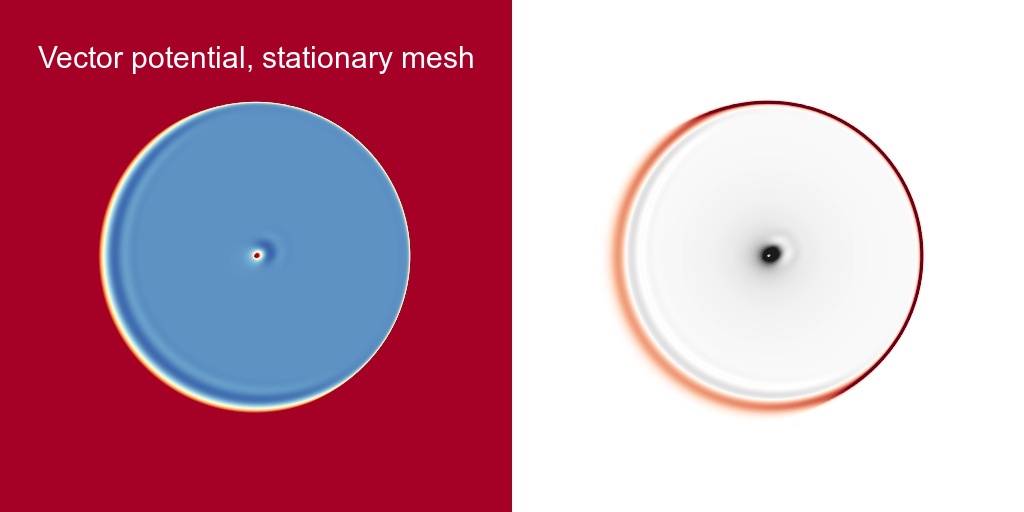}}\\
\resizebox{7.9cm}{!}{\includegraphics{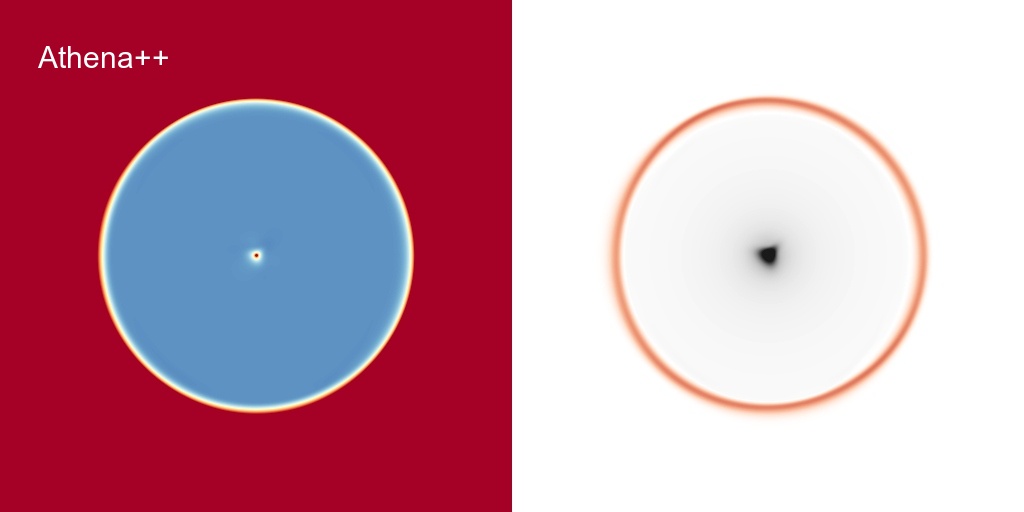}}\\
\resizebox{7.9cm}{!}{\includegraphics{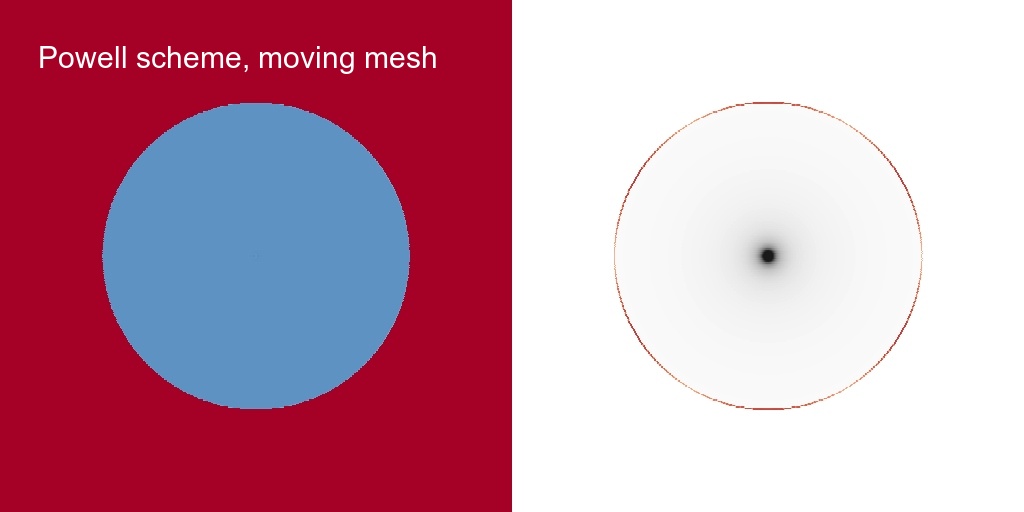}}\\
\resizebox{7.9cm}{!}{\includegraphics{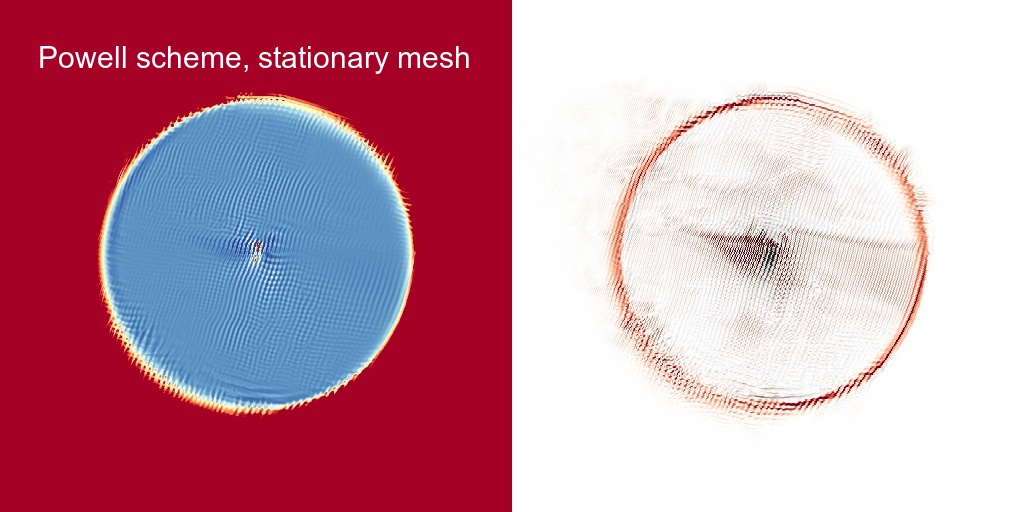}}\\
\resizebox{3.95cm}{!}{\includegraphics{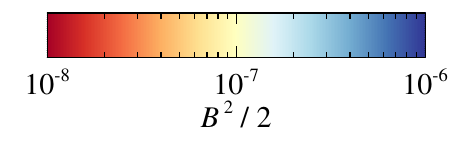}}%
\resizebox{3.95cm}{!}{\includegraphics{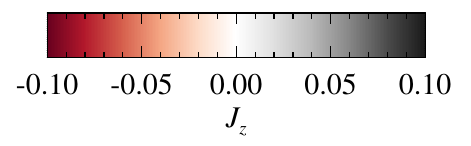}}\vspace*{-0.45cm}
\end{center}
\caption{Flux loop advection test. We show images of the magnetic energy density (left column) and the current density (right column) when the flux loop is advected once through the box. From top to bottom, we show results for our new vector potential formulation when applied to a moving mesh (top row) or a stationary mesh (second row). In the middle row, we show the {\small ATHENA++} code for a stationary Cartesian mesh. Finally, we include our old Powell 8-wave implementation for a moving mesh (fourth row), and for a fixed mesh (bottom row). 
\label{FigFluxLoopAdvection}}
\end{figure}

\begin{figure*}
\resizebox{5.5cm}{!}{\includegraphics{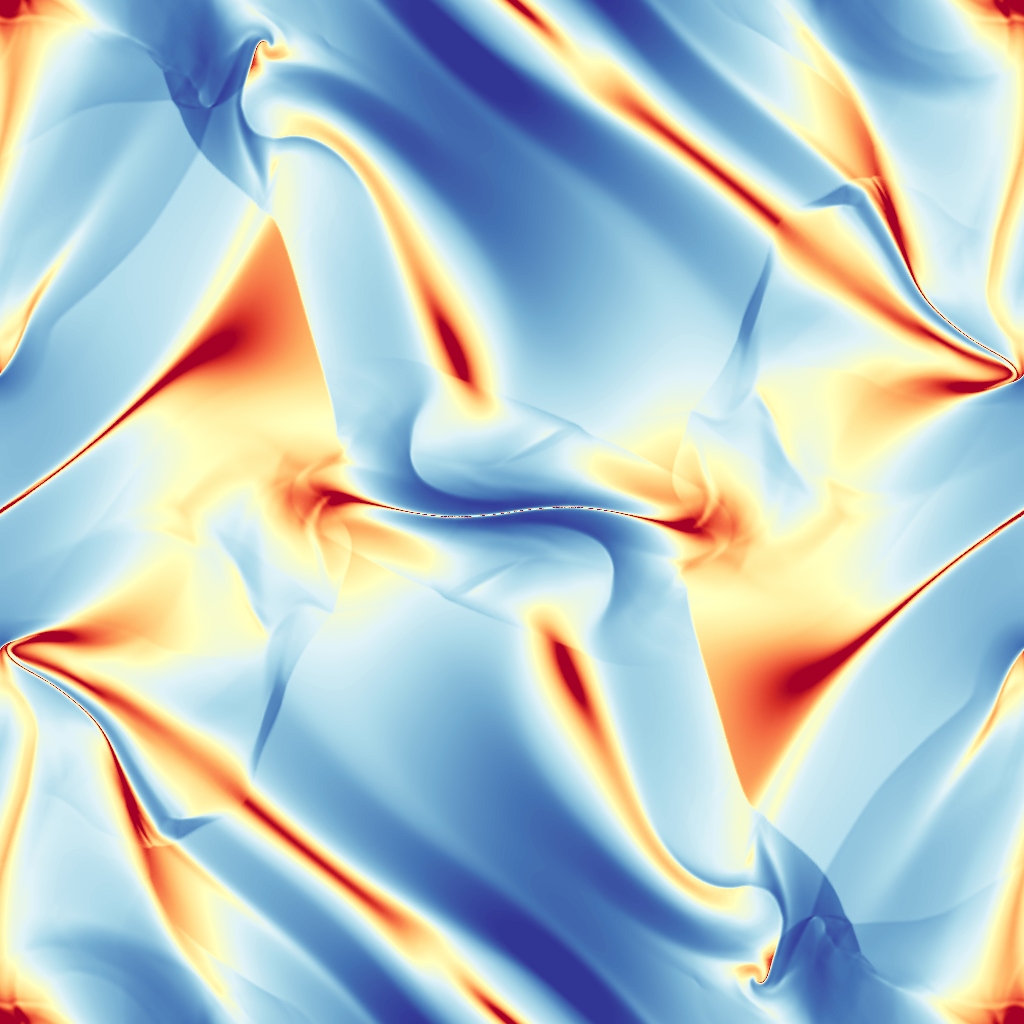}}\hspace*{0.2cm}%
\resizebox{5.5cm}{!}{\includegraphics{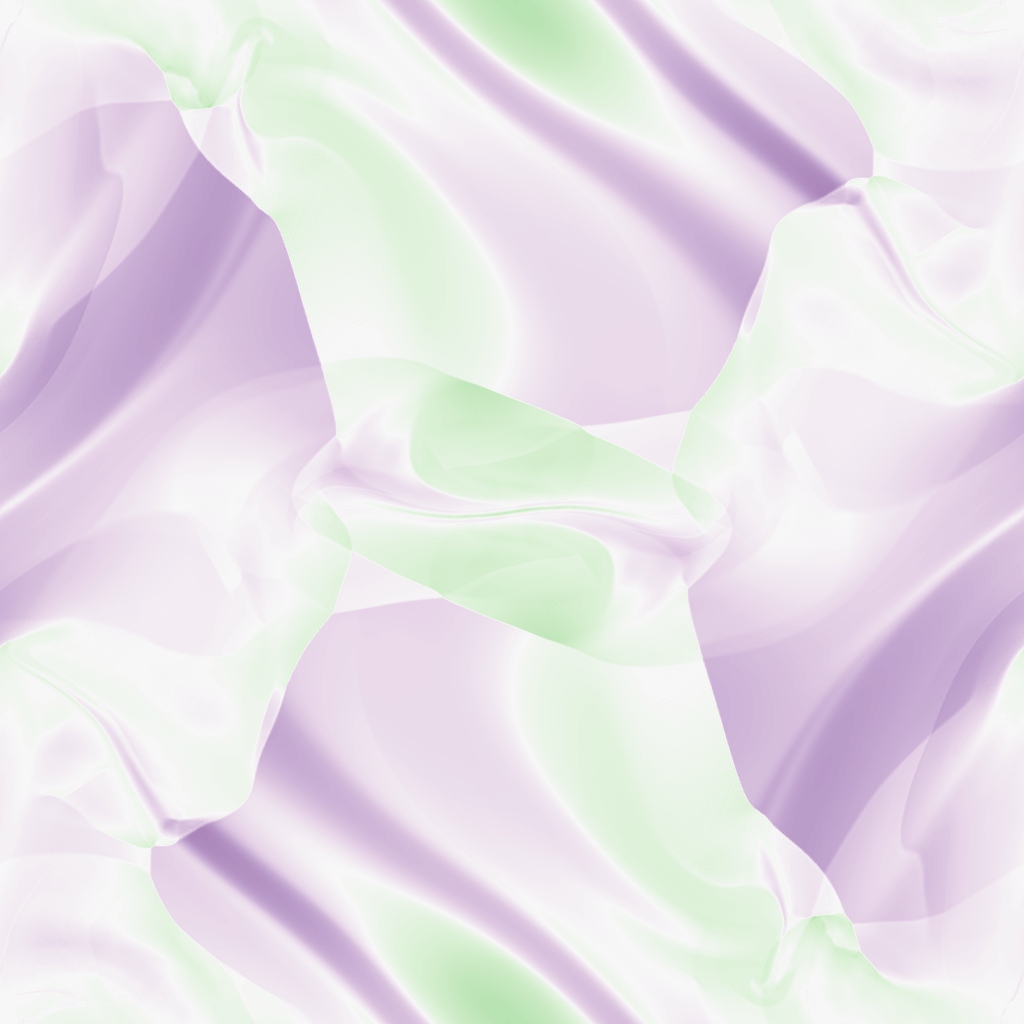}}\hspace*{0.2cm}%
\resizebox{5.5cm}{!}{\includegraphics{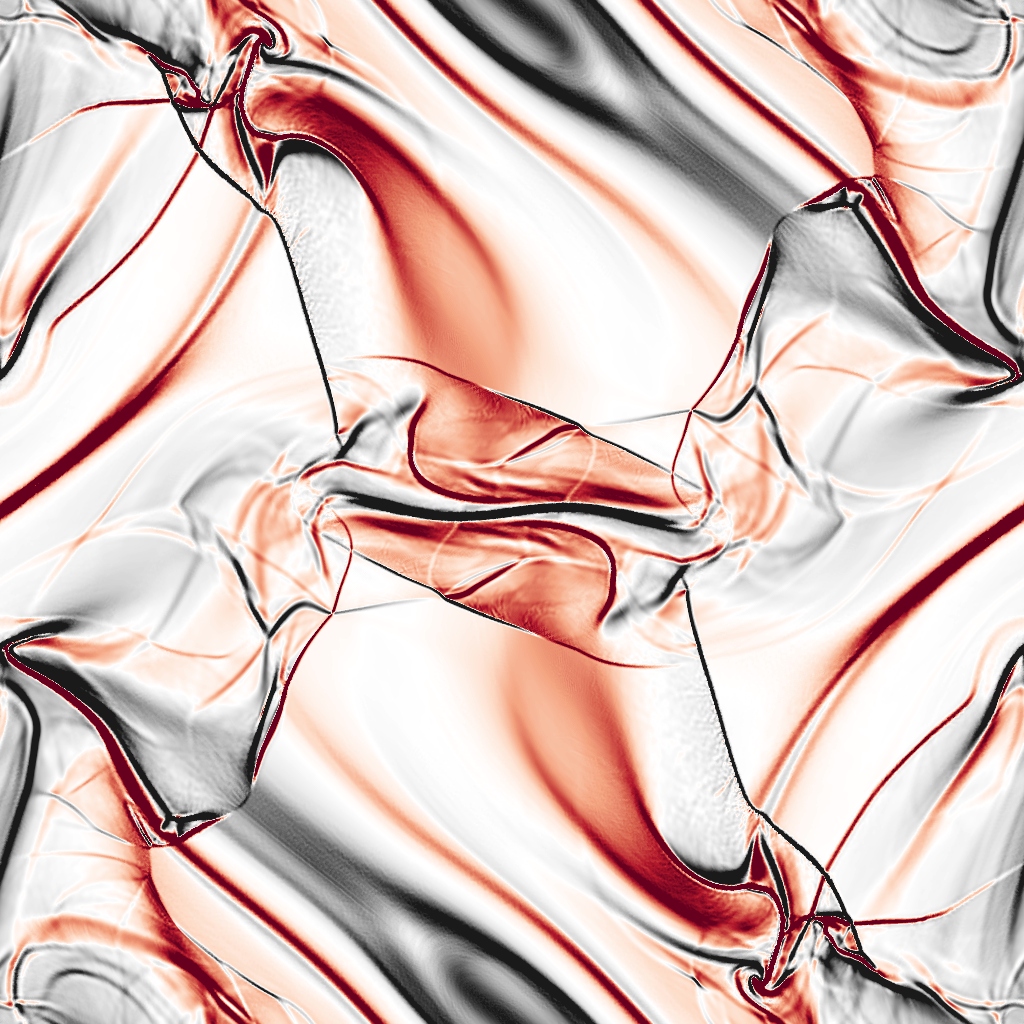}}\vspace*{0.2cm}\\
\resizebox{5.5cm}{!}{\includegraphics{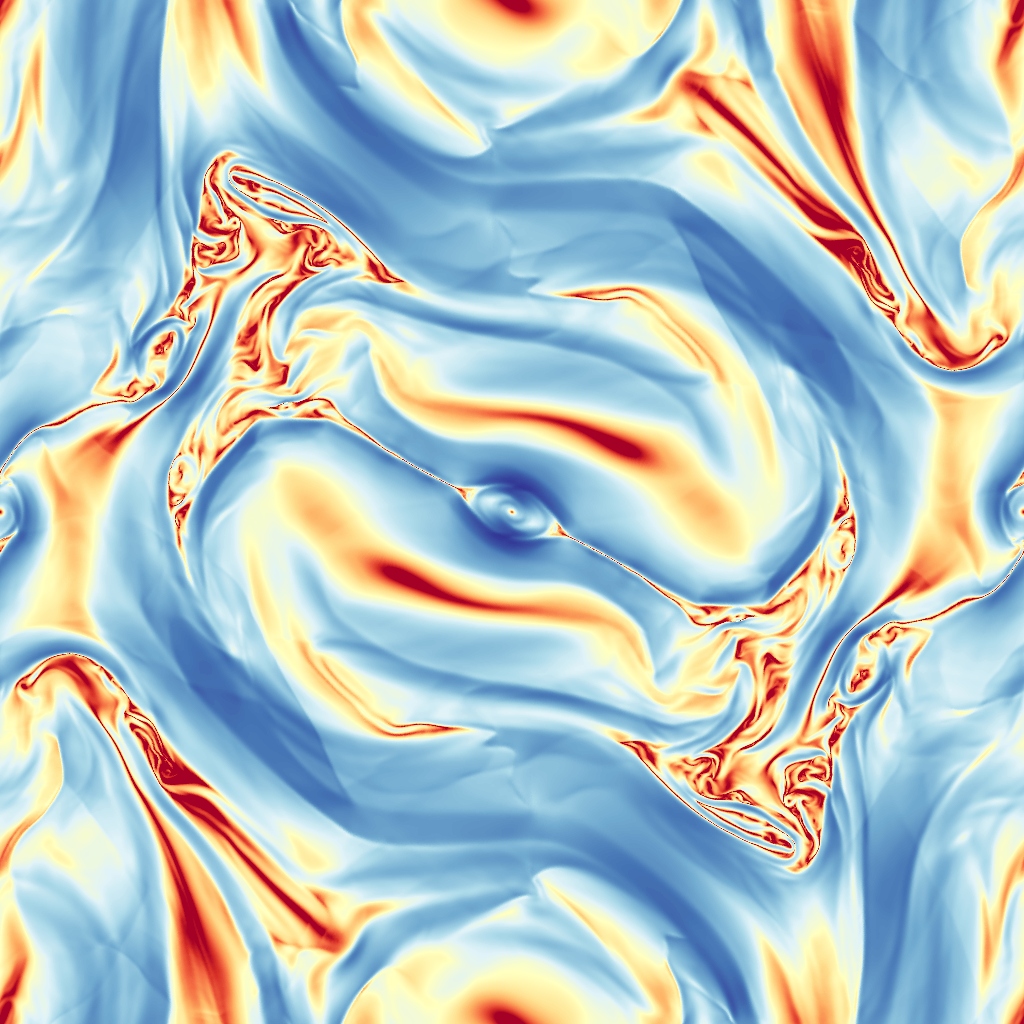}}\hspace*{0.2cm}%
\resizebox{5.5cm}{!}{\includegraphics{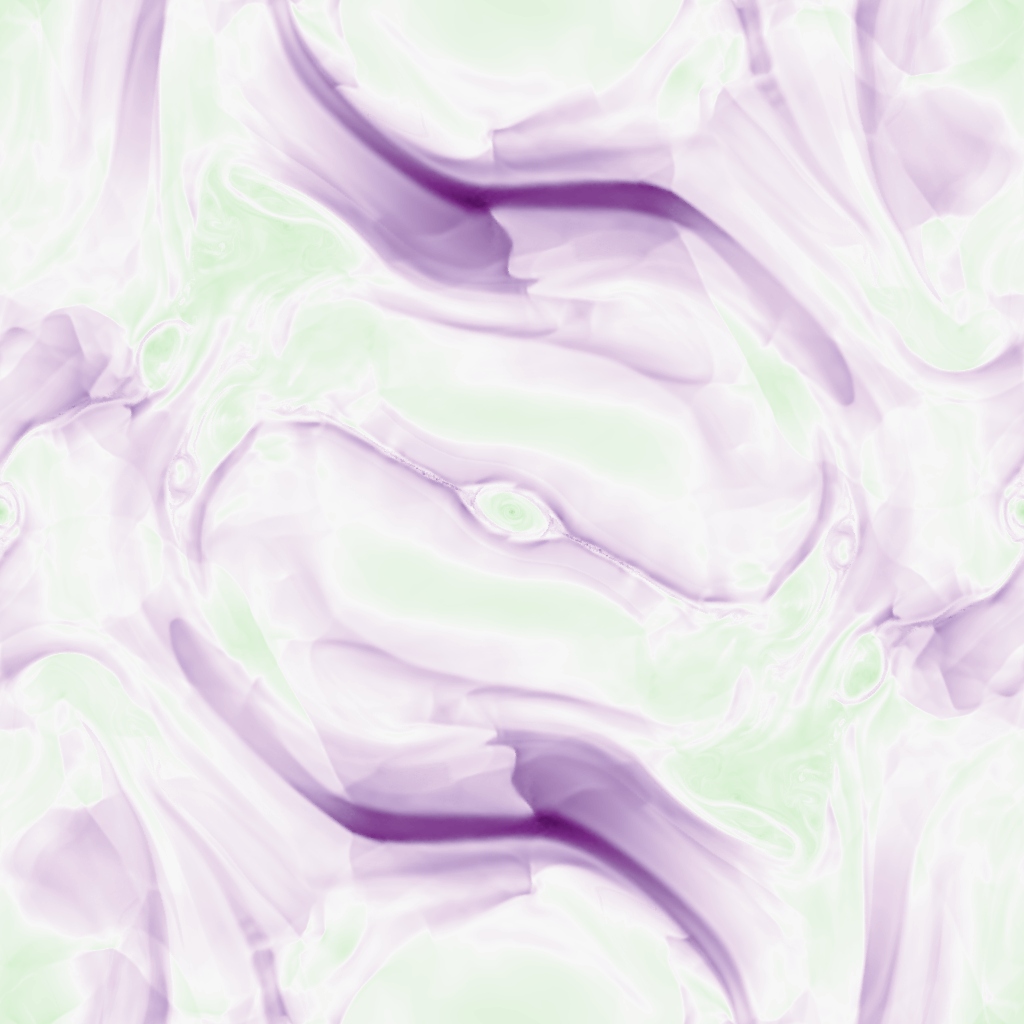}}\hspace*{0.2cm}%
\resizebox{5.5cm}{!}{\includegraphics{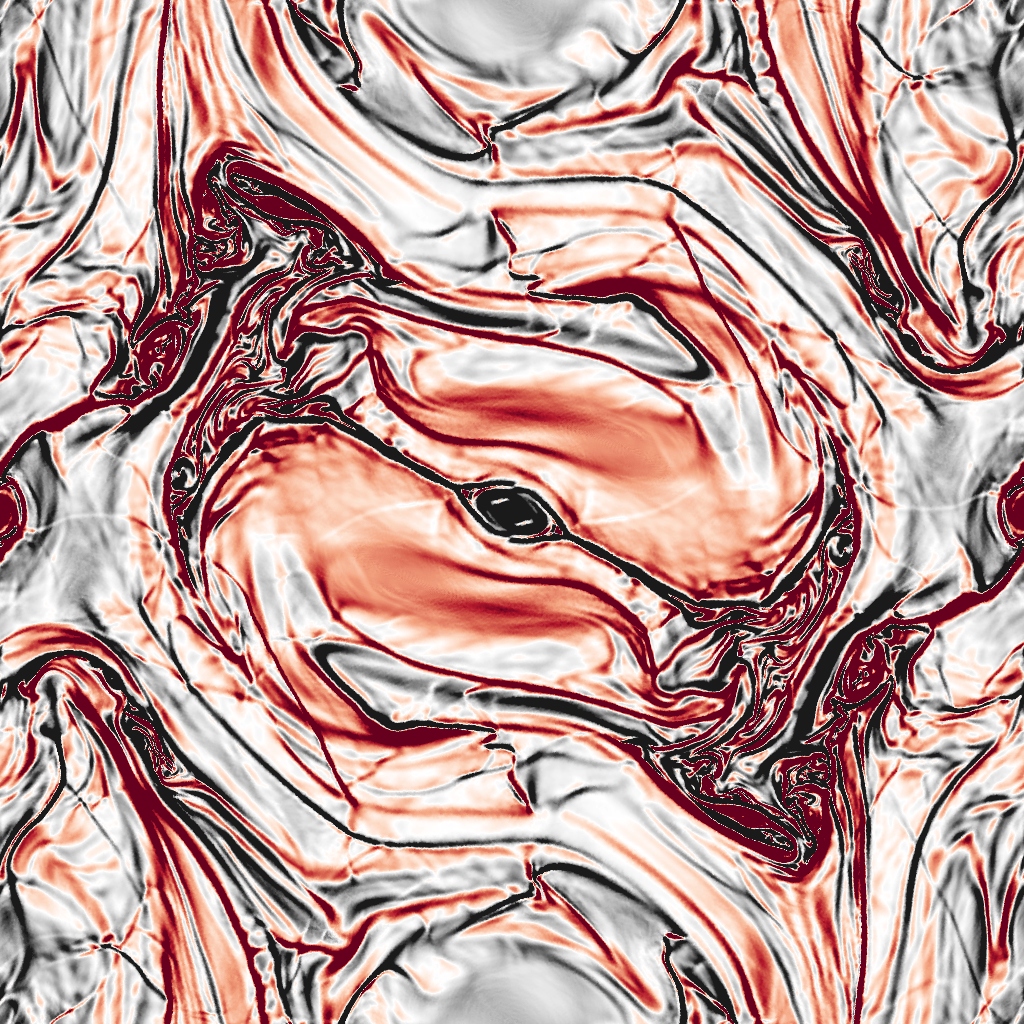}}\vspace*{0.2cm}\\
\resizebox{5.5cm}{!}{\includegraphics{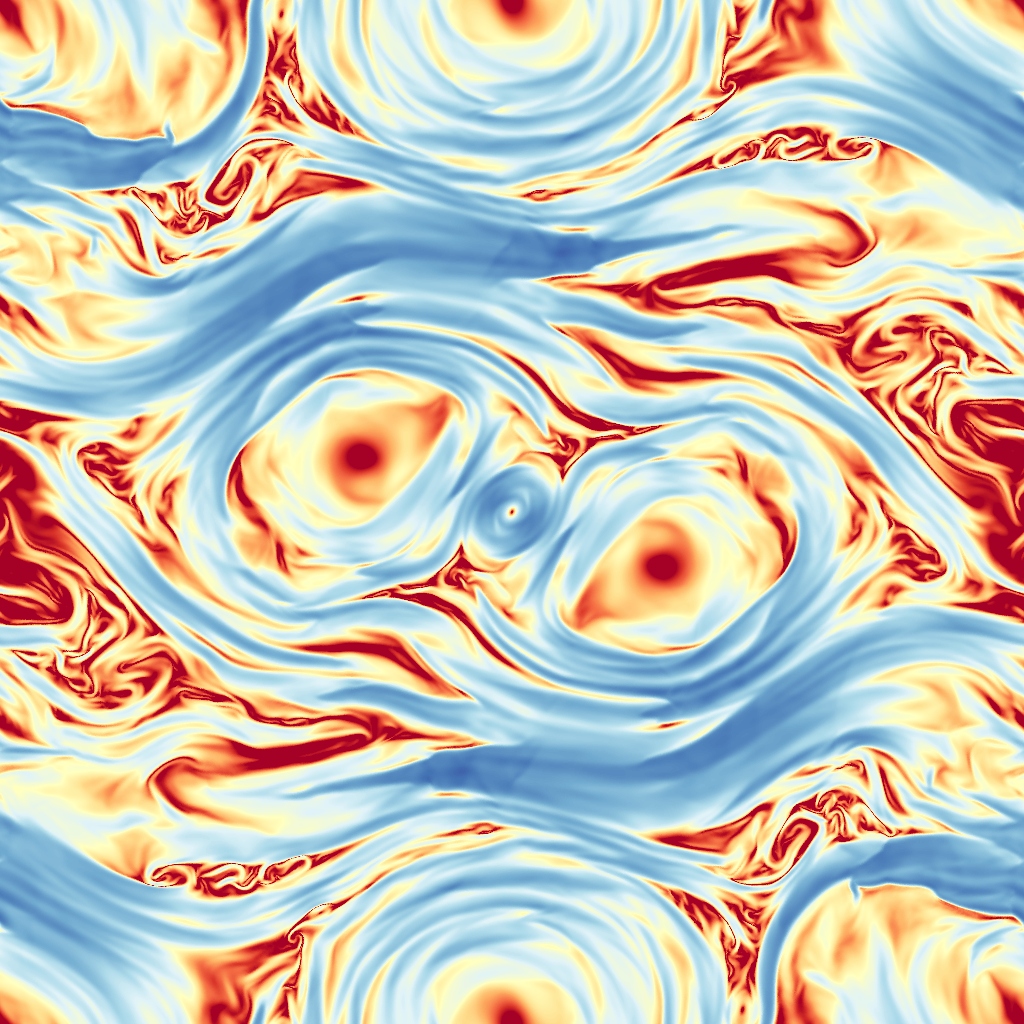}}\hspace*{0.2cm}%
\resizebox{5.5cm}{!}{\includegraphics{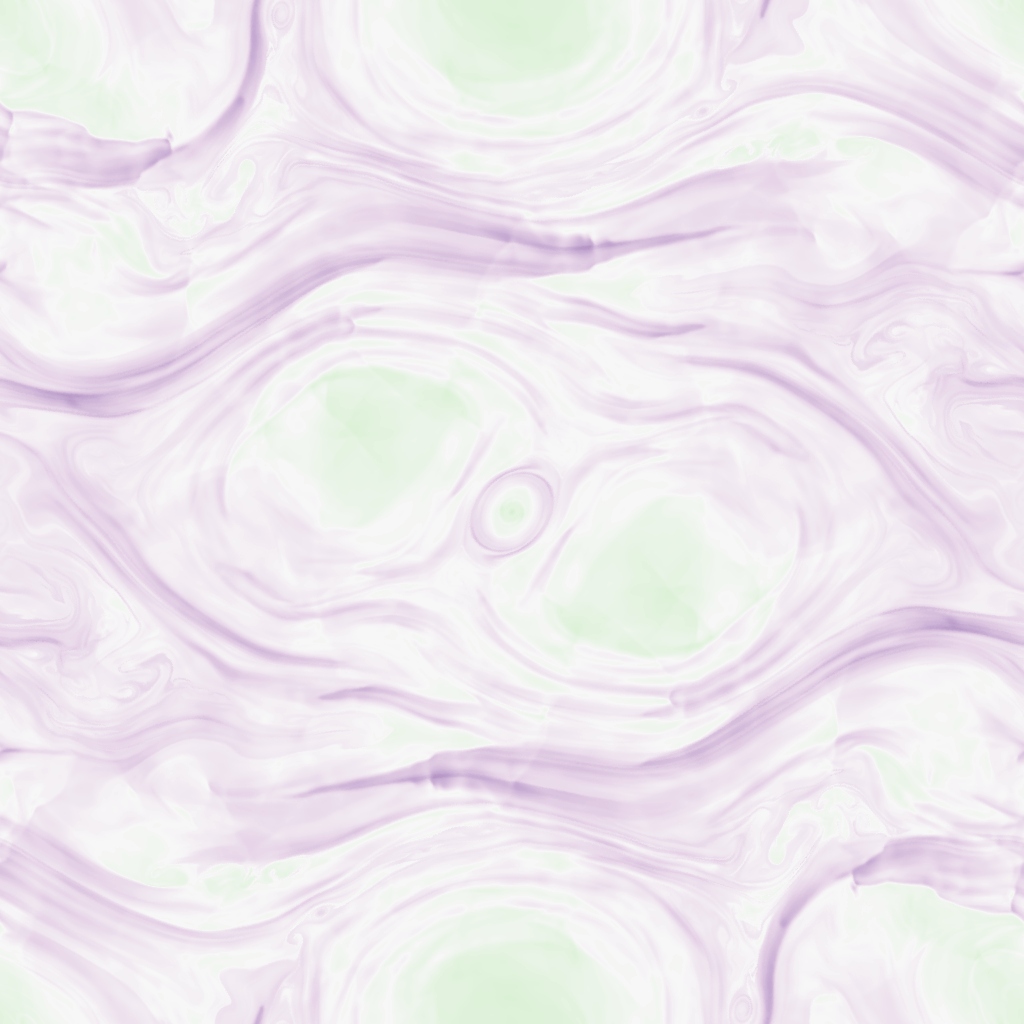}}\hspace*{0.2cm}%
\resizebox{5.5cm}{!}{\includegraphics{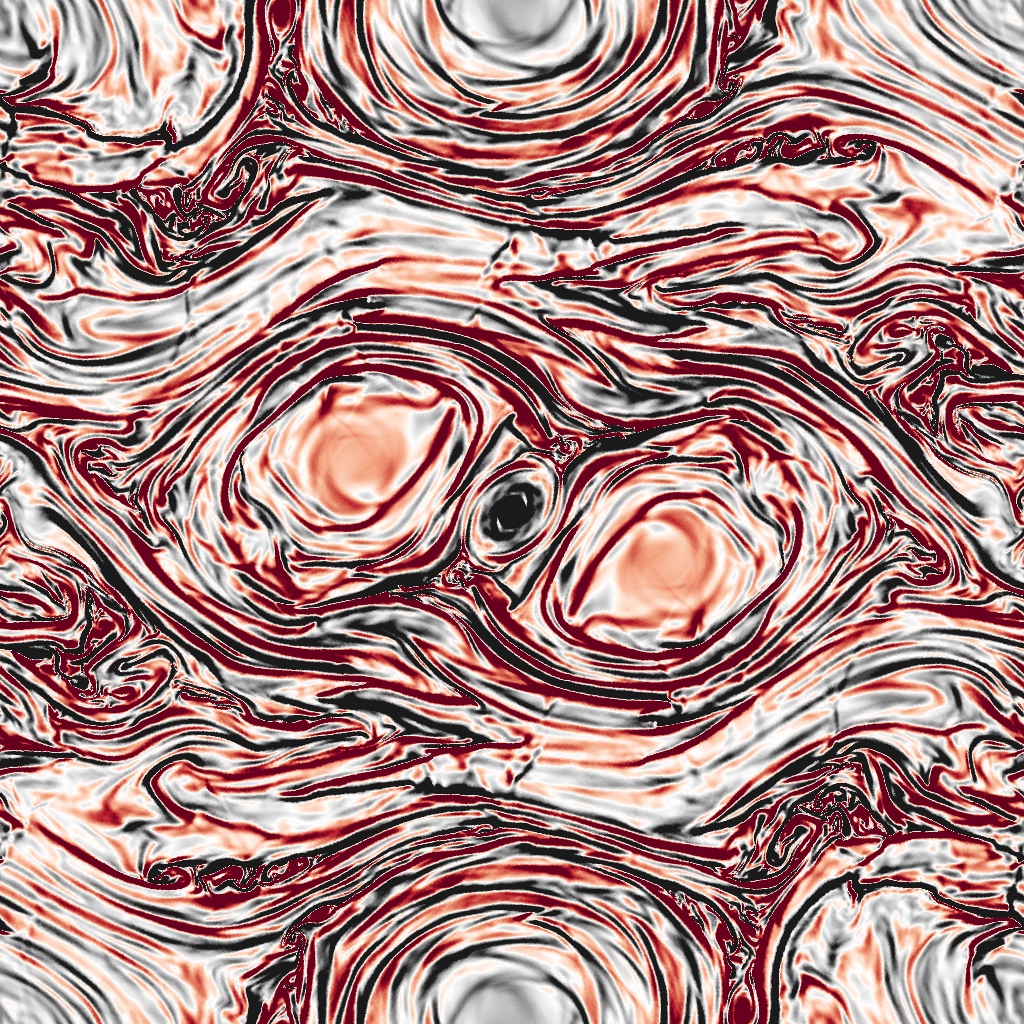}}\\
\resizebox{5.5cm}{!}{\includegraphics{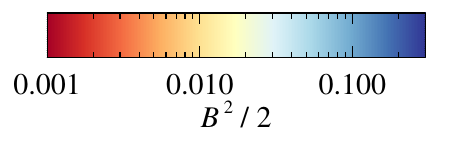}}\hspace*{0.2cm}%
\resizebox{5.5cm}{!}{\includegraphics{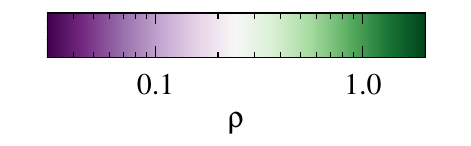}}\hspace*{0.2cm}%
\resizebox{5.5cm}{!}{\includegraphics{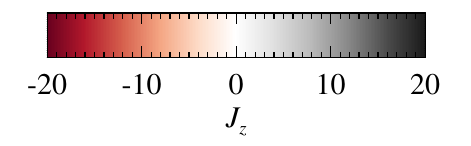}}
\caption{Orszag-Tang vortex flow test at times 0.5, 1.0, and 2.5, from top to bottom, computed with a moving-mesh, starting from an initially Cartesian mesh with $1024^2$ cells. We show maps of the magnetic energy (left), gas density (middle), and current density (right). Our scheme very robustly evolves this problem. Also note that our simulation maintains perfect symmetry despite the use of a moving Voronoi mesh, which is highly non-trivial and requires great care in the proper ordering of floating point operations and the treatment of spatial coordinates in the mesh construction to obtain exactly symmetric round-off errors. \label{FigOrszangTang}}
\end{figure*}

\subsection{Flux loop advection in 2D}

We here consider the advection of a magnetic field loop in 2D, with a magnetic field strength negligibly small compared to the thermal pressure. We use a box of unit side length, unit density, and unit thermal pressure $P=1$, and define the initial magnetic field through the vector potential
\begin{equation}
A_z = \left\{
\begin{array}{cc}
0.001/\sqrt{4\pi} \times (0.3 - r) & \mbox{for}\;\; r \le 0.3 ,\\
0 & \mbox{otherwise},
\end{array}
\right.
\end{equation}
where $r$ is the distance to the box centre. The resulting magnetic field is so small that it does not affect the dynamics. Instead, the test focuses on the question how well it is preserved during passive, supersonic advection. As advection velocity, we pick $\vec{v} = (2, 1, 0)$, following previous versions of this test in the literature.

In Figure~\ref{FigFluxLoopAdvection}, we show images of the magnetic energy density and the current of the flux tube at $t=1$, when it has returned to its starting position, comparing different numerical codes and schemes at a resolution of $512 \times 512$ cells. For the runs with a Voronoi mesh, a hexagonal mesh is used, whereas for {\small ATHENA++} \citep{Stone2008}, a Cartesian mesh is employed. The top row shows the moving-mesh result, obtained with our new MHD implementation. Here no advection errors occur at all, reflecting the Galilean invariance of the scheme. When we instead evolve the problem with a stationary mesh, as shown in the second row, clear advection errors become apparent, which are slightly asymmetric on the upwind and downwind sides of the flux loop. This can be fixed by adopting a more conservative (and more dissipative) minmod slope-limiter, at the price of more smoothing of the discontinuities. In the third row, we show the {\small ATHENA++} result, which does a better job in preserving the symmetry of the flux loop by providing equal amounts of dissipation on the up- and downwind sides without any signs of ringing.

Finally, we show the results of the Powell 8-wave cleaning approach as implemented in {\small AREPO} by \citet{Pakmor2013}. The fourth row gives the Powell moving-mesh result, which is again maintaining the initial state perfectly, thanks to Galilean invariance. In fact, here the magnetic field discontinuity at the outer edge of the flux loop can be represented even more sharply than in the vector potential formulation, as the magnetic field is obtained directly without a numerical differentiation. However, the Powell 8-wave method does not  handle the stationary mesh case particularly well, as evidenced by the result in the bottom row. Here some oscillatory artifacts develop that with time deteriorate the solution more and more, and can even cause instability in long-term evolutions of this sensitive test problem.

\begin{figure*}
Vector potential, moving mesh:\\
\resizebox{17cm}{!}{\includegraphics{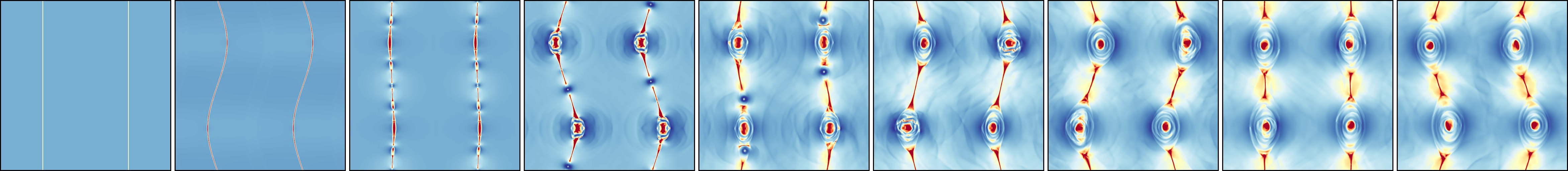}}\\
\resizebox{17cm}{!}{\includegraphics{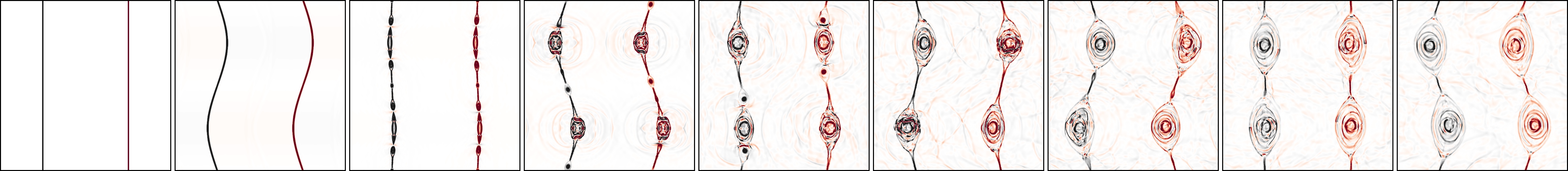}}\\
\ \\
Vector potential, stationary mesh: \\
\resizebox{17cm}{!}{\includegraphics{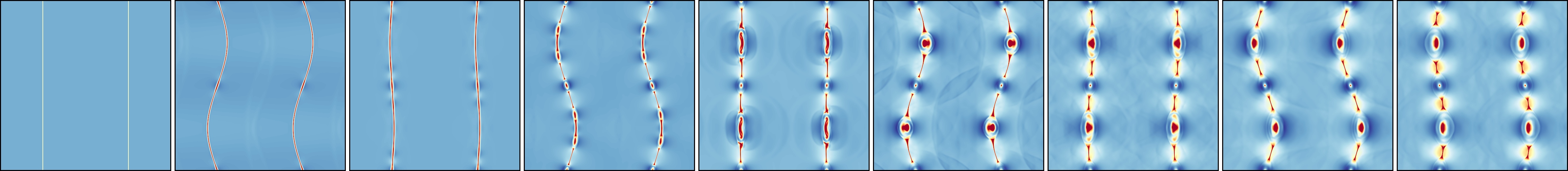}}\\
\resizebox{17cm}{!}{\includegraphics{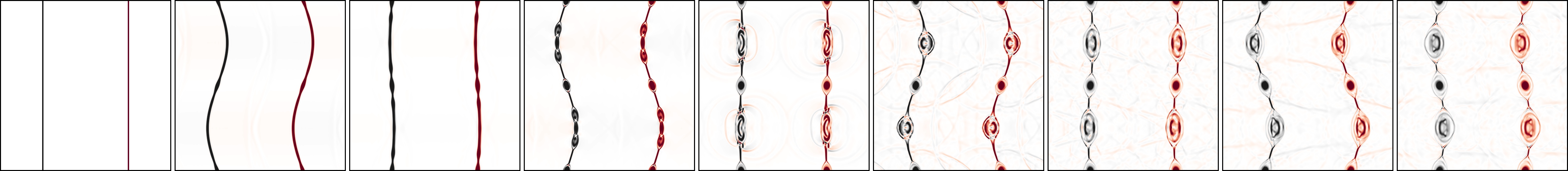}}\\
\ \\
Constrained transport, stationary mesh ({\small ATHENA++}): \\
\resizebox{17cm}{!}{\includegraphics{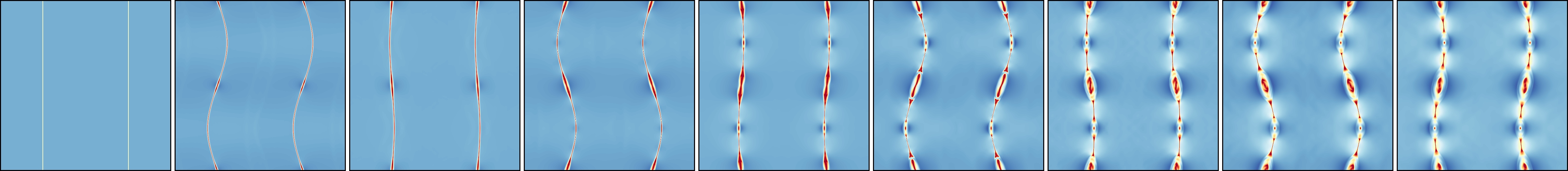}}\\
\resizebox{17cm}{!}{\includegraphics{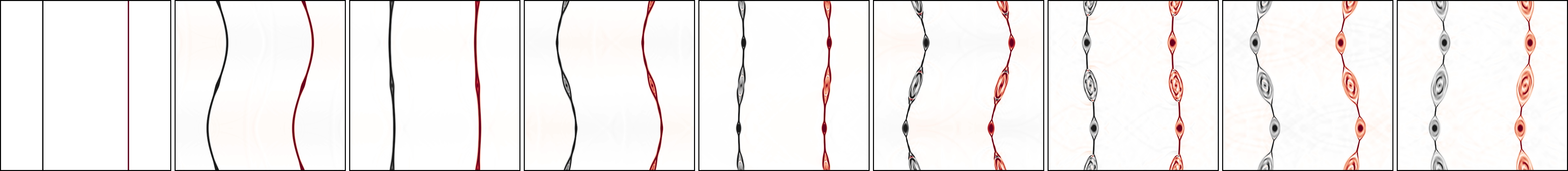}}\\
\ \\
Powell 8-wave divergence control, moving mesh: \\
\resizebox{17cm}{!}{\includegraphics{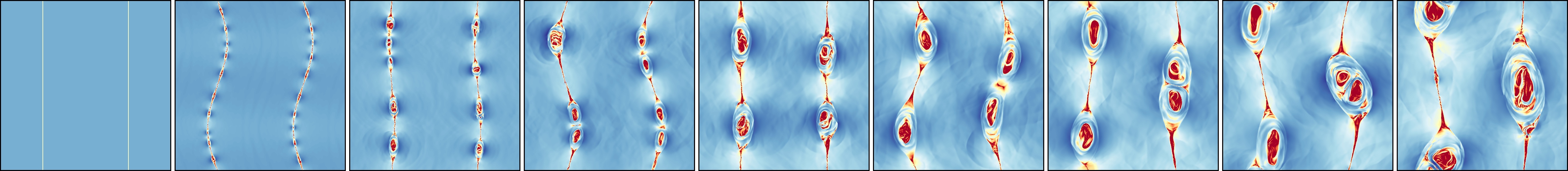}}\\
\resizebox{17cm}{!}{\includegraphics{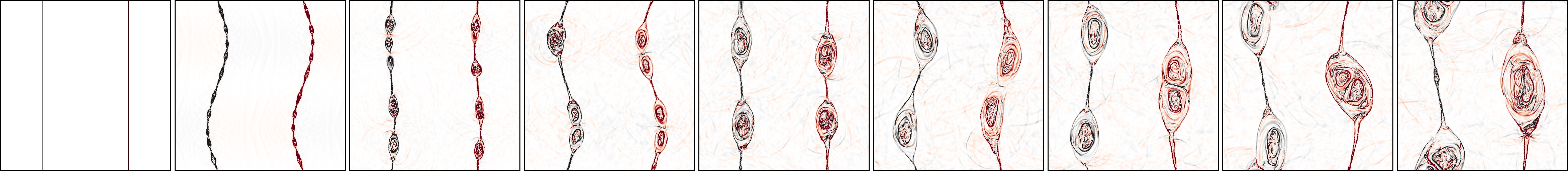}}\\
\ \\
\resizebox{4cm}{!}{\includegraphics{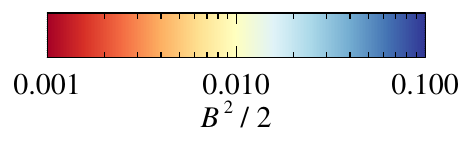}}\hspace*{4cm}%
\resizebox{4cm}{!}{\includegraphics{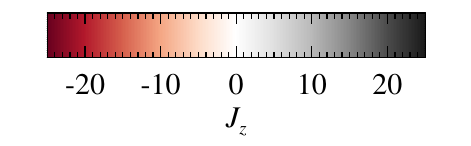}}
\caption{Current sheet test for four different simulation methods. Shown in rows from top to bottom are our new MHD technique for a moving and a stationary mesh, the {\small ATHENA++} constrained transport code, and finally our older Powell cleaning method with a moving mesh in {\small AREPO}, as labelled. In each case, the images placed horizontally show the time evolution of magnetic energy density and current density in terms of snapshots  spaced $\Delta t = 1.0$ apart in time, with the left-most image giving the initial conditions at $t=0$, and the right-most image displaying the final state at $t=8.0$.   \label{FigCurrentSheet}}
\end{figure*}

\begin{figure}
\begin{center}
\resizebox{8cm}{!}{\includegraphics{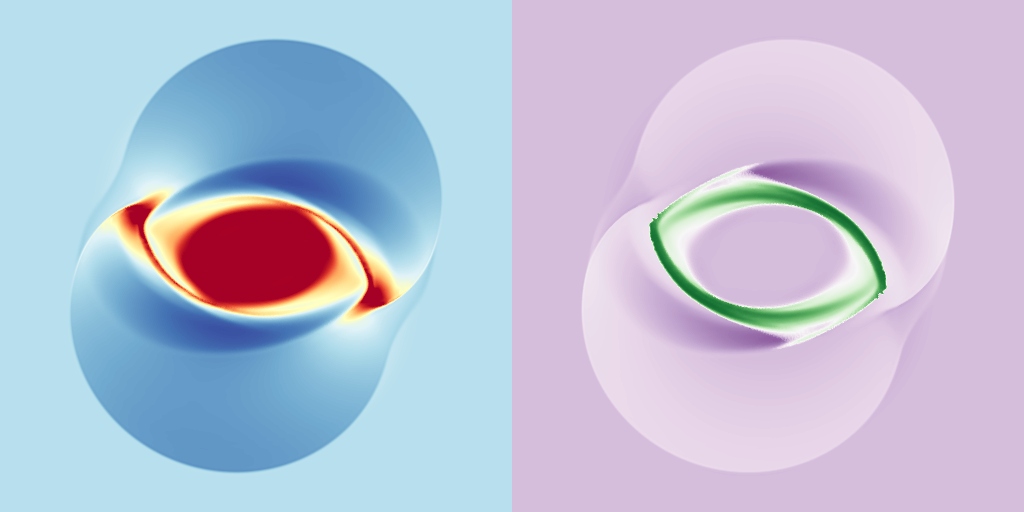}}\\
\resizebox{8cm}{!}{\includegraphics{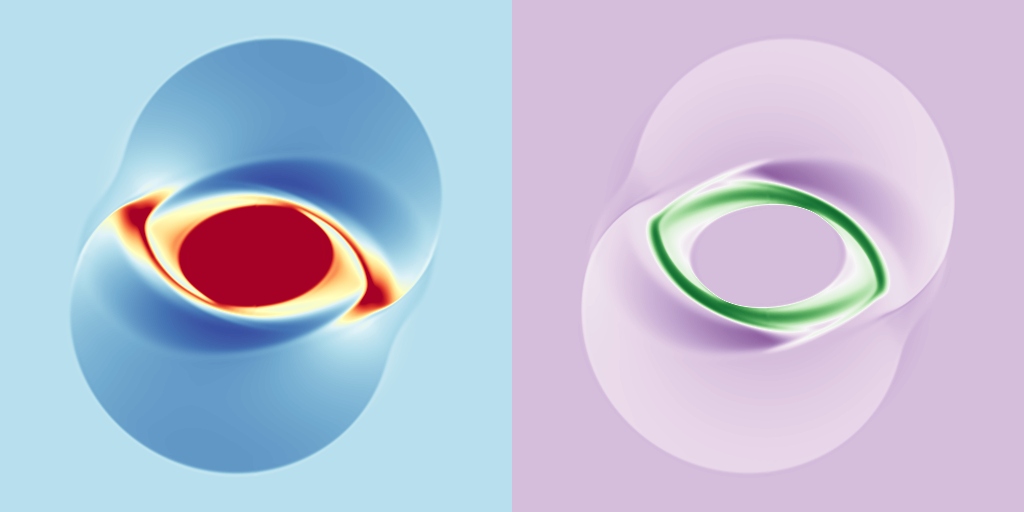}}\\
\resizebox{8cm}{!}{\includegraphics{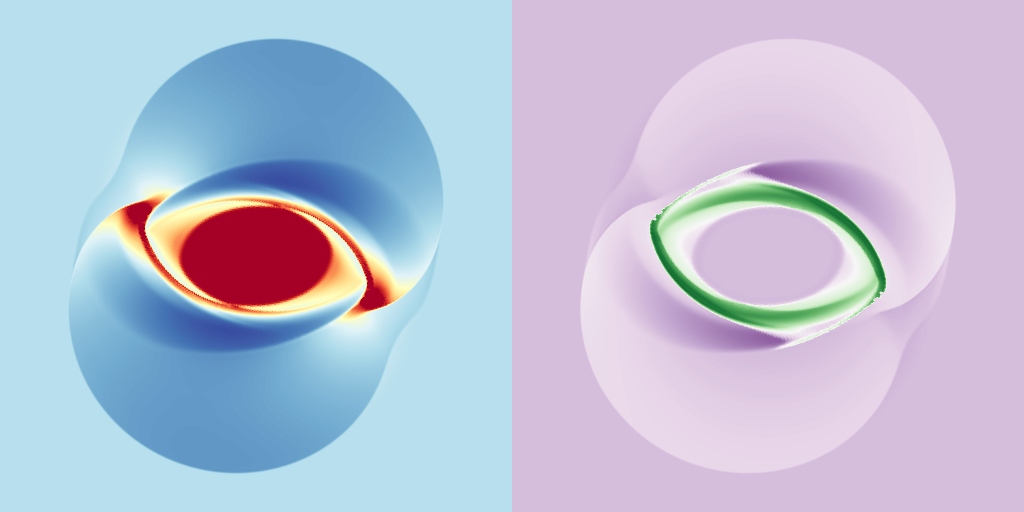}}\\
\resizebox{4cm}{!}{\includegraphics{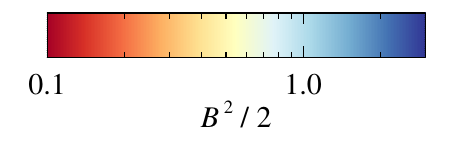}}%
\resizebox{4cm}{!}{\includegraphics{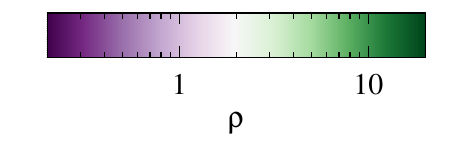}}
\end{center}
\caption{Magnetic rotor test. From top to bottom, we show  results at $t=1.15$ for our new vector potential formulation on the moving mesh (top), for a fixed mesh instead (middle), and for comparison for the Powell 8-wave cleaning method (bottom). The runs used $512\times 512$ Cartesian cells in the initial conditions. All methods have no problem with this test and produce results that closely align, both for the magnetic energy density (left column), and the gas density (right column). \label{FigMagneticRotor}}
\end{figure}

\subsection{Orszag-Tang Vortex in 2D}

The \citet{Orszag1979} vortex is a commonly employed MHD test problem in 2D, characterized by smooth initial conditions that give rise to a variety of shocks and a decaying turbulent state, and featuring regions where the plasma beta drops below unity, i.e.~where the magnetic energy density dominates over the thermal energy density. The initial state for a $\gamma=5/3$ gas is given by $\rho=\gamma^2 / (4\pi)$, $P = \gamma / (4\pi)$, velocity field 
$\vec{v}= [-\sin(2\pi\,y), \sin(2\pi\,x),0]$, and initial magnetic field $\vec{B}=1.0 / (4 \pi)^{1/2} [-\sin(2\pi y), \sin(4\pi x, 0)]$, which we however initialize via the vector potential $A_z = 1.0 / (4 \pi)^{3/2}  [ 2  \cos( 2\pi y)  + \cos( 4 \pi x)]$. The simulation domain is a periodic box of unit length, covering $[0, 1.0]^2$. 

In Figure~\ref{FigOrszangTang} we show the result for a moving-mesh simulation with $1024\times 1024$ cells, initially arranged as a Cartesian grid. We show images of the magnetic field, density field, and current density, at times $t=0.5$, $1.0$ and $2.5$. We note that this 2D problem can be run both with and without a smoothed large-scale velocity field. In the latter case, one can set $\vec{u}=\vec{v}$, which is possible in 2D without instabilities because here the term $\vec{A}\cdot\nabla\vec{v}$ always vanishes. In this case (which is shown in Fig.~\ref{FigOrszangTang}), our code  actually maintains {\it perfect} axial point symmetry around the centre, even for the case of a moving-mesh, something that is generally elusive in other Lagrangian codes \citep[e.g.][]{Karapiperis2026}. When we instead employ our default scheme with a smoothed velocity $\vec{u}$, we obtain a qualitatively very similar result but eventually lose axial symmetry due to floating point round-off differences originating in the computation of the exact value of the smoothed velocity field $\vec{u}$ applied at mesh centres.

\subsection{Current sheet}

This test, introduced by \citet{Hawley1995}, considers a tangential discontinuity of the magnetic field caused by a current sheet, and a superimposed transverse sinusoidal velocity perturbation that tries to bend the field, triggering oscillations of the current sheet that invoke an instability of the system to reconnection of the opposing field lines. We set up the test following \citet{Hawley1995}, using a 2D simulation domain and initial vector potential
\begin{equation}
A_z=B_0 \left\{
\begin{array}{ll}
x+0.5 & \mbox{for}\;\; x\le -0.25,\\
-x & \mbox{for}\;\;   -0.25 < x\le 0.25,\\
x-0.5 & \mbox{for}\;\;   x > 0.25.
\end{array}
\right.
\end{equation}
This yields a magnetic field $B_y = B_0$ in the region $<-0.25<x<0.25$ and $B_y = -B_0$ elsewhere, except right at the locations $x=\pm 0.25$ of the current sheets, where the cells left and right of these interfaces (assuming a Cartesian mesh) have half the field strength each, due to the fact that our gradient estimate (which is needed to derive $\vec{B}$ from $\vec{A}$) does not allow arbitrarily sharp discontinuities in the magnetic field but always broadens them over $\sim 2-3$ cells. The density is $\rho=1$, the adiabatic index is $\gamma = 5/3$, and the total pressure is set to $P_{\rm tot} = (1 + \beta P_{\rm mag})$, so that there is initial pressure equilibrium. We choose $\beta = 0.1$, making the magnetic energy dominant at the beginning. For still lower $\beta$, the problem becomes progressively harder, especially for codes that only evolve the total energy equation.

The initial velocity perturbation is set as $v_x = A\sin(2\pi y)$, with an amplitude of $A=0.1$ as chosen by \citet{Hopkins2016}. As the transverse motion bends the magnetic field lines, their tension force causes oscillations  of the current sheet with an Alfvén frequency until reconnection sets in. This creates plasmoids (magnetic islands) with locally closed field lines, and is associated with the emission of strong magnetosonic waves. The magnetic energy lost by reconnection will appear as heat. This problem is a good robustness test for any numerical MHD implementation, although the non-linear reconnection process depends sensitively on details of the numerical implementation for this setup.  

In Figure~\ref{FigCurrentSheet}, we show results for the current sheet test at a resolution of $512\times 512$ cells, started from a Cartesian grid. We present the time evolution of the problem in terms of maps of the magnetic energy density and the current density, for four different simulations. In each of the cases (given by the different rows, as labelled), we show the evolution until $t=8.0$, with intermediate results spaced $\Delta t= 1.0$ apart. In the top two rows, we give results for our new method for a moving and a stationary grid, respectively. The latter may be, in particular, compared to the fixed grid result obtained with the {\small ATHENA++} code which is shown in the third row, while the moving mesh result may be contrasted with an alternative moving-mesh outcome on the bottom row obtained with the older Powell cleaning method in {\small AREPO}. 

Interestingly, at time $t=1.0$ (second column), the current sheet is still stable in our new moving-mesh implementation, whereas first signs of the onset of reconnection are already visible for the two fixed-mesh results. In contrast, in the Powell scheme, some small-scale reconnection has already happened at this time, and also in the subsequent evolution the plasmoid formation proceeds to a more evolved state, in which the smaller plasmoids along the current sheet have coalesced into only one large reconnection island on each of the sheets. While such merging also takes place in our new moving-mesh scheme, the symmetry is maintained much better and still two plasmoids remain on each of the current sheets. 

The two fixed mesh results show a qualitatively similar outcome at the final time of $t=8.0$, but the reconnection occurs somewhat earlier and proceeds slightly more violently, as indicated by the larger perturbations in the background, in the {\small AREPO} simulation compared to the {\small ATHENA++} run. The detailed morphology of the plasmoids appears just interchanged between {\small AREPO} and {\small ATHENA++}, perhaps reflecting the offset in the storage locations of the vector potential in the two methods, which differs by half a grid spacing. Overall, all four codes handle the reconnection robustly, but the details of the non-linear evolution are quite different and show significant sensitivity to numerical details of the solvers. The slower progression of reconnection in the fixed mesh simulations is consistent with the expected higher numerical dissipation in these codes when the current sheet moves. Compared to the Powell method, our new moving-mesh formulation appears better in suppressing spurious noise, as evidenced by its ability to delay the onset of reconnection much longer.

\subsection{Magnetic rotor}

Another popular 2D test problem for MHD is the rotor by 
\citet{Balsara1999}, which tests, in particular, the propagation of nonlinear shear Alfvén waves. We use the first setup described in \citet{Toth2000} for this test \citep[see also][]{Stone2008, Hopkins2016}. A 2D domain is filled with gas at unit pressure and adiabatic index $\gamma = 7/5$. The initial magnetic field is uniform along the $x$-axis, with $B_x = 5/\sqrt{4\pi}$. An inner disk with radius $r_0 = 0.1$ is given a density of $\rho = 10$, and is set into solid body rotation with $\vec{v} = v_0/r_0[-y, x, 0]$ and $v_0=2$. A transition region is defined in a ring between $r_0=0.1$ and $r_1=0.115$ around the disk, where the density declines according to $\rho(r) = 1 + 9 f(r)$, with $f(r) = (r_1-r)/(r_1-r_0)$. Similarly, the rotational velocity declines as $\vec{v}(r)= v_0 f(r)/r\, [-y, x, 0]$ in this region. Outside this ring, the initial density is $\rho=1$, and the gas is at rest. 

In Figure~\ref{FigMagneticRotor}, we show the outcome of this magnetic rotor test at time $t=1.15$, for three different simulation methods computed with an initial resolution of $512\times 512$ cells setup as a Cartesian mesh. In the top row, we give results for our new MHD method with a moving mesh, in the middle row we show corresponding results if the mesh is kept fixed instead, and finally, the bottom row shows the result when the Powell 8-wave method with a moving mesh is used, for comparison. In each case, we show maps of the magnetic energy density (left) and the gas density (right). All three codes have no problem with this test and produce results that are extremely similar.

\begin{figure}
\begin{center}
\resizebox{8cm}{!}{\includegraphics{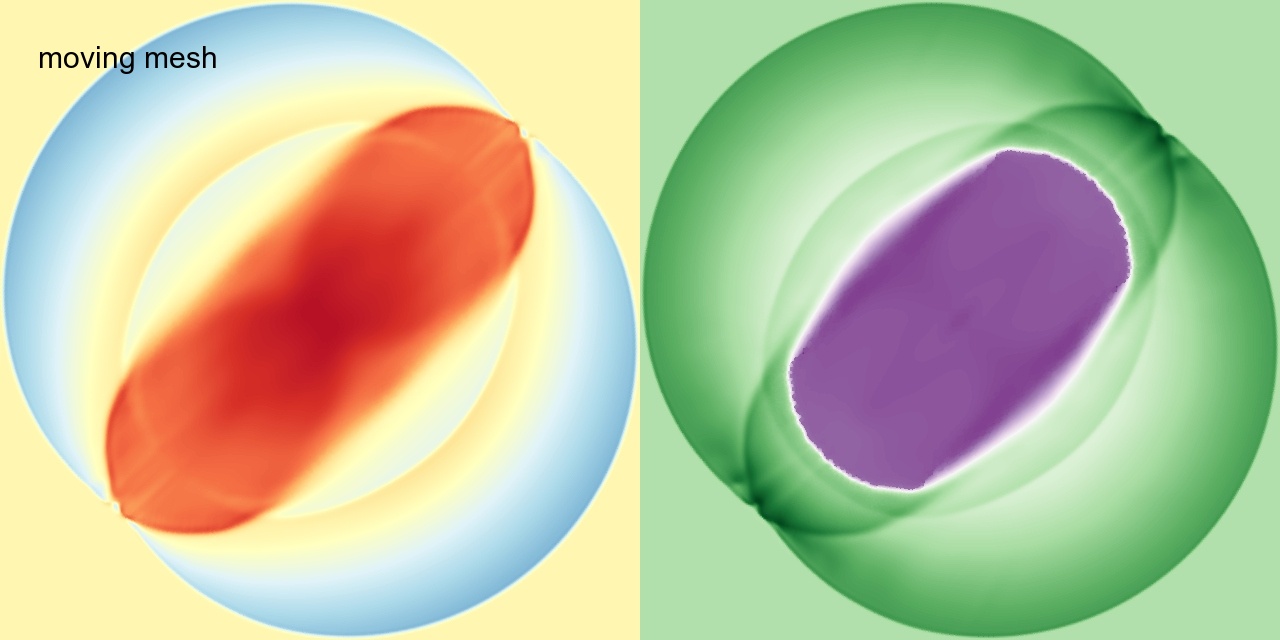}}\\
\resizebox{8cm}{!}{\includegraphics{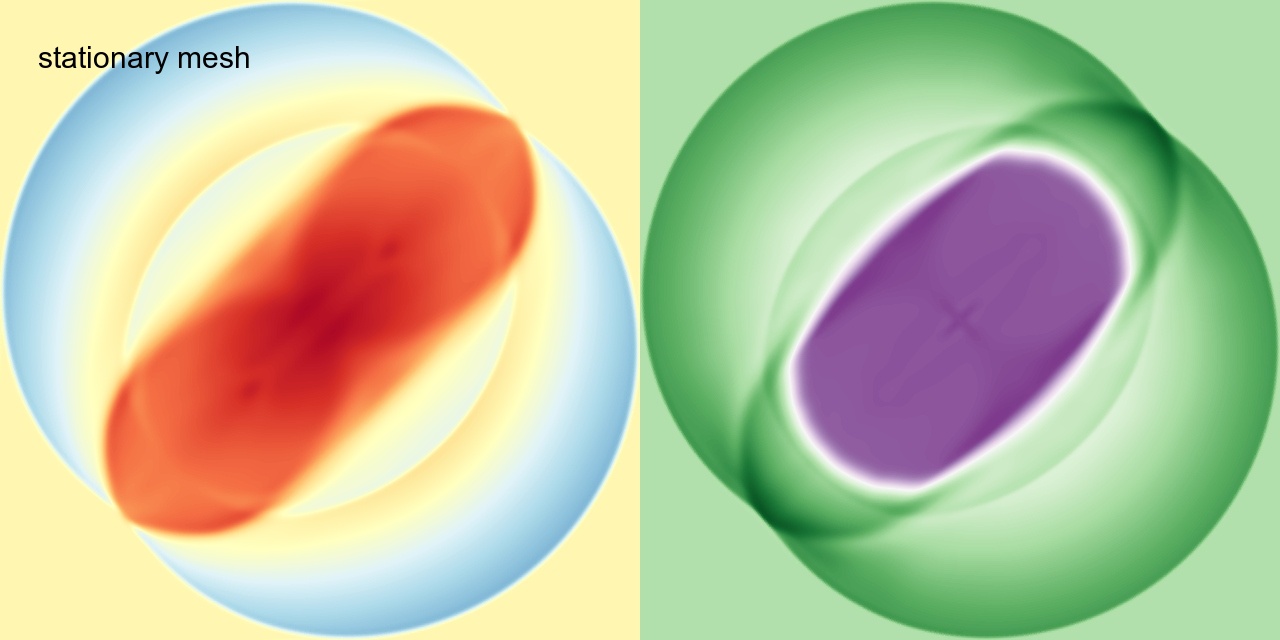}}\\
\resizebox{4cm}{!}{\includegraphics{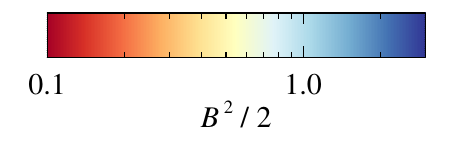}}%
\resizebox{4cm}{!}{\includegraphics{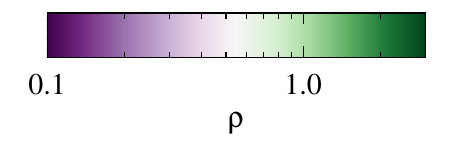}}%
\end{center}
\caption{MHD blast wave test in 2D at time $t=0.2$. In the top row, we show results for the moving mesh, in the bottom for the stationary mesh. The simulations have here been started from a hexagonal mesh with $2\times (256\times 147)$ cells. Use of a Cartesian mesh with  $256\times 256$ cells instead yields  very similar results. \label{FigBlast2D}}
\end{figure}

\subsection{MHD blast wave in 2D and 3D}
\label{secTestBlast2D}

Another test problem considered by many authors \citep[e.g.][]{Gardiner2008, Pakmor2011} is a point-like explosion in a magnetized medium, which involves strong shocks, strong rarefactions, and their interaction with magnetic fields. Various variants of this test exist. We shall first consider a typical 2D setup, followed by a further test in 3D.

In our 2D test, we follow \citet{Hopkins2016} and initialize a 2D periodic domain of unit size with uniform gas density $\rho=1$ initially at rest, adiabatic index $\gamma=5/3$, and a uniform magnetic field strength equal to $|\vec{B}| = 1$ oriented along the lower-left to upper-right diagonal. Inside a central circle of radius $R=0.1$, the thermal pressure is set to $P=10$, and outside to $P=0.1$. 

In Figure~\ref{FigBlast2D}, we compare results for the magnetic energy density and gas density at time $t=0.2$ obtained with our new method for the moving and the stationary mesh cases. We here start the simulations from a hexagonal grid, created by combining two rectangular grids of mesh generating points with $256\times 147$ points each, and where the second grid is shifted by half a particle spacing in each dimension relative to the first grid in a body centred fashion. Results started instead from a $256\times 256$ Cartesian grid are extremely similar.

For our 3D version of the blast wave test we follow \citet{Gardiner2008}, based on the parameters of \citet{Londrillo2000}. The test is carried out in a 3D box of unit side-length, with initially uniform gas density $\rho=1$ and uniform magnetic field $\vec{B} = 10/\sqrt{2} [1, 0, 1]$ along the $xz$-diagonal. The thermal pressure inside a central spherical region of radius $R=0.125$ is set to $P=100$, yielding a plasma beta of $\beta=2$. Outside this region, the pressure is much lower, $P=1$, with $\beta = 0.02$, i.e.~the magnetic pressure is dominating in the background. The gas has $\gamma=5/3$ and is initially at rest, and we evolve the system until $t=0.02$, to allow comparison with \citet{Gardiner2008}.

\begin{figure}
\begin{center}
\resizebox{4cm}{!}{\includegraphics{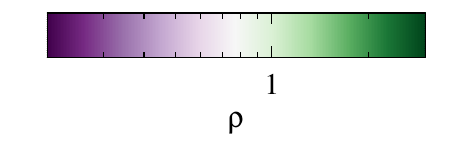}}%
\resizebox{4cm}{!}{\includegraphics{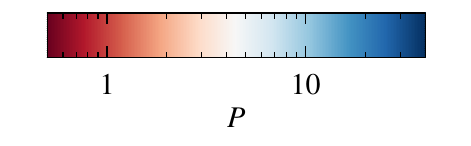}}\\%
\resizebox{8cm}{!}{\includegraphics{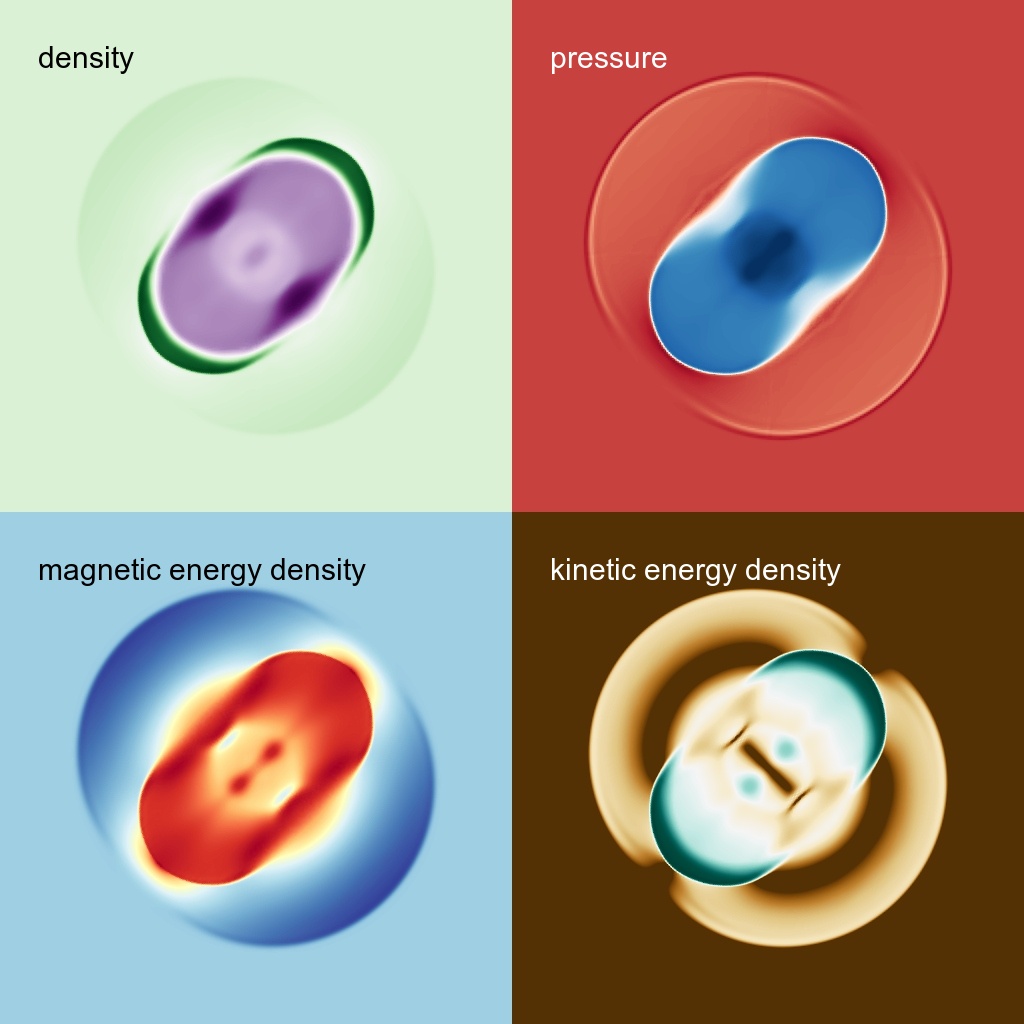}}\\
\resizebox{4cm}{!}{\includegraphics{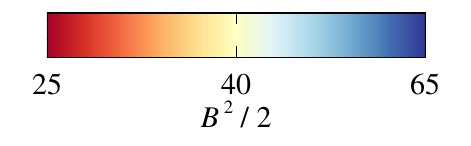}}%
\resizebox{4cm}{!}{\includegraphics{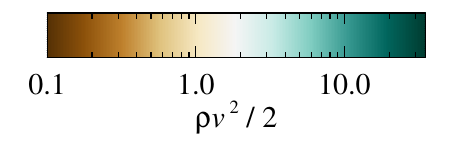}}\\%
\end{center}
\caption{MHD blast wave test in 3D at time $t=0.02$, following the setup of \citet{Gardiner2008}. We use a moving mesh, starting from a Cartesian mesh with $256^3$ cells.  \label{FigBlast3D}}
\end{figure}

In Figure~\ref{FigBlast3D} we show results for carrying out the test on a moving grid with initially $256^3$ cells arranged initially as a Cartesian mesh. The four panels give maps of the density, pressure, magnetic energy density, and kinetic energy density in the $xz$-plane through the $y$-midpoint of the box at time $t=0.02$. The code stably integrates this problem (including with local timesteps), irrespective of whether we use a moving or a stationary mesh. An exact quantitative comparison with previous results is not readily possible, but the maximum and minimum values reported by \citet{Gardiner2008} for the four fluid quantities displayed in the figure are very close to what we find.

\begin{figure}
\begin{center}
\resizebox{4.0cm}{!}{\includegraphics{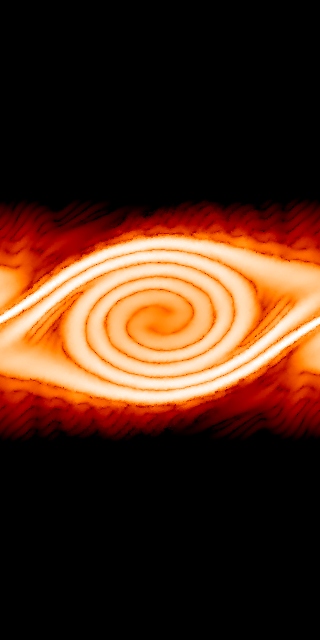}}\hspace*{0.4cm}%
\resizebox{4.0cm}{!}{\includegraphics{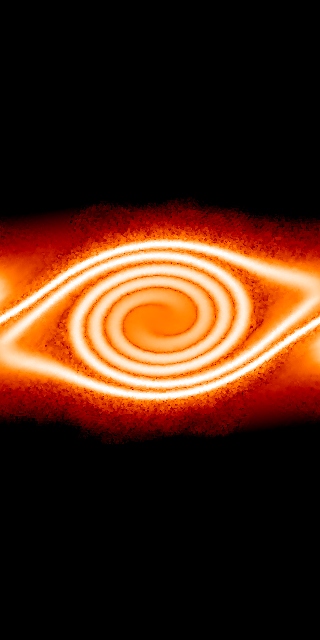}}\\%
\resizebox{7.0cm}{!}{\includegraphics{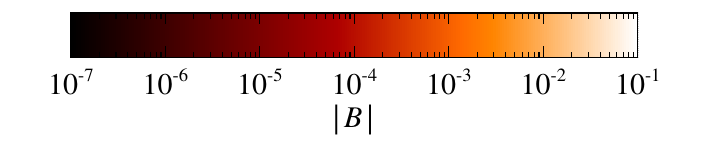}}%
\end{center}
\caption{Map of the magnetic field strength at $t=3.0$ in the localized field Kelvin-Helmholtz instability test on a moving-mesh. On the left, we show a result computed in 2D with a resolution of $128\times 256$ cells, and on the right in 3D with $128\times 256\times 8$ cells. There is no build-up of field outside of the initial field region despite the use of $\nabla \cdot\vec{A}$ cleaning, and the results are consistent with constrained transport results of \citet{Tomida2026} for a stationary mesh. \label{FigKHimage}}
\end{figure}

\begin{figure}
\begin{center}
\resizebox{8.0cm}{!}{\includegraphics{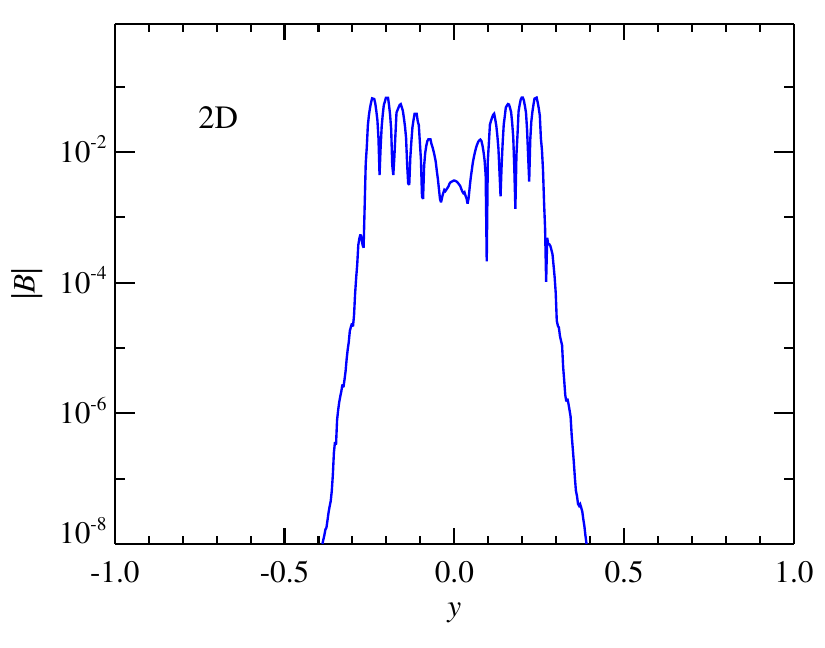}}\\%
\resizebox{8.0cm}{!}{\includegraphics{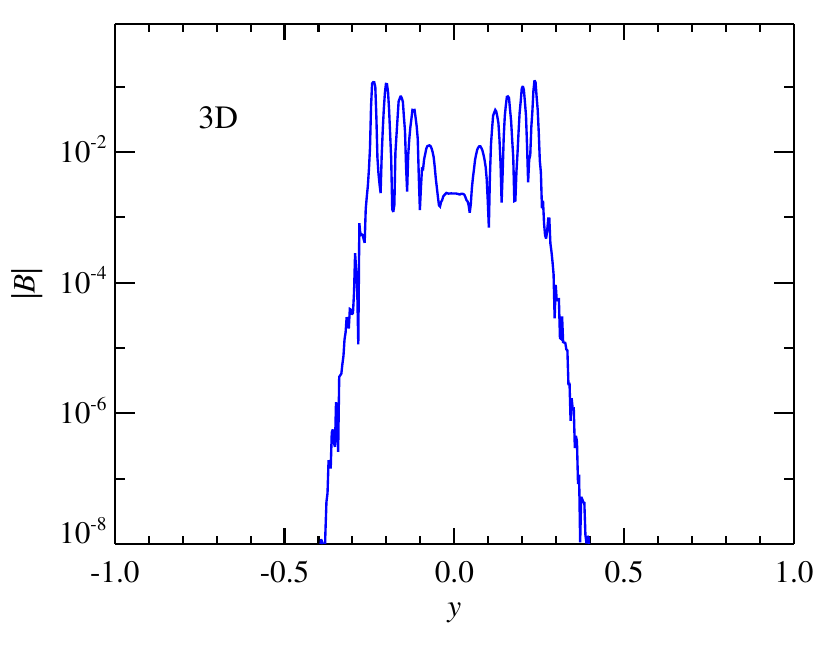}}\vspace*{-0.5cm}\\%
\end{center}
\caption{Profile of the magnetic field strength along the $y$-direction through the mid-plane of the two simulations shown in Fig.~\ref{FigKHimage}, with the top giving the 2D calculation (where effectively no cleaning is done), and the bottom the 3D calculation where $\nabla\cdot\vec{A}$ cleaning is active. \label{FigKHBprofile}}
\end{figure}

\begin{figure*}
\begin{center}
\resizebox{8.5cm}{!}{\includegraphics{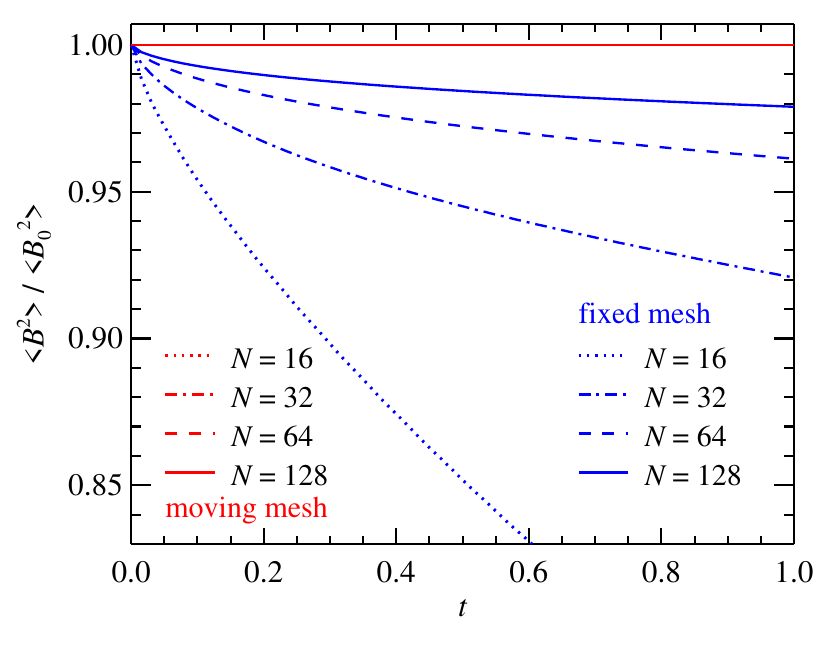}}\hspace*{1.0cm}%
\resizebox{8.5cm}{!}{\includegraphics{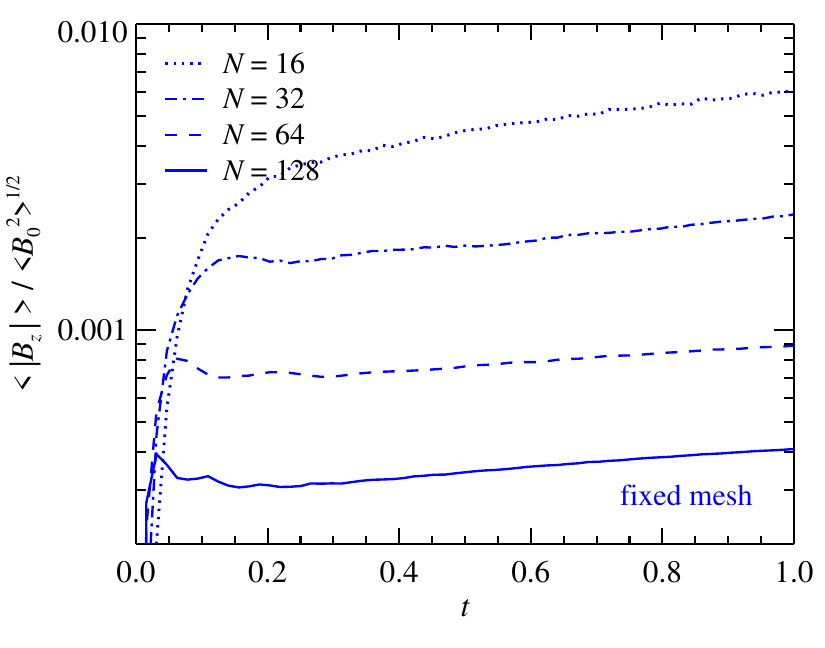}}\\
\end{center}
\caption{Time evolution of the magnetic field energy density (left panel), and the volume-averaged magnitude of the magnetic field component (right panel) along the cylindrical tube in the oblique 3D flux tube advection test. We show results for a stationary fixed mesh at different resolutions $N$, as well as matching results for a moving mesh. The magnetic energy is normalized to its value at the beginning, and the magnetic field component to the root-mean-square of the magnetic field at the initial time. The advection errors in the fixed mesh case, as well as the developing residual $B_z'$-components become quickly smaller with better resolution in the fixed mesh case. The absolute sizes of these errors are comparable to (actually slightly smaller) than those reported in \citet{Gardiner2008}.    \label{FigFluxLoop3D}}
\end{figure*}

\subsection{Kelvin-Helmholtz instability with localized B-field}

Recently, \citet{Tomida2026} compared constrained transport schemes with Dedner cleaning methods for the magnetic field divergence. They argued, in particular, that the cleaning method may propagate magnetic strength into areas of a simulation where the magnetic field should be absent or extremely weak, thereby creating a particular type of systematic error inherent in such cleaning methods. The particular system they analysed to demonstrate this error is a Kelvin Helmholtz instability with a localized, dynamically unimportant magnetic field initially oriented along the direction of the shear flow. We here repeat a calculation of their setup in order to test whether our vector potential method is affected by the corresponding problem.

We adopt the same setup as\citet{Tomida2026}, using a box of dimension $[-0.5, 0.5]\times [-1,1]$ in the $x$- and $y$-directions, filled with gas of unity density, $\rho=1$, and constant pressure $P=10$. The initial velocity field is setup with a smoothed jump of the form
\begin{equation}
v_x = v_0 \tanh\left({\frac{y}{a}}\right),
\end{equation}
together with a perturbation
\begin{equation}
v_y = A \sin(2\pi x) \exp\left(-\frac{y^2}{\sigma^2}\right),
\end{equation}
and $v_z=0$. The constants are chosen as $v_0=1$, $a=0.05$, $A=0.01$, and $\sigma = 0.2$. The problem is translationally invariant in the $z$-direction, hence it can be computed in 2D, or equivalently in 3D. We have done both, since only in 3D our cleaning method can make a non-trivial difference (unlike when Dedner cleaning is done for the magnetic field). 

A localized magnetic field is imposed as,
\begin{equation}
B_x = B_0 \exp\left( -\frac{y^2}{b^2} \right),
\end{equation}
with $B_0 = 0.01$ and $b=0.1$. We realize this field with a vector potential of the form
\begin{equation}
A_z = \frac{\sqrt{\pi} }{2} b\, B_0 \,\textrm{erf}\left(\frac{y}{b} \right).
\end{equation}
We adopt periodic boundary conditions in the $x$-direction. In the $y$-direction, this is not readily possible for the primary computational domain. However, by doubling the size of the computational in the $y$-direction and filling it with a $xy$-mirrored copy of the primary setup, we can avoid the need for special boundary conditions in the $y$-direction and instead also apply periodic boundaries there (at the price of doubling the computational cost, but this is fine for this test problem). When we quote mesh resolutions they refer to the primary computational domain only and do not include the cells for the mirrored copy.

In Figure~\ref{FigKHimage}, we give maps of the magnetic field strength obtained at time $t=3.0$, comparing a 2D simulation resolved with $128\times 256$ cells initially arranged as a Cartesian grid with an equivalent 3D simulation with $128\times 256\times 8$  cells. The moving mesh algorithm is used in both cases. A rolled-up Kelvin-Helmholtz billow has formed as a result of the excited perturbation mode, and the stretching of the magnetic field in the billow has amplified the field somewhat above its initial value. The results we obtain are qualitatively similar to those shown by \citet{Tomida2026} for the {\small ATHENA} code, although our 2D outcome looks slightly sharper with more retained small-scale structure. We interpret this as a reflection of the reduced advection errors in our moving-mesh approach. Note that in both cases we show a two-dimensional cut through the simulation, which in the 3D case leads to pixels in corners of 3D Voronoi cells, which is responsible for the small-scale grid noise seen in the image of the 3D run.  Importantly, no magnetic field is developing in the regions above and below, unlike in the solutions with Dedner cleaning  examined by \citet{Tomida2026}.

This is quantified more explicitly in Figure~\ref{FigKHBprofile}, where we show the $\vec{B}$-field strength profile of both simulations. This confirms that there is no significant propagation of field outside the narrow region in which the field is placed initially. If Dedner cleaning in the magnetic field is used instead, \citet{Tomida2026} found that significant tails of magnetic field strength  build up outside the billow region. In contrast, our results are very similar to  what they found when  constrained transport is employed.

\subsection{Flux tube advection in 3D}

As discussed by \citet{Gardiner2008} and \citet{Stone2008}, extending the advection problem of a weak flux loop to a cylindrical flux tube in 3D makes up for further interesting test possibilities that are in fact quite demanding for numerical MHD. In particular, in conservative methods that are not satisfying the $\nabla \cdot \vec{B} = 0$ constraint sufficiently accurately, the advection can easily generate an erroneous and strongly growing magnetic field transverse to the flux loop (i.e.~along the cylinder).

We follow  \citet{Gardiner2008} and extend a 2D flux loop in the transverse direction to a flux tube, and then incline it relative to the grid, and advect it at an oblique angle to it. We consider a periodic box of size $1 \times 1 \times 2$ that is resolved with $N\times N  \times 2N$ cells. Similarly to the circularly  polarized Alfvén wave test, the flux loop is setup in a coordinate system $x'$, $y'$, $z'$ that is then rotated to grid coordinates $x$, $y$, $z$ via
\begin{equation}
\left(
\begin{array}{c}
x\\ y \\ z   
\end{array}
\right) = 
\left(
\begin{array}{ccc}
2/\sqrt{5} &  0 &   -1/\sqrt{5} \\
0 & 1  & 0 \\
1/\sqrt{5}           &   0        &  2/\sqrt{5}
\end{array}
\right)
\left(
\begin{array}{c}
x'\\ y' \\ z'   
\end{array}
\right).
\end{equation}
The flux tube is oriented along $z'$, and its initial magnetic field is setup in this coordinate system by specifying a vector potential
\begin{equation}
A_z' = \left\{
\begin{array}{cc}
0.001 \times (0.3 - R) & \mbox{for}\;\; R \le 0.3 ,\\
0 & \mbox{otherwise},
\end{array}
\right.
\end{equation}
where $R=\sqrt{x'^2 +y'^2}$ is the distance to the nearest periodic image of the origin in the $x'y'$-plane, where periodicity here occurs at distances $2/\sqrt{5}$ in the $x'$ direction, and at distances of $1$ in the $y'$-direction, due to the tilt relative to the simulation coordinates.

We specify a uniform velocity of $\vec{v} = (1, 1, 2)$ and set the density and pressure to $\rho = 1$ and $P = 1$, respectively. We then evolve the system until time $t=1$, at which point an ideal solution should return to its initial state, given that the magnetic pressure is negligible over this timescale.

In the left panel of Figure~\ref{FigFluxLoop3D}, we show measurements of the volume averaged magnetic energy density as a function of time, normalized to its initial value. We compare results for resolutions $N=16$, $N=32$, $N=64$, and $N=128$ when our code is run for a stationary fixed mesh, and for a moving mesh. At low resolution, the magnetic energy is dissipated by advection errors in the fixed mesh case, an effect that becomes progressively smaller with improved resolution, as expected. The moving mesh approach trivially reproduces an invariant solution, independent of resolution.

In the right panel of Figure~\ref{FigFluxLoop3D}, we show the time evolution of the volume average value of the $|B_z'|$ component of the magnetic field, normalized to the volume-averaged rms-value of the initial magnetic field. Here the fixed mesh result shows the development of a small $B_z'$-component early in the evolution, which very slowly grows further, but overall stays negligibly small. The volume affected by this declines rapidly with resolution. The moving mesh results are orders of magnitude below these results and are therefore not shown. 

When comparing our results to those reported by \citet[][see their Fig.~1]{Gardiner2008} at matching resolution, we actually find slightly smaller errors. We think this can likely be attributed to our different time integration approach, where the vector potential is advected directly and not built-up and then destroyed again by the induction term in the path of the flux tube. It is good to see that our approach is at least competitive, if not advantageous, compared to the standard approach in constrained transport schemes. Arguably more importantly, the good performance of our approach in this test shows explicitly that maintaining a zero flux sum for cell surfaces to machine precision is not a prerequisite for accurate MHD results. Rather it appears that the key benefits of `divergence-free' treatments are already realized for the most part if  the vector potential is used as the independent variable (which is implicitly also the case for constrained transport schemes, as the staggered storage of magnetic fluxes can be replaced by storing the vector potential at mesh corners), whereas the difference operator to derive the magnetic field from it is if secondary importance.

\subsection{Orszag-Tang Vortex in 3D}

\begin{figure}
\begin{center}
\resizebox{8.5cm}{!}{\includegraphics{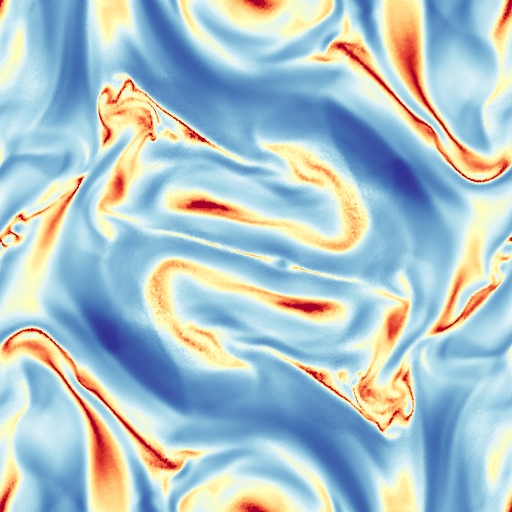}}\\
\resizebox{5cm}{!}{\includegraphics{plots/OT_bfield_legend.pdf}}\vspace*{-0.3cm}\\
\end{center}
\caption{Magnetic energy density in an Orszag-Tang vortex simulation in 3D at time $t=1.0$ through the $z$-midplane, for a resolution of $256^3$ cells. \label{FigOT-3D}}
\end{figure}

\begin{figure*}
\begin{center}
\resizebox{8.0cm}{!}{\includegraphics{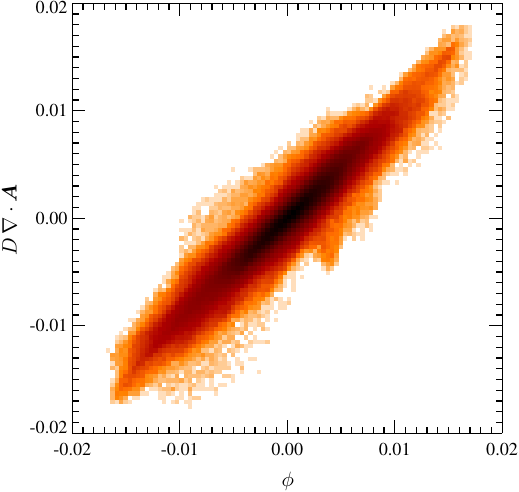}}\hspace*{1cm}%
\resizebox{8.0cm}{!}{\includegraphics{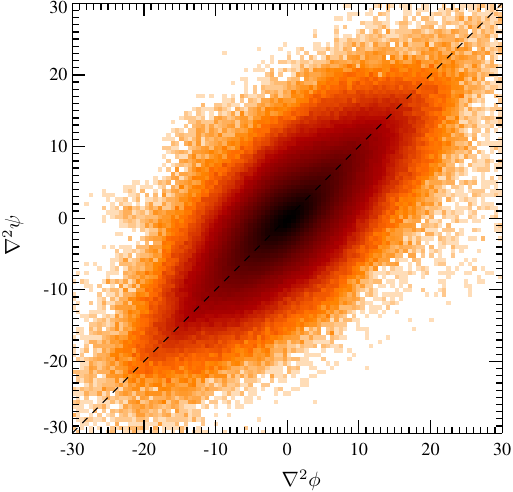}}
\end{center}
\caption{Local scaling properties of the scalar field $\phi$ used for divergence control of the vector potential in a 3D Orszag-Tang simulation. The left panel shows the correlation between $\phi$ and the product of `diffusivity' $D$ and $\nabla\cdot \vec{A}$. The right panel relates the Laplacian of the cleaning scalar to the Laplacian of $\psi = \vec{u}\cdot\vec{A}$, our primary gauge field that establishes numerical Galilean invariance of our scheme. Both measurements are taken at $t=2.0$ for a simulation of the 3D Orszag-Tang vortex problem at $256^3$ resolution. \label{FigOT-3D-Cleaning}}
\end{figure*}

\begin{figure}
\begin{center}
\resizebox{8.5cm}{!}{\includegraphics{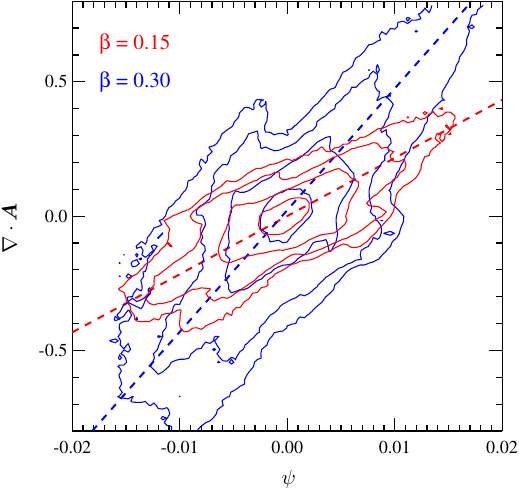}}
\end{center}
\caption{Relation between the residual divergence in the vector potential and the value of the gauge scalar. We show individual cell measurements (contours) for a 3D Orszag-Tang vortex test simulation at time $t=2.0$ and resolution $256^3$, for two different values of the cleaning parameter $\beta$, as labelled. The dashed lines mark the theoretical expectations for the relation based on equation~(\ref{eqndivAbeta}). \label{FigOT-3D-DivA}}
\end{figure}

The classic two-dimensional Orszag-Tang problem is a popular and useful test to assess the basic stability of a numerical MHD scheme. However, MHD in 2D is comparatively simple compared to 3D, primarily because the vector potential has only one non-trivial component ($A_z$) which physically is simply advected with the fluid. Furthermore, the divergence of the vector potential  vanishes in 2D, and so the robust Coulomb gauge is automatically realized. In comparison, genuinely three-dimensional magnetic field dynamics is much more complex and offers several numerical failure modes that simply do not exist in 2D.

It is therefore useful to extend the Orszag-Tang vortex problem and generalize it to a more difficult, genuine 3D problem. To realize this, we follow \citet{Helzel2011} and consider a modified initial velocity field that has $z$-dependent perturbations in the $x-$ and $y$-components, and an additional velocity component in the $z$-direction, namely
\begin{equation}
    \vec{v}= 
\left(
 \begin{array}{c}
    - \left[ 1 + \epsilon \sin(2 \pi \, z)\right] \,\sin(2\pi\,y)\\
  \left[ 1+ \epsilon \sin (2\pi \,z) \right] \, \sin (2\pi \, x)\\
    \epsilon \, \sin (2\pi\,z)
\end{array}
\right),
\end{equation}
with $\epsilon = 0.2$. Otherwise, the setup is as for the classic 2D Orszag-Tang vortex, with  density $\rho=\gamma^2 / (4\pi)$, pressure $P = \gamma / (4\pi)$, and $\gamma = 5/3$. The initial magnetic field is given by $\vec{B}=1.0 / (4 \pi)^{1/2} [-\sin(2\pi y), \sin(4\pi x, 0)]$, with a corresponding vector potential $A_z = 1.0 / (4 \pi)^{3/2}  [ 2  \cos( 2\pi y)  + \cos( 4 \pi x)]$, and the simulation domain is a periodic box covering $[0, 1]^3$. 

In Figure~\ref{FigOT-3D}, we show the result for the magnetic energy density in the $z$-midplane, at time $t=1.0$, computed at a resolution of $256^3$ cells that are initially arranged  as a Cartesian mesh. Our results compare favourably to those reported in \citet{Tu2022} for a Lagrangian meshless simulation with the same number of resolution elements. Notably our result does not show the noise artifacts visible in their vector potential implementation (see their Figure~8), and it also reveals much more fine detail at the same nominal resolution compared both to their vector potential formulation and the constrained gradient scheme \citep{Hopkins2016CG} implemented in the {\small GIZMO} code that they considered for comparison.

The 3D Orszag-Tang problem is also useful for testing whether the $\nabla \cdot\vec{A}$ control works as expected based on the discussion in Section~\ref{secA}. To this end, we first consider the correlation between the cleaning scalar $\phi$ and the divergence of $\vec{A}$. According to equation~(\ref{eqnphinablaA}), we expect $\phi \propto D \,\nabla\cdot\vec{A}$, where $D$ is the diffusivity of a cell. In the left panel of Figure~\ref{FigOT-3D-Cleaning}, we plot $\phi$ on the $y$-axis, and $D\,\nabla\cdot\vec{A}$ on the $x$-axis, measured in the simulation at $t=2.0$. We indeed find a clear correlation between these quantities, along the 1:1 relation. Next, we test whether -- as expected from equation~(\ref{eqnlabplacephilaplacepsi}) -- the Laplacian of the cleaning scalar approximately balances the Laplacian of our gauge field $\psi$, which is given through the large scale velocity field, $\psi = \vec{u}\cdot\vec{A}$. To this end, we plot $\nabla^2\phi$ against $\nabla^2\psi$ in the right panel of Figure~\ref{FigOT-3D-Cleaning}, again on a cell-by-cell basis as measured in our Orszag-Tang simulation. Reassuringly, the results meet the expectation of equation~(\ref{eqnlabplacephilaplacepsi}), albeit with substantial scatter.

Finally, we want to check whether the absolute size of the residual $\nabla\cdot\vec{A}$ varies in the expected way with the numerical parameters that control the cleaning scalar. In particular, according to equation~(\ref{eqndivAbeta}) we should be able  to lower the residual value of the divergence by increasing the diffusivity parameter. We here realize such a variation by lowering $\beta$ by a factor of 2 from $\beta=0.3$ to $\beta=0.15$, and then compare two simulations where everything is equal except for this parameter. In Figure~\ref{FigOT-3D-DivA}, we compare the distributions of $\nabla\cdot\vec{A}$ against $\psi$ in the corresponding simulations. Based on equation~(\ref{eqndivAbeta}), we would expect that a diffusivity enlarged by a factor of two should reduce the values of $\nabla \cdot\vec{A}$ by about the same factor, for comparable values of $\psi$. This is indeed borne out by our measurements.

\begin{figure}
\begin{center}
\resizebox{8.5cm}{!}{\includegraphics{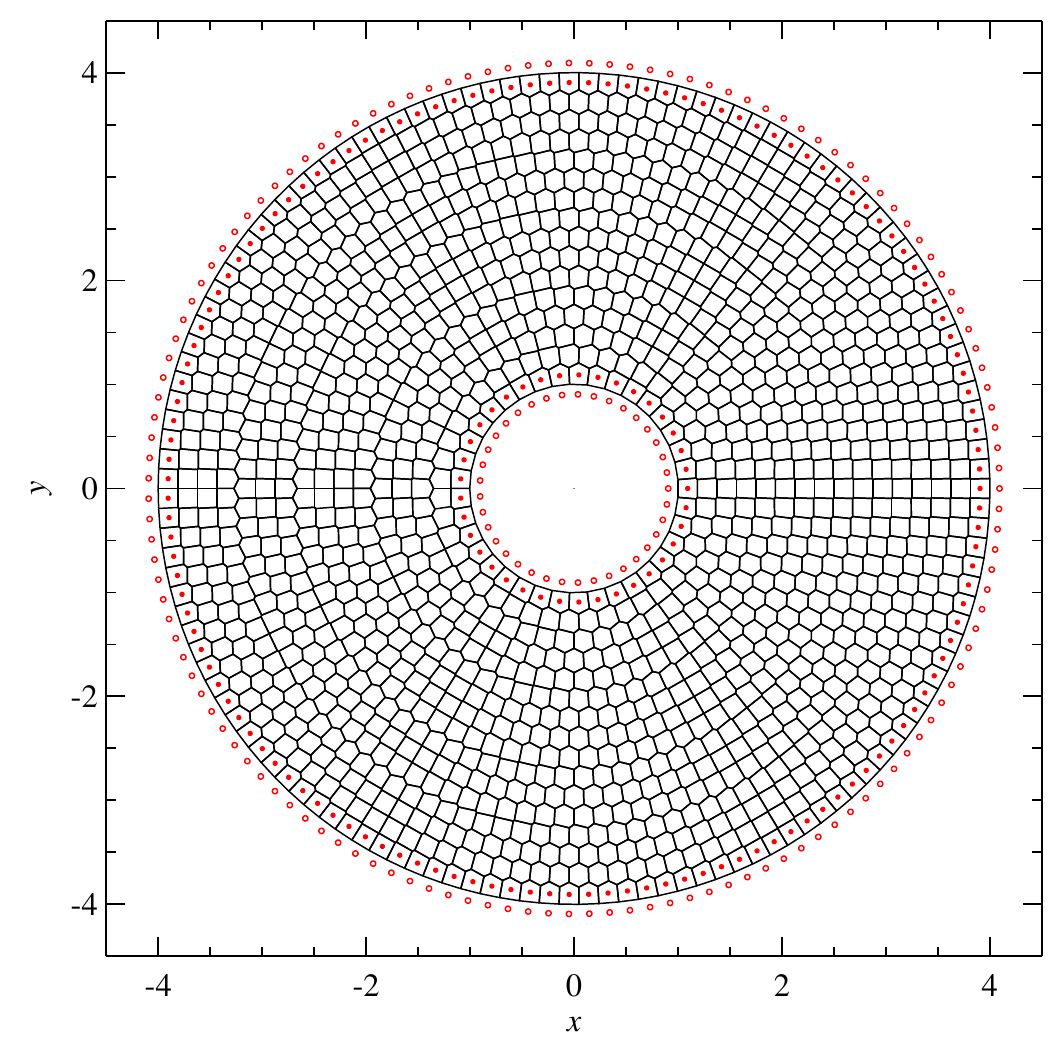}}\vspace*{-0.3cm}
\end{center}
\caption{Mesh structure used in our 3D MRI test, here shown in a slice through the $xy$-plane and at low resolution for visual clarity with 16 radial cells in the domain covered by the fluid, $1 < R < 4$. The inner and outer boundaries of the cylindrical simulation domain are realized by two rings of paired mesh generating points (filled and hollow circles) that are moved as a solid body during the actual simulation. Note that no Voronoi cells for the points outside of the simulation domain (hollow circles) are actually computed by the simulation code. \label{FigMRImesh}}
\end{figure}

\begin{figure}
\begin{center}
\resizebox{8.5cm}{!}{\includegraphics{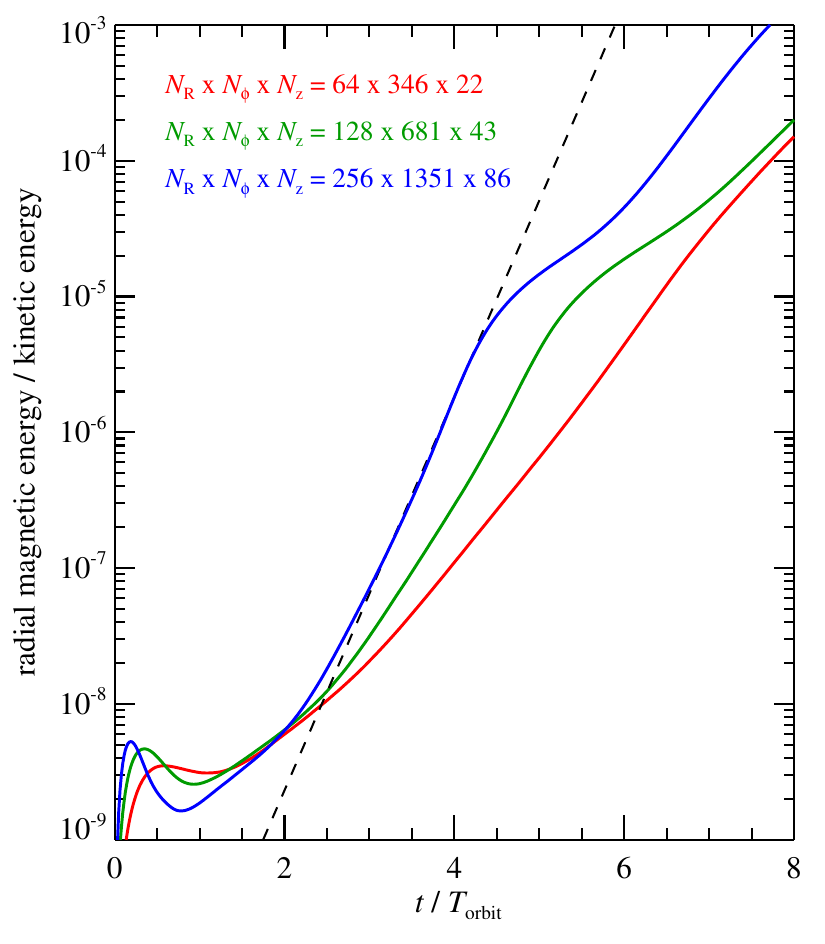}}\vspace*{-0.3cm}
\end{center}
\caption{Growth of the MRI measured in terms of the evolution of the radial magnetic field energy, shown for three different mesh resolutions, as labelled. The dashed line shows the maximum expected exponential growth rate $B^2\propto \exp(2\,\gamma_{\rm MRI}\, t)$ for the critical modes at $R_0$.  \label{FigMRIgrowth}}
\end{figure}

\begin{figure}
\begin{center}
\resizebox{8.0cm}{!}{\includegraphics{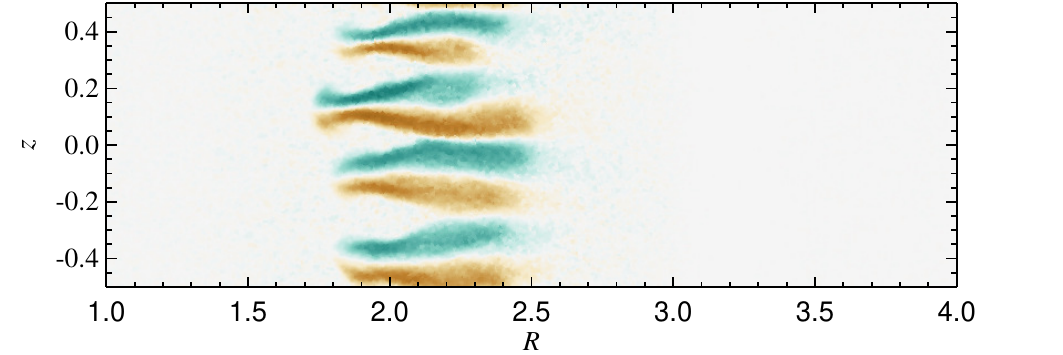}}%
\hspace*{-0.5cm}\rotatebox{90}{\hspace*{0.3cm}\resizebox{2.5cm}{!}{\includegraphics{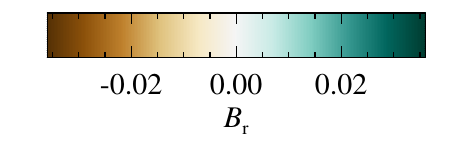}}}\\
\end{center}
\caption{Map of the strength of the radial magnetic field in a slice cut through the $R$-$z$ plane for our high-resolution MRI simulation at $t=8$, for an arbitrary azimuthal angle, here picked as $\phi = 0$. We have chosen the vertical magnetic field strength such that the growth of the $n=4$ mode around $R=2$ should be favoured according to analytic theory \citep{Balbus1991}, and this is indeed what is found in our simulation which was started from white noise perturbations in the field.  \label{FigMRIimage}}
\end{figure}

\subsection{Magneto-rotational instability}

Differentially rotation gas disks exhibit a magneto-rotational instability (MRI) in which radial and azimuthal field components grow exponentially, and comparatively rapidly \citep{Balbus1991, Hawley1991}. This instability is thought to be of critical importance for creating turbulence and an effective viscosity in accretion disks, ultimately mediating  angular momentum transport in such disks.

Being able to correctly compute the magneto-rotational instability is therefore an important requirement for a MHD code. We will here test the instability in a full 3D simulation, following the setup of \citet{Flock2010}, which has subsequently also been studied by \citet{Pakmor2013} and \citet{Duffel2016}. We consider an unstratified gas disk at constant unit density, adiabatic sound $c_s=0.1$, rotating in a cylindrical potential with potential $\propto -1/R$. The angular rotation profile is Keplerian with $\Omega(R) = 1/R^{3/2}$. We only consider the radial region between $R\in [1,4]$, and simulate the zone $z\in [-0.5, 0.5]$ in the vertical direction. Inside the ring-like region $2<R<3$, the disk is threaded with a uniform vertical magnetic field $B_0$ in our initial conditions. 

To seed the MRI, we impose random velocity perturbations in every cell in the radial and vertical directions, equal to $v = 5.0\times 10^{-5}$. In order to favour the growth of a certain mode
\begin{equation}
    \lambda_{\rm crit} = L_z / n,
\end{equation}
we pick $B_0$ such that $\lambda_{\rm crit}$ becomes equal to the critical wavelength for fastest growth \citep{Balbus1991}.
This is the case for
\begin{equation}
  B_0 = \frac{1}{2\pi}\sqrt{\frac{15} {16}} \sqrt{\rho} \, \Omega  \frac{L_z}{n}.
\end{equation}
Using $n=4$ as in \citet{Flock2010}, and focusing on the location $R=2$ where we expect the strongest growth rate, we obtain $B_0=0.05448 / n$ (in Lorentz-Heaviside units). When the MRI sets in, the magnetic field associated with the corresponding mode is then expected to be able to grow exponentially, $B\propto \exp(\gamma_{\rm MRI} t)$, with a rate as fast as $\gamma_{\rm MRI} = 0.75\, \Omega$. 

For numerically realizing this setup, we impose an inner and outer boundary on our fluid domain by means of a pairwise ring of cells. This is sketched in Figure~\ref{FigMRImesh}, which illustrates our initial mesh geometry in the $xy$-place at low resolution. In the $z$-direction, we simply stack these cells. The inner and outer boundary conditions are realized as zero gradient boundaries, while in the vertical direction periodic boundaries are used. Note that in our new mesh construction methodology in {\small AREPO-2}, there is no need to fill the corners of the computational domain with mesh cells. Furthermore, cells corresponding to the boundary points marked with hollow circles are never actually constructed. These points are only needed to ensure that there is a closing wall at the desired location for the outermost points (filled circles) in the actual fluid domain. Each of the two rings of pairwise boundary points is rotated during the simulation as a solid body with prescribed angular frequency (following the $\Omega(R)$ profile) to keep the cylindrical geometry of the boundaries exactly in place. The inner mesh generating points are allowed to move freely. 

To realize the initial magnetic field, we can use the technique of a non-vanishing mean magnetic field described in subsection~\ref{secBmean}, combined with a vector potential that eliminates the field in the regions $1<R<2$ and $3<R<4$. However, since in this problem the initial conditions are strictly translationally invariant in the $z$-direction, and since we are not actually applying periodic boundaries in the $x$- and $y$-directions, we can actually simply directly realize the initial $\vec{B}$-field with a vector potential that only depends on $x$ and $y$. We choose $\vec{A} = f(R) (-yB_0, x B_0, 0)$ with
\begin{equation}
    f(R) = \left\{ 
    \begin{array}{cl}
    0 & \mbox{for $R < R_0$},  \vspace*{0.15cm}\\ 
    \frac{R^2 - R_0^2}{2 R^2}  & \mbox{for $R_0 \le R < R_1$}, \vspace*{0.15cm}\\
    \frac{R_1^2 - R_0^2}{2 R^2}  & \mbox{for $R_0 \le R$}, \\
    \end{array}
    \right.
\end{equation}
and $R_0=2$ and $R_1=3$
for this purpose, with $R=\sqrt{x^2+y^2}$. Note that the vector potential needs to be continuous at $R_0$ and $R_1$. In the particular realization we have chosen here, it initially vanishes for $R < 2$.

We consider three simulations of this problem, at resolutions $64\times 22$, $128\times 43$, and $256\times 86$ in the radial and vertical directions, respectively. The cells in the azimuthal direction vary with radius and are chosen such that all cells have approximately equal size. The total number of cells in the simulations is then 487322, 3749299, and 29752130, respectively. We evolve the tests over 8 orbits, where we define the orbital time as $T_{\rm orbit}= 2 \pi/\Omega(R_0)$, i.e.~using the inner edge of the simulated region. Note that we do not need an orbital advection algorithm to efficiently simulate this problem as our moving-mesh approach allows us to absorb the Keplerian velocity into the mesh motion, so that the bulk velocity does not reduce the permissible timestep sizes.

In Figure~\ref{FigMRIgrowth} we show the amplification of the radial magnetic energy as a function of time. After an initial transient region, the exponential growth of the MRI clearly sets in after about 2 orbits, and is comparable to the maximum rate expected analytically. In fact, for our highest resolution run we reach this rate exactly, although it is only maintained exactly for about two orbits, and then gives way to slightly slower growth subsequently. The shape of our amplification curves is qualitatively and quantitatively in good consistence with \citet{Flock2010} and the simulations of \citet{Duffel2016}. It is also reasonably similar to \citet{Pakmor2013}, but we note that in the latter case the amplification sets in notably earlier.

Finally, in Figure~\ref{FigMRIimage} we visualize the radial magnetic field strength in the $R$-$z$ plane, using a section at azimuthal angle $\phi=0$ at time $t=8$ in our high resolution simulation. This shows a clear pattern of alternating maxima and minima around $R\sim 2$, where we expect the strongest amplification. The number of maxima and minima is consistent with the $n=4$ mode that we aimed to preferentially excite with our chosen value for $B_0$. Overall, the test results reassuringly confirm that our new method is well capable of following the MRI.

\begin{figure}
\begin{center}
\resizebox{4cm}{!}{\includegraphics{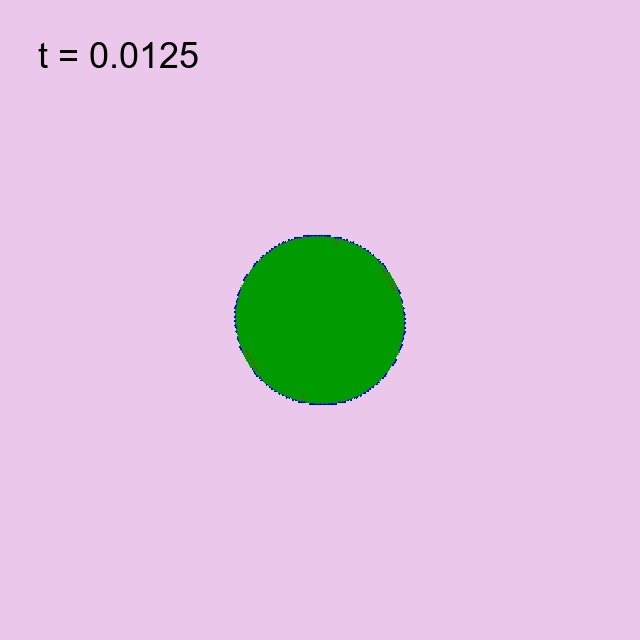}} %
\resizebox{4cm}{!}{\includegraphics{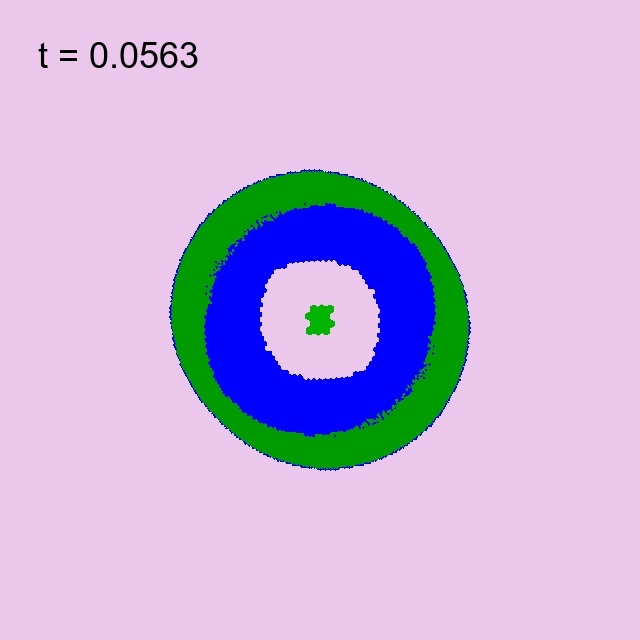}}\\
\resizebox{4cm}{!}{\includegraphics{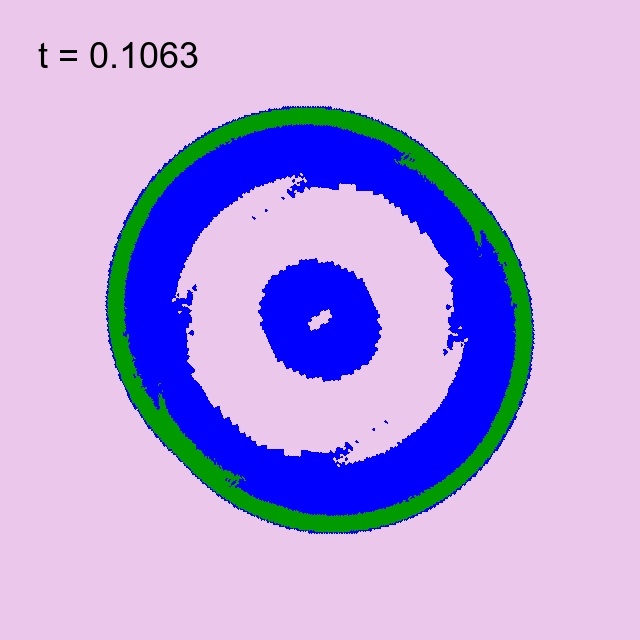}} %
\resizebox{4cm}{!}{\includegraphics{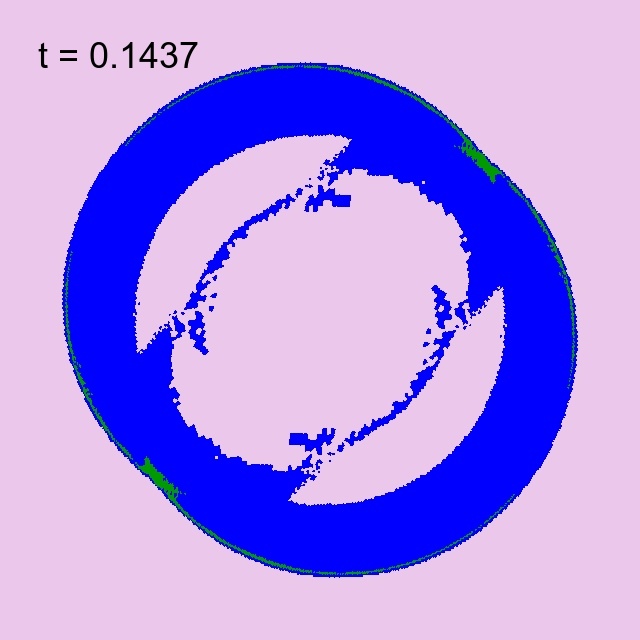}}\\
\resizebox{5cm}{!}{\includegraphics{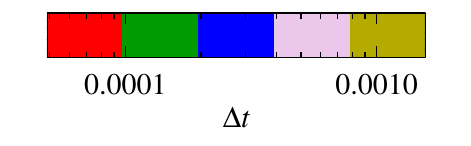}}%
\end{center}
\caption{Local timesteps in the 2D MHD blast wave test, at different times as labelled, for a moving mesh with $2\times (256\times 147)$ cells in the initial conditions. In this particular case, the computational effort is reduced by a factor of 3 when switching from global to local timestepping, with no discernible quality reduction compared to the fixed timestepping result shown in Figure~\ref{FigBlast2D}.  \label{FigBlast2DTimesteps}}
\end{figure}

\begin{figure*}
\begin{center}
\resizebox{7cm}{!}{\includegraphics{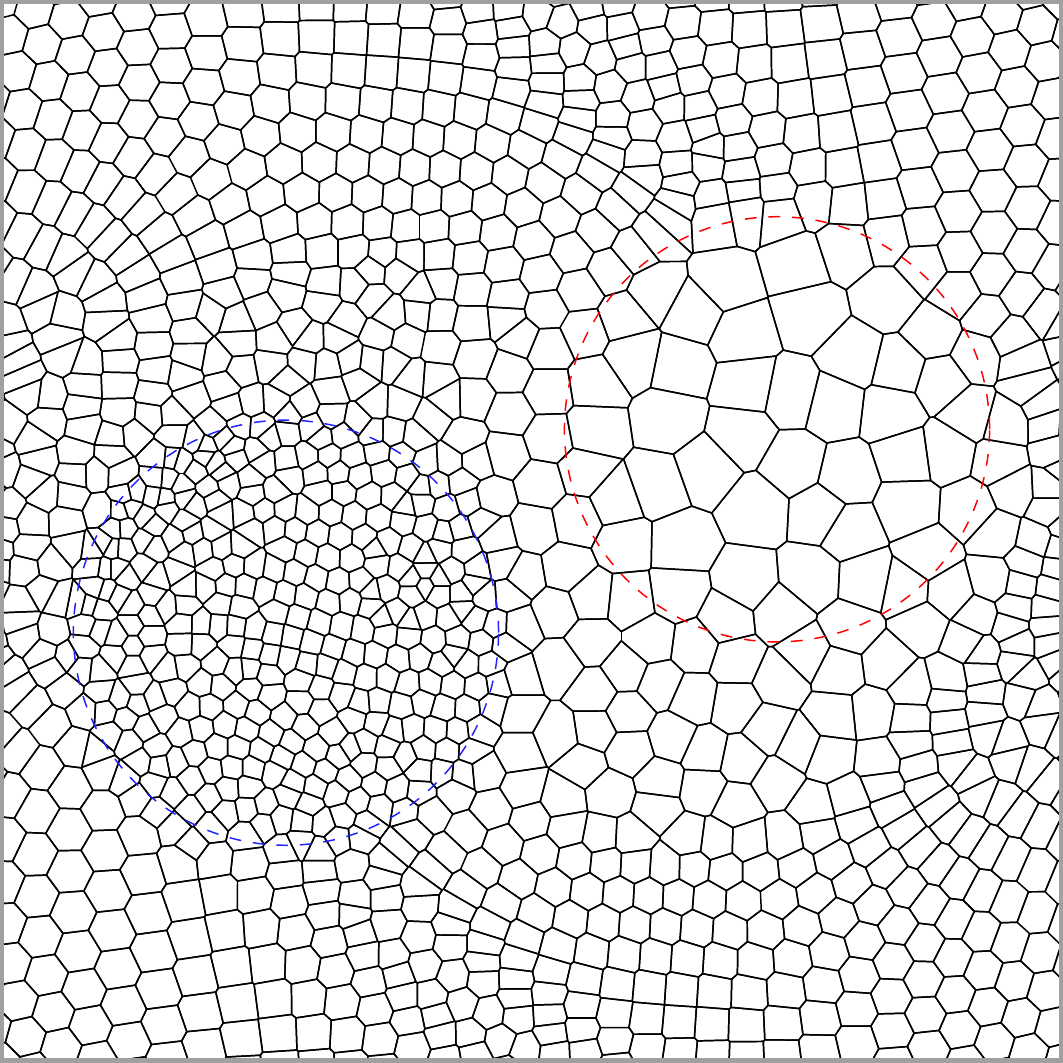}}\hspace*{0.5cm}%
\resizebox{7cm}{!}{\includegraphics{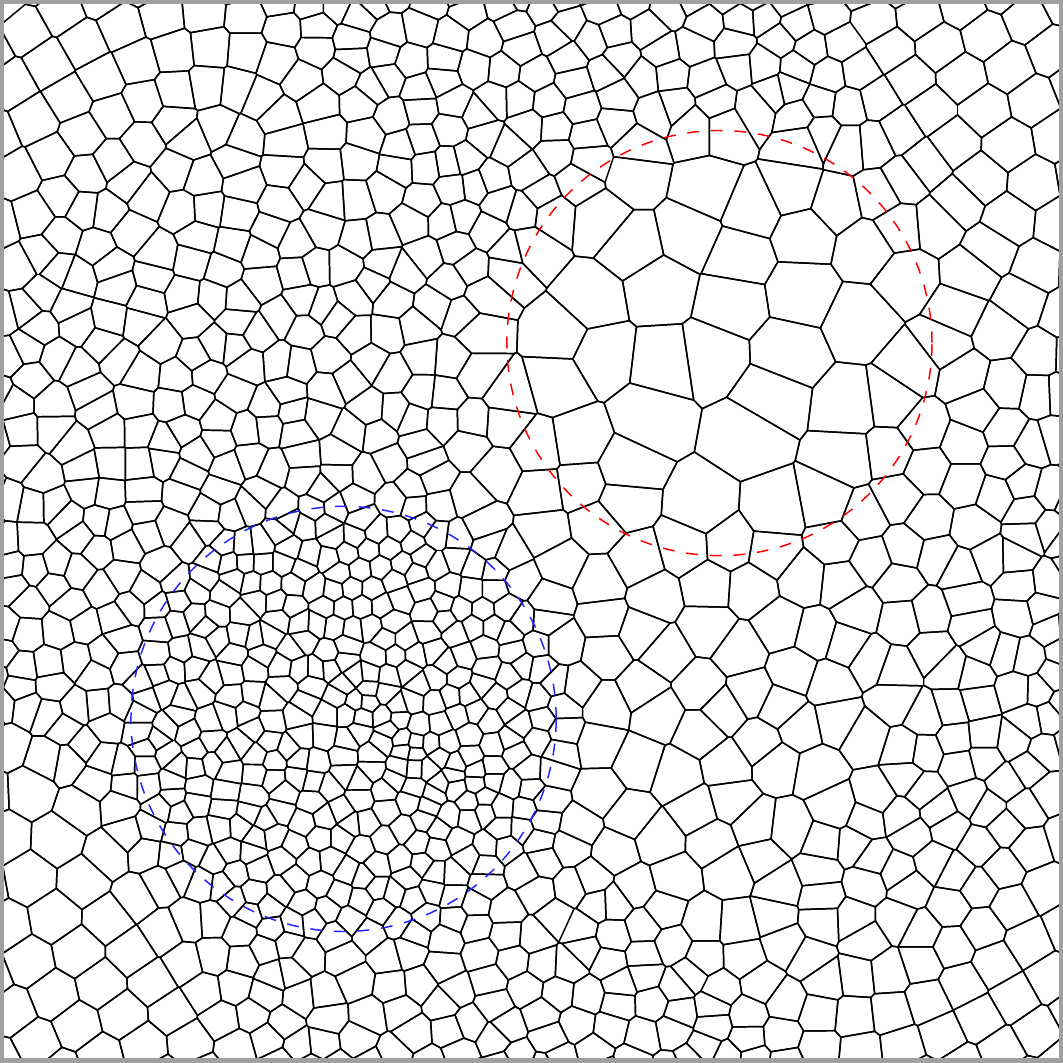}}\vspace*{0.5cm}\\
\resizebox{7cm}{!}{\includegraphics{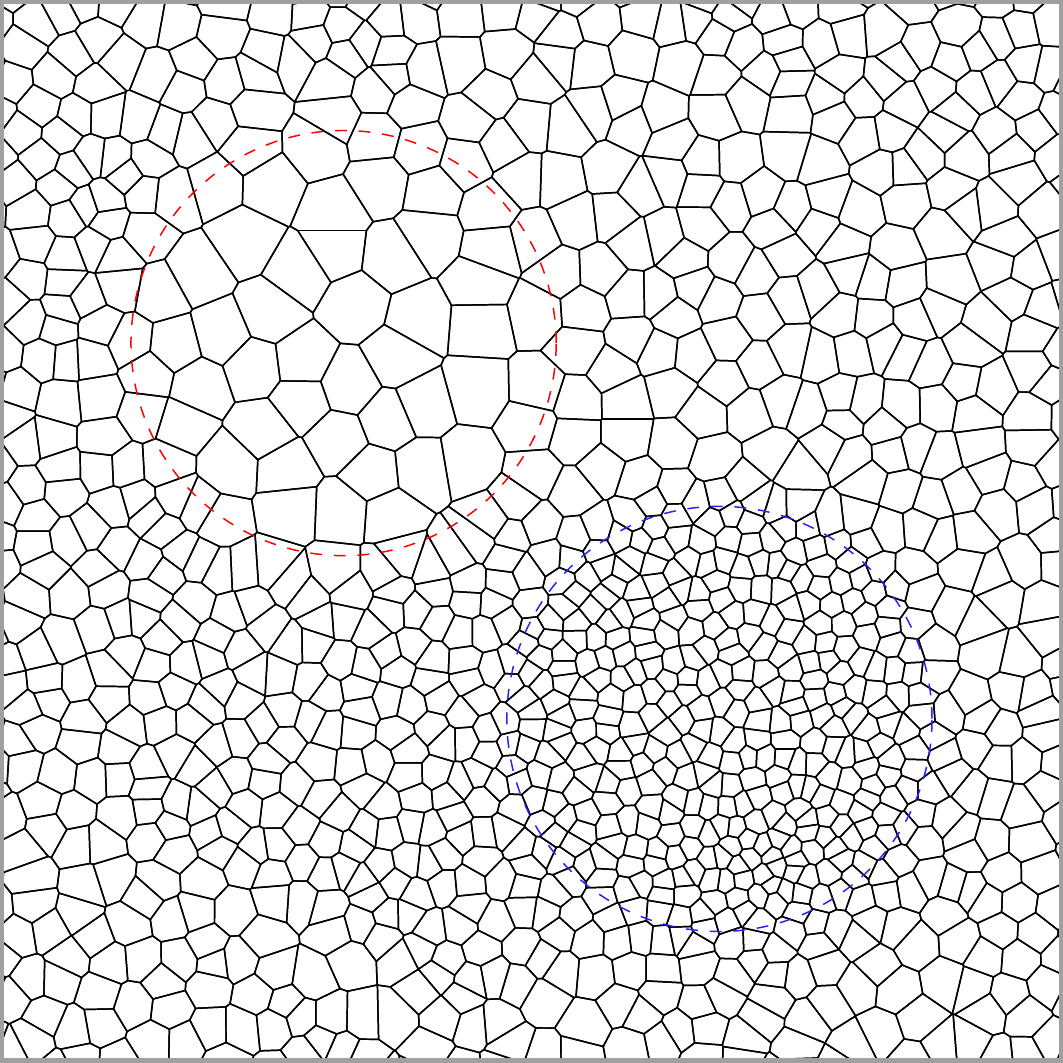}}\hspace*{0.5cm}%
\resizebox{7cm}{!}{\includegraphics{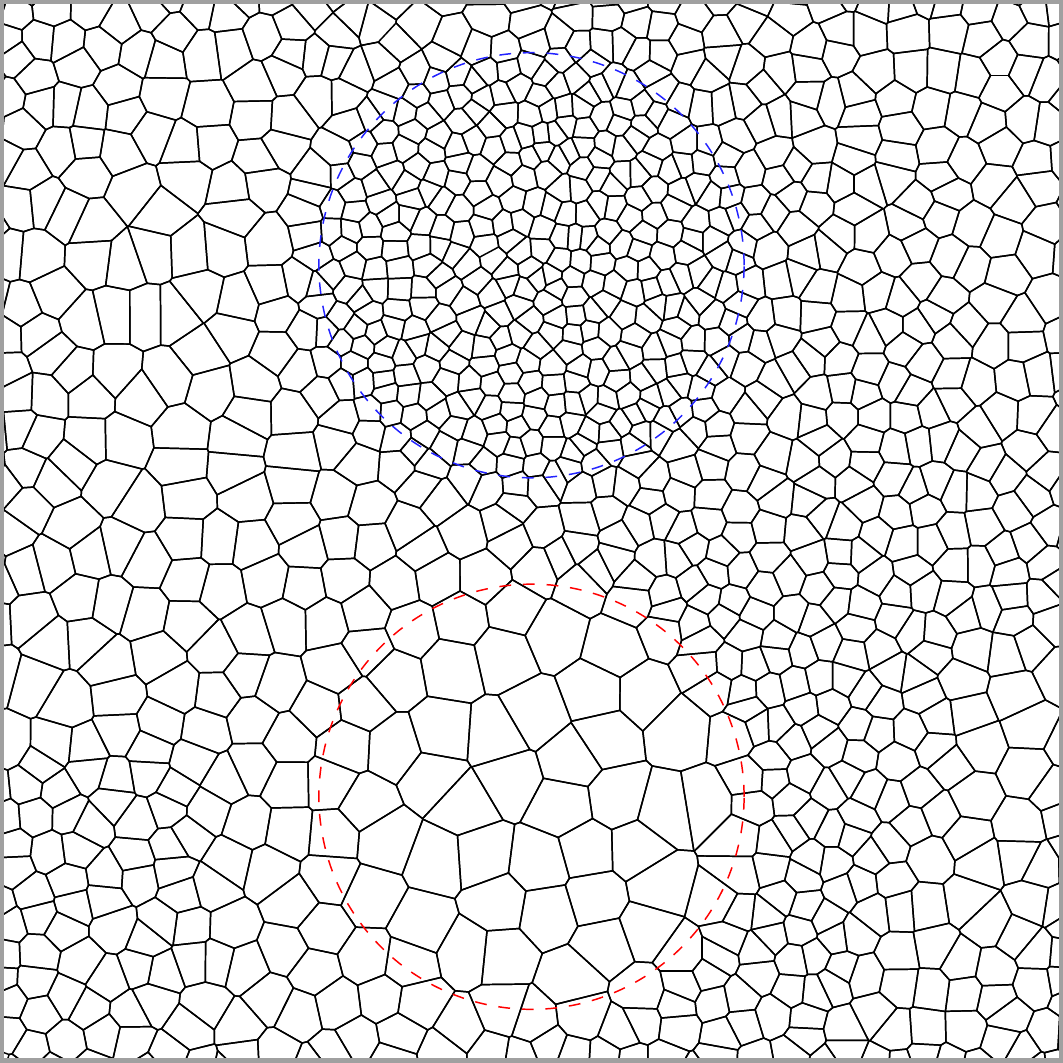}}\\
\end{center}
\caption{Mesh geometry for an illustrative refinement and de-refinement test, at different simulation times equal to $t=0.128$, $0.256$, $0.768$ and $1.572$ (from top left to bottom right), respectively. The problem that is simulated here is the Orszag-Tang vortex in 2D, started from an initial hexagonal background mesh with resolution $2\times (32\times 18)$. This mesh geometry can still be recognized in parts of the volume at early times. Besides the distortion of the mesh due to the gas motion, the imposed refinement criteria force the resolution to be higher than the default in a spherical region marked by a blue dashed circle, and to be lower than the default in the dashed red circle. Both circles are made to move counter clock-wise on a circular trajectory around the centre of the box with an orbital time $T=2.048$. Despite the constant on-the-fly mesh refinement and de-refinement operations, the Voronoi mesh realized by the code maintains a fairly regular geometry, and the magnetohydrodynamical evolution itself stays accurate and stable. \label{FigBlast2DRefinementMesh}}
\end{figure*}

\begin{figure*}
\begin{center}
\resizebox{5.2cm}{!}{\includegraphics{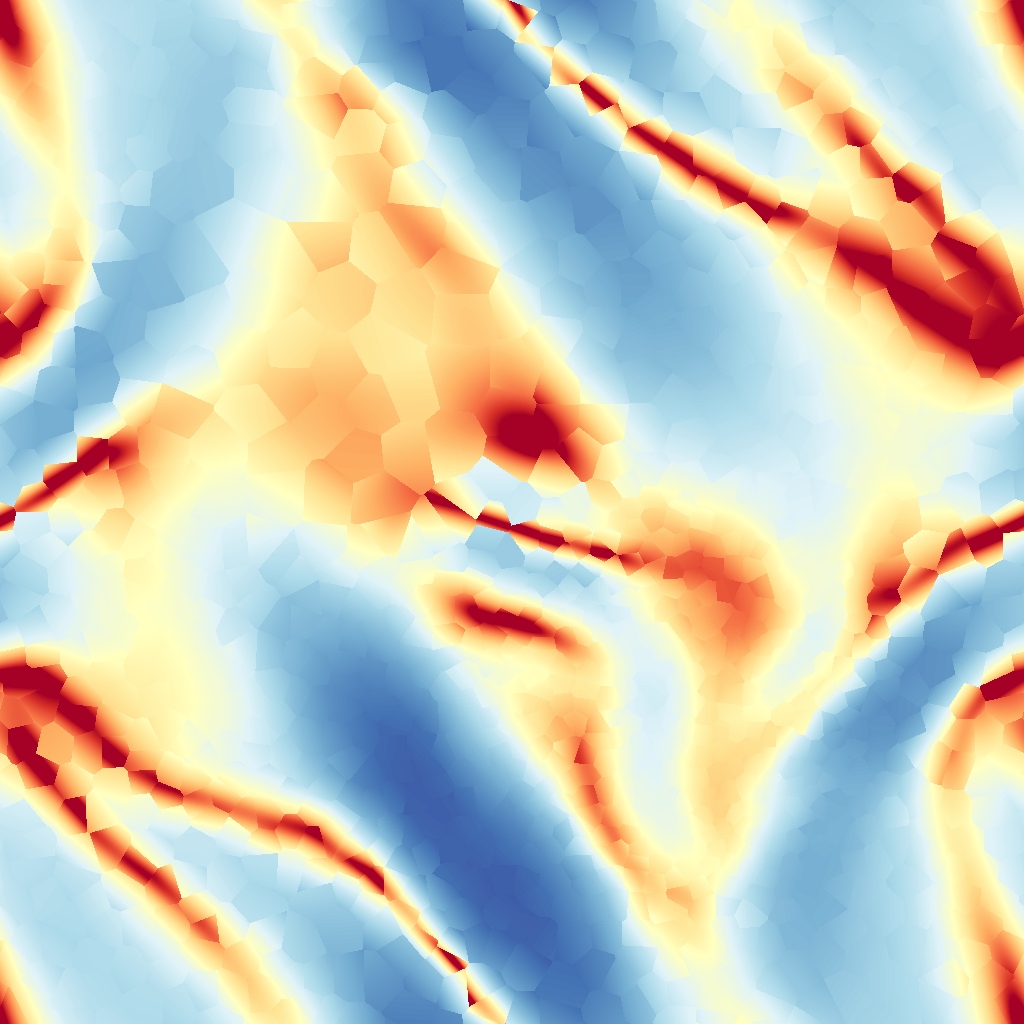}}\hspace*{0.2cm}%
\resizebox{5.2cm}{!}{\includegraphics{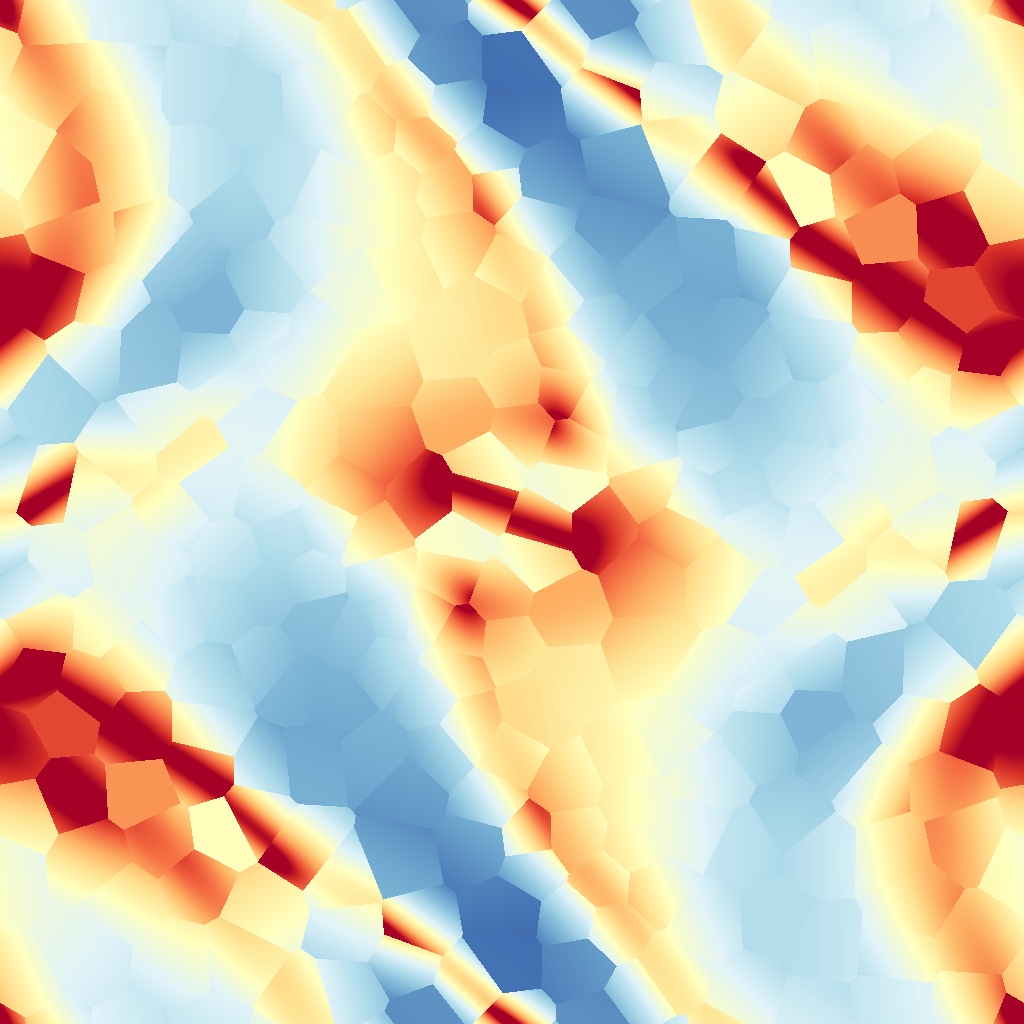}}
\resizebox{5.2cm}{!}{\includegraphics{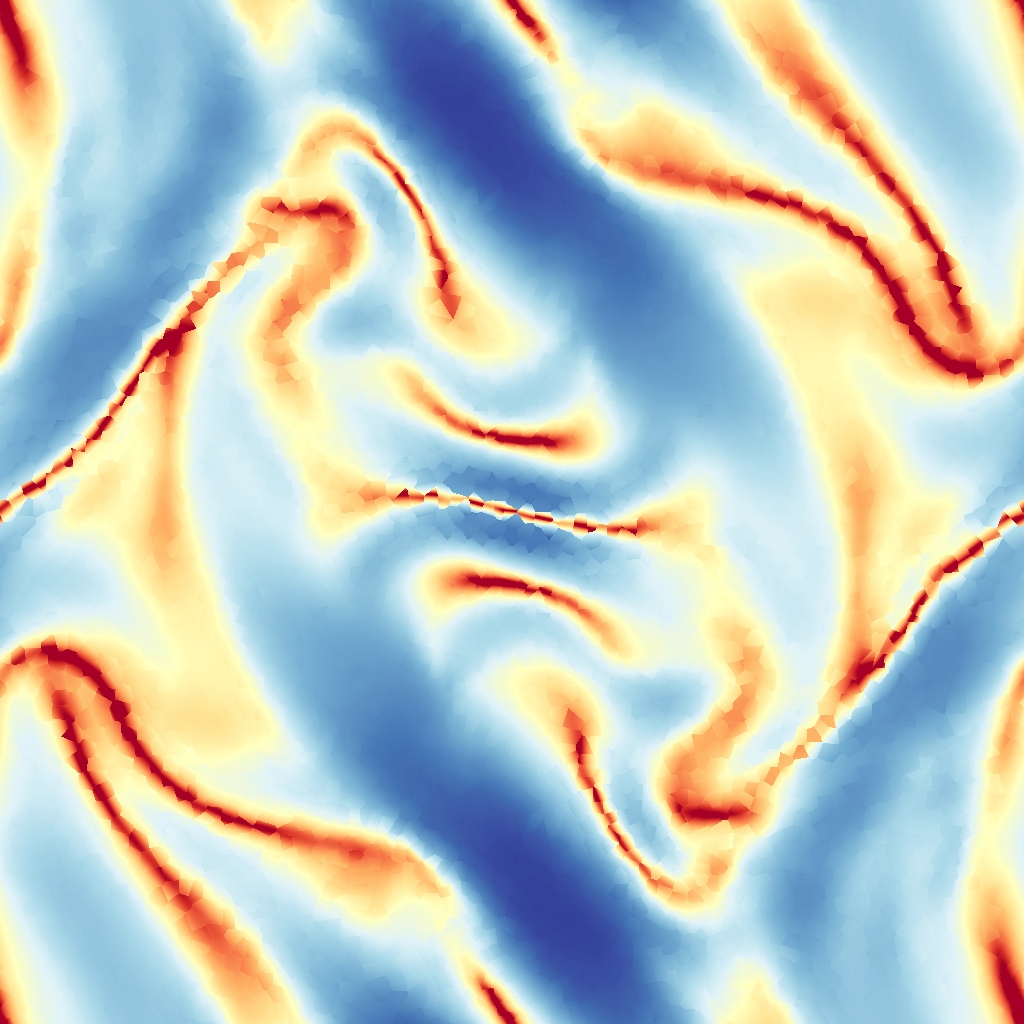}}\hspace*{0.3cm}%
\rotatebox{90}{\resizebox{4.2cm}{!}{\hspace*{2.0cm}\includegraphics{plots/OT_bfield_legend.pdf}}}
\end{center}
\caption{Maps of the magnetic energy density in the 2D Orszag-Tang problem at time $t=0.768$, computed with our fiducial variable resolution setup (left), with very low resolution without refinement (middle), and at higher fixed resolution (right panel). Note that the mesh geometry for the left panel is displayed in the lower left panel of Figure~\ref{FigBlast2DRefinementMesh}. \label{FigBlast2DRefinementB2}}
\end{figure*}

\subsection{Local timestep test}

To illustrate and test our local timestepping procedure, we here repeat the MHD blast wave test of subsection~\ref{secTestBlast2D}, but this time with local timestepping enabled. The deposition of thermal energy in a central region makes the timestep required there much smaller than in the background volume. Note that background cells are ``informed'' about the arriving blast wave and reduce their timesteps sufficiently early by the tree-based timestep scheme of {\small AREPO} \citep{Springel2010}, which determines in a computationally efficient way for every cell the first time a signal from any of the other cells can arrive. This  information is then included in determining the next timestep size adopted during the simulation for any local cell. 

Visually, the outcome of the calculation with local timesteps shows no discernible differences to the result reported in Figure~\ref{FigBlast2D} for a fixed global timestep, as desired. This demonstrates that our method for local timestepping works robustly and accurately when extended to the vector potential based MHD implementation introduced here. Previously, this was already known for ordinary hydrodynamics combined with the Powell cleaning method, as described in \citet{Pakmor2013}. 

In Figure~\ref{FigBlast2DTimesteps}, we show maps of the timestep sizes at different times for the moving mesh case with an initial hexagonal mesh with $2\times (256\times 147)$ cells. At the earliest time shown in the top left, only a couple of cells  are requiring a timestep corresponding to the red colour shown in the legend. Note that without local timestepping all cells would have to adopt this small timestep, whereas for the vast majority of the cells a timestep a factor of 8 larger is sufficient. This already illustrates the inherent inefficiency of global timestep schemes for problems with a wide dynamic range of timescales. In this particular test problem, the total number of cell updates required over the full simulation up to time $t=2.0$ goes down by a factor of 3 when using local instead of global timestepping, which is approximately the speed-up that can be realized in a well optimized code. Note, however, that this efficiency gain is highly problem dependent, and that this simple test problem is not particularly demanding in terms of its dynamic range. In applications of interest such as galaxy formation, the efficiency gain from local timestepping can be much larger, so large in fact that global timestep schemes are prohibitively expensive in comparison.

\subsection{Refinement and derefinement test}
\label{secTestrefinement}

We next consider a test of our refinement and derefinement approach. In order to allow a visual assessment of the quality of the resulting mesh we settle on a low-resolution fiducial problem  in 2D in which we prescribe particular resolution requirements. In particular, for definiteness, we consider the Orszag-Tang problem once more (starting out with a resolution of $32\times 32$ cells), and we prescribe a function
\begin{equation}
V_{\rm res} (\vec{r}, t)= 
\left\{
\begin{array}{cc}
1/16^2 & \mbox{for $[x -  0.25 \cos(2\pi t/T)]^2 $}\\
       & \mbox{$+ [y -  0.25 \sin(2\pi t/T)]^2 < 0.2^2$},\vspace*{0.1cm}\\
1/64^2 & \mbox{for $[x -  0.25 \cos(2\pi t/T) + \pi]^2 $}\\
       & \mbox{$+ [y -  0.25 \sin(2\pi t/T + \pi)]^2 < 0.2^2$},\vspace*{0.1cm}\\
1/32^2 & \mbox{otherwise,}
\end{array}
\right.
\end{equation}
specifying the desired target resolution at time $t$ and position $\vec{r}=(x,y)$ relative to the centre of the simulation box. Our refinement criterion is chosen as indicating that a cell $i$ should be refined if its current volume fulfils $V_i > 2 V_{\rm res}(\vec{r}_i,t)$, whereas it should be removed if its current volume fulfils $V_i < \frac{1}{2} V_{\rm res}(\vec{r}_i,t)$, where $\vec{r}_i$ is the coordinate of the mesh-generating point of cell $i$. With this choice, we hence have two spherical regions of radius $0.2$ that circle counter-clock wise and with opposite position angles at a distance of 0.25 around the centre of the box, completing one orbit over our adopted simulation time of $T=2.048$. In one of the spherical regions, the mesh resolution should be increased by a linear factor of 2 compared to the background grid, whereas in the other it should be decreased by this factor. Note that the actual motion of the gas is clock-wise in this problem. We thus expect that mesh refinement and de-refinement operations must be continuously carried out by the code to comply with the prescribed desired mesh resolution. It will be particularly interesting to see whether our vector potential approach for the magnetic field is robust and stable in such a situation. 

In Figure~\ref{FigBlast2DRefinementMesh} we illustrate the obtained mesh geometry at different times. The different resolution regions can be clearly recognized, and a video of the time evolution of the mesh\footnote{Available as accompanying online material at \url{https://wwwmpa.mpa-garching.mpg.de/~volker/arepo2_mhd}} demonstrates the high mesh quality sustained during the evolution, including the moments right after new cells are inserted.

Reassuringly, also the magnetic field results make sense for this test. In Figure~\ref{FigBlast2DRefinementB2}, we show the magnetic field energy density at time $t=0.768$, which corresponds to the lower left panel in Fig.~\ref{FigBlast2DRefinementMesh}. Of course, the symmetry is now broken due to the circling zones with different resolution, but the overall result is still similar to when the simulation is carried out with uniform resolution. We show this explicitly by comparing with results obtained for a simulation with the very low resolution $2\times(16\times 9)$ (middle panel), and with one that has a still moderate but substantially higher $2\times(64\times 36)$ resolution (right panel). These two resolutions bracket the lowest and highest resolutions occurring in our spatially and temporarily adaptive test simulation shown on the left. Reassuringly, the results of the fiducial test take on an intermediate role, with no visible artifacts in the magnetic field from the frequent refinement and de-refinement operations in the fiducial simulation.

\begin{figure*}
\resizebox{5.5cm}{!}{\includegraphics{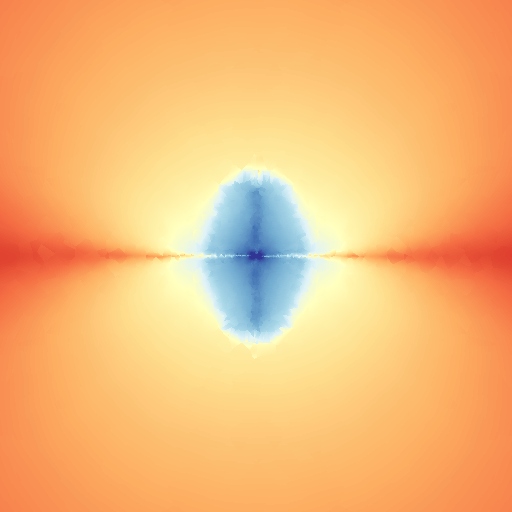}}\hspace*{0.2cm}%
\resizebox{5.5cm}{!}{\includegraphics{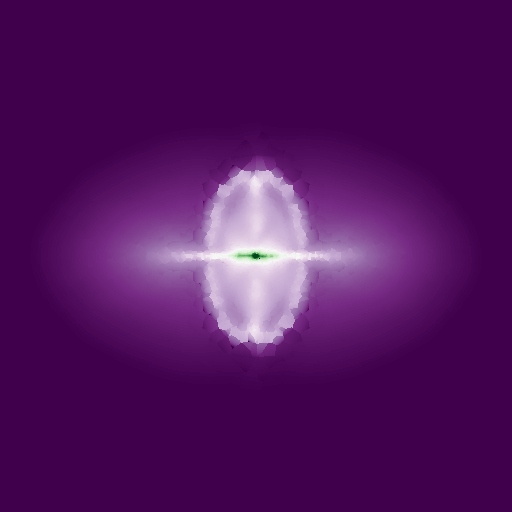}}\hspace*{0.2cm}%
\resizebox{5.5cm}{!}{\includegraphics{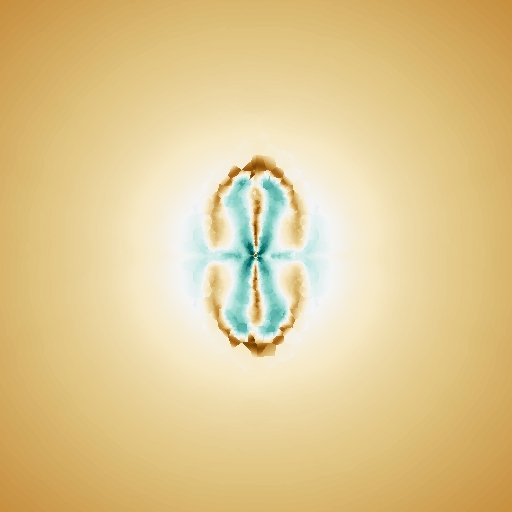}}\vspace*{0.2cm}\\
\resizebox{5.5cm}{!}{\includegraphics{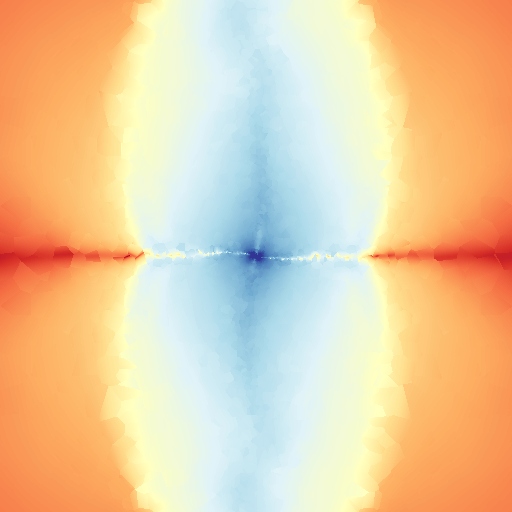}}\hspace*{0.2cm}%
\resizebox{5.5cm}{!}{\includegraphics{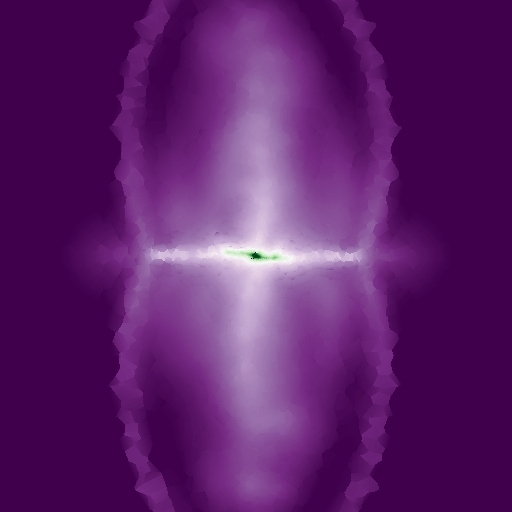}}\hspace*{0.2cm}%
\resizebox{5.5cm}{!}{\includegraphics{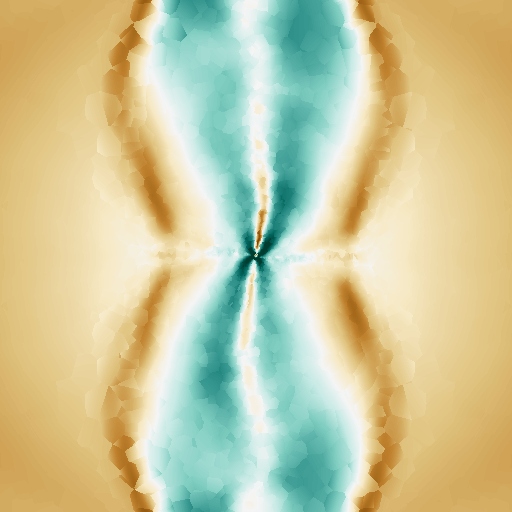}}\vspace*{0.2cm}\\
\resizebox{5.5cm}{!}{\includegraphics{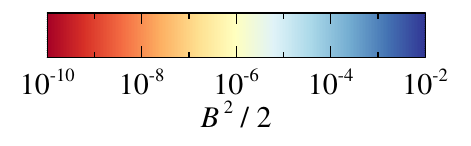}}\hspace*{0.2cm}%
\resizebox{5.5cm}{!}{\includegraphics{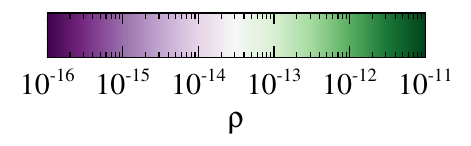}}\hspace*{0.2cm}%
\resizebox{5.5cm}{!}{\includegraphics{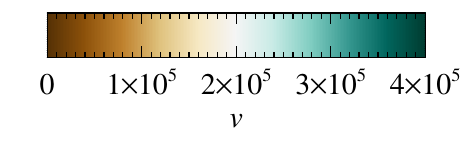}}
\caption{Collapse of a magnetized cloud. The panels show slices through the rotation axis of the cloud, and have an extension of $2.0\times 10^{16}\,{\rm cm}$, corresponding to half the initial cloud radius. The upper row shows magnetic energy density, mass density, and velocity magnitude at time $t=1.05\,t_{\rm ff}$, while the lower row gives the same quantities at time   $t=1.10\,t_{\rm ff}$. \label{FigCollapse}}
\end{figure*} 

\subsection{Collapse of a magnetized cloud}

Finally, we test our code by applying it to the collapse of an idealized magnetized cloud, following a setup similar to \citet{Hennebelle2008} and \citet{Pakmor2011}. This problem is meant to resemble typical situations in protostellar collapse, and is thus of particular relevance for applications in star and planet formation.

We consider a magnetized cloud of radius $R_c = 4.0\times 10^{16}\,{\rm cm}$ in solid body rotation, with  mass $M_c =  1.0\,{\rm M}_\odot = 1.989\times 10^{33}\,{\rm g}$ and homogenous density $\rho_c = 7.42\times 10^{-18}\, {\rm g}\,{\rm cm}^{-3}$. The sphere is set into solid body rotation, with a ratio $\alpha= 0.01$ of rotational $E_{\rm rot} = \frac{1}{5} M_c R_c^2\,{\Omega}^2$ to gravitational energy $E_{\rm grav}= \frac{3}{5}{G M_c^2}/{R_c}$, which yields an angular velocity of $\Omega = 7.42\times 10^{-18}\,{\rm s}^{-1}$. We impose a uniform magnetic field parallel to the rotation axis, with flux $\Phi = B_z\times \pi R_c^2$ and a mass-to-flux ratio of $\mu = 10$ in units of the critical  mass-to-flux ratio \citep{Mouschovias1976},  adopted as $(M/\Phi)_{\rm crit}= \frac{0.53}{3\pi} \left(\frac{5}{G}\right)^{1/2}$. This corresponds to an initial field strength of $B_z = 81.3\,\mu{\rm G}$.

We embed the cloud into a box of size $L_{\rm box} = 2.5 \times R_c = 1.0\times 10^{17}\,{\rm cm}$ and sample this volume with an initially uniform grid of $128^3$ cells. The gas outside the sphere is at rest and set to a density 30 times lower than that of the cloud, while its initial magnetic field strength is the same as that in the sphere. No attempt to smooth out the velocity field transition at the edge of the sphere is made.

We impose periodic boundary conditions at the box boundary but compute the self-gravity of the material in the box with non-periodic boundaries. The gravitational softening of the cells is adaptive, but we impose a minimum softening length of $\epsilon_{\rm grav}= L_{\rm box}/50000 = 2.5\times10^{12}\,{\rm cm}$ in order to put a lower floor on the maximum collapse scale. The initial free fall time is $t_{\rm ff} = [3 \pi / (32 G \rho_c)]^{1/2} = 7.71\times 10^{11}\,{\rm s} = 2.44\times 10^4\,{\rm yr}$. We adopt the same barotropic equation of state as used in \cite{Hennebelle2008} and \citet{Pakmor2011}, which is given by
\begin{equation}
P(\rho) = \rho\, c_0^2  \left[ 1 + \left(\frac{\rho}{\rho_0}\right)^{4/3}\right]^{1/2},
\end{equation}
where $c_0=2.0\times 10^4\,{\rm cm}\,{\rm s}^{-1}$ and $\rho_0 = 10^{-13}\, {\rm g}\,{\rm cm}^{-3}$. The adiabatic index used for the Riemann solver is chosen as $\gamma = 1.0001$.

When the evolution is started, the cloud collapses under self-gravity and forms a small accretion disk that winds up the magnetic field and eventually drives a magnetically powered outflow. In Figure~\ref{FigCollapse}, we show the state of the cloud at times $t = 1.05\, t_{\rm ff}$ and $t = 1.10\, t_{\rm ff}$.  The magnetic field along the $z$-direction has strongly increased due to the collapse, and in the azimuthal plane, very high field strengths are reached. The magnetic field dominates over the thermal pressure in extended regions, reaching low plasma beta values that cause the formation of strong, jet-like outflows and drive shocks into the low-density atmosphere around the collapsed cloud. Overall, the results of this highly non-linear test are in good qualitative agreement with those shown, for example,  in \citet{Hennebelle2008}, \citet{Pakmor2011},  \citet{Hopkins2016}, \citet{Marinacci2018} and \citet{Mayer2025, Mayer2026}.


\section{Discussion}
\label{sec:discussion}

In this section, we briefly discuss how our specific formulation of numerical MHD compares to other approaches in the literature that are based on the vector potential. We aim to focus on the main conceptual differences without going into the details, although we caution that the latter can sometimes also be surprisingly important in numerical MHD.

\subsection{Comparison to constrained transport on a Cartesian mesh} 

It is instructive to briefly recall the basic principles of constrained transport (CT) methods in order to discuss their differences with respect to our MHD implementation on an unstructured moving mesh. In CT, the magnetic field is typically represented in a staggered fashion on the cell faces of a Cartesian grid, with its component normal to the corresponding face. This is equivalent to storing the magnetic fluxes through the cell faces. In setting the initial conditions (ICs), one needs to make sure that the sum of these fluxes over the faces that make up any given cell is zero, something that is in practice accomplished by specifying the vector potential of the ICs and carrying out line integrals along the edges surrounding the faces. The flux sum realizes one specific discrete version of the divergence operator for each computational cell, and demanding that this sum vanishes can be viewed as a  discrete analogue of the $\nabla \cdot\vec{B} = 0 $ condition. CT implementations aim to update the surface magnetic fields such that the vanishing flux sum is preserved to machine precision in every timestep.

In practice, this can be achieved by estimating electric fields at the corners of the grid, and by using them in contour integrals along the edges surrounding facets to update their magnetic fluxes. Since each edge is traversed twice in opposite directions for the facets surrounding a given cell, the total magnetic flux emerging from a cell will not change in such an update step, as desired. 

As we have mentioned earlier, one can equivalently store the vector potential as primary independent variable \citep{Toth2000, Mocz2017}. This does not require any fundamental change in the CT algorithm, and does not change the results either. CT methods are therefore vector potential methods.

A primary difference to our approach is how the magnetic field is obtained from the vector potential. In the case of CT, this is done via using the Stokes theorem to obtain the surface magnetic fluxes through a line integral over the vector potential. Averaging the normal magnetic fields from opposite sides of a cell, one then obtains the cell-centred magnetic field. In our method, we instead obtain the magnetic field directly by computing the curl of the vector potential through a finite difference operator. Our tests do not indicate that this is inferior to the integral construction, suggesting that the origin of the accuracy of CT primarily lies in its (implicit) use of the vector potential, whereas a vanishing flux sum to machine precision is not required for this success.

The staggered representation of magnetic flux (or equivalently the vector potential) in CT compared to the cell-centred fluid quantities is another key difference to our approach, which uses a cell-centred storage for the vector potential as well. Through the staggering, CT gains additional spatial resolution without any memory or compute overhead, something that is a significant benefit compared to our approach. 

Finally, arguably the most important conceptual difference is that CT implementations use the Weyl gauge, whereas our method employs a special advective gauge that renders the evolution of the vector potential Galilean invariant if the mesh motion is tied to the fluid velocity. This has important practical advantages especially in simulations where large bulk velocities appear, because it allows advection errors to be eliminated or greatly reduced. The downside is that working in this gauge is numerically quite subtle, and special measures are required to prevent numerical instabilities. We have introduced them in the form of an auxiliary large-scale velocity field that is continuous and smoothly differentiable, and a method to suppress the build-up of large irrotational contributions in the vector potential.

An interesting side effect of our gauge and the special measures accompanying it is that they can also be applied for stationary meshes, and then provide an alternative to the commonly used technique of stabilizing the integration of the induction equation in the Weyl gauge with an ad-hoc upwinding scheme. In our case, advective motions are instead absorbed into a bona-fide advection term for the vector potential, which is arguably more natural. This term can be treated  with standard methods for advection, offering improved stability as it does not require extra ad-hoc upwinding, and in some tests is also slightly more accurate.

\subsection{The vector potential method of Mocz at al.~(2016)}
\label{SecCompareWithMocz}

It is interesting to compare our new approach, as summarized in Section~\ref{SecNewApproach}, with an earlier implementation of a vector potential method \citep{Mocz2016} in the moving-mesh code {\small AREPO}.  A first key difference concerns the construction of the $\vec{B}$-field from the vector potential. To this end, \citet{Mocz2016} associate the vector potential of a cell with the location of its mesh-generating point. These points are the vertices of the Delaunay tessellation that is dual to the Voronoi mesh. For each Delaunay tetrahedron, one can then straightforwardly compute magnetic fluxes for their four facets through an integral construction, as done in ordinary CT. These magnetic fluxes add to zero for each tetrahedron, thereby recovering a discretized analogue of $\nabla \cdot \vec{B} = 0$, again simply because each edge enters twice in opposite directions in the magnetic flux sums. Furthermore, there is a unique $\vec{B}$-field value for the interior of each tetrahedron that reproduces all surface magnetic fluxes. It is then suggestive to define the volume-average magnetic field of each Voronoi cell by averaging the contributions of Delaunay tetrahedra according to their spatial overlap.

While this approach is elegant and parameter free, it also has a number of disadvantages.  For the hydrodynamical evolution, Riemann problems need to be solved for the facets of the Voronoi tessellation, but for them the above reconstruction does not yield continuous normal components of the $\vec{B}$-field, and the sum of the magnetic fluxes through the surfaces of Voronoi cells is not guaranteed to be vanishing either (which does, however, not necessarily impact the accuracy of the overall scheme negatively, because such a local $\nabla\cdot \vec{B}$ `error' is in part a reflection of the chosen discretization of the divergence operator, and it can in any case not grow as the evolution is formulated in terms of the vector potential). However, the reconstructed $\vec{B}$-field is not unique for degenerate mesh configurations, simply because in this case the Delaunay tessellation is not unique. Even though the above construction appears to sidestep the need to explicitly differentiate the vector potential, such a spatial derivative is still needed. In fact, since we need the spatial derivative of $\vec{B}$ to construct a second-order accurate integration scheme, one effectively even needs second derivatives of $\vec{A}$. How they should be estimated in a way that makes them sufficiently consistent with the above geometric reconstruction is not fully clear.

The second key difference lies in the update scheme for $\vec{A}$ itself. The approach of \citet{Mocz2016} interprets the vector potential stored for a cell $\vec{A}_i$ as a cell-average, in fact, the vector potential integrated over the cell volume is evolved as the independent variable for the magnetic field. This quantity is updated based on advective fluxes when the mesh moves, and with a source terms reflecting the electromagnetic forces evaluated at cell centres. This formulation has a few conceptual disadvantages compared with the new approach we propose in this paper. First of all, it is not clear that the cell averaging of the $\vec{A}$-field is actually advantageous. While this is inspired by how conserved quantities are treated in finite-volume methods, we note that  $\vec{A}$ is actually not a conserved field. Importantly, it needs to be differentiated to get the physical quantity of interest. But the cell-averaging provides extra smoothing that could easily introduce avoidable numerical diffusivity into the scheme, and worse, any advection error will be amplified by the differentiation operator needed to get the physical $\vec{B}$-field. 

A more serious issue concerns the evaluation of the source term in the evolution equation for $\vec{A}$. In the scheme of \citet{Mocz2016}, this is based on cell-centred values without directly taking the solutions of the Riemann problem into account. This thus potentially misses sufficient `upwinding' that is by experience necessary for achieving numerical stability in all situations. We suspect that this is the primary reason why the scheme's practical implementation has frequently shown stability problems in general fluid flows, to the point that it cannot be used reliably at high resolution in several problems of interest \citep{Pakmor2026}. A final issue is related to the gauge choice. This only matters in 3D, but here the Weyl gauge used by CT and \citet{Mocz2016} alike is not allowing manifest Galilean invariance at the discretized level, unlike the method based on the advective gauge we introduce here.

\subsection{Comparison to SPH methods using the vector potential}

There have been a number of attempts to construct Lagrangian treatments of MHD using the vector potential, but this has often been met with substantial numerical stability problems. In particular, \citet{Price2010} concluded after extensive numerical tests of a range of SPH discretization schemes that the use of the vector potential appears not viable in SPH. A revised formulation by \citet{Tricco2023} that adopted an integral form of the induction equation could likewise not cure the prevalent numerical instabilities.

Interestingly, however, \citet{Stasyszyn2015} reported a successful implementation of MHD in SPH based on the vector potential. Their method adopted a cleaning approach in the vector potential, limiting $\nabla\cdot\vec{A}$ to small values. Even though the formulation was using the Weyl gauge, thus giving up manifest Galilean invariance that is otherwise one of the strong features of SPH, the fact that a prevention of the pollution of $\vec{A}$ with irrotational modes made a significant difference to the stability is a key insight of this study. Note that this is something we have likewise seen in our moving mesh method.

Finally, a recent study by \citet{Tu2022} reported a successful implementation of a vector potential method in the meshless finite mass code {\small GIZMO}. The vector potential is evolved in the Weyl gauge in their numerical scheme, and thus should already provide long-term protection against the build-up of divergence errors. However, the study  additionally applied Powell and Dedner cleaning in the $\vec{B}$-field, arguing that this is needed to provide numerical stability across magnetic shocks. While the method passed many standard tests satisfactorily, it also shows noise features in high-resolution tests of the 3D Orszag-Tang problem, suggesting that it is not yet fully free of instabilities in all types of flows.


\section{Summary and Conclusions}
\label{sec:conclusions}

In this work, we have presented a new MHD discretization for the moving-mesh code {\small AREPO}. It is based on evolving a cell-centred vector potential, which by construction guarantees a divergence-free magnetic field. We have formulated the scheme using a novel variant of the advective gauge in order to produce Galilean invariant solutions at the discretized level, and we have outfitted the method with local timestepping and flexible, on-the-fly refinement and de-refinement techniques applied to our fully dynamic, unstructured Voronoi mesh.

We argue that the use of the vector potential provides the fundamental basis for the stability and accuracy of numerical MHD in CT and our scheme alike, whereas maintaining a vanishing magnetic flux sum to machine precision for all computational cells is not strictly required. Instead of constructing the magnetic field via  line integrals of the vector potential around the edges of each face, which first yields magnetic fluxes and magnetic field normal components that can then then be averaged from opposite sides of cells to yield the cell-centred field, one can alternatively directly apply a finite-difference curl-operator. The integral construction used in CT corresponds merely to one among many possible definitions of discrete operators to evaluate the cell-centred field from the vector potential, but it is not unique. A case in point is that CT methods that maintain a vanishing flux sum do in general not guarantee that the cell-centred magnetic fields have a vanishing divergence if the latter is evaluated as sum of the local derivatives of $\vec{B}$, where these derivatives are the ones that are used for extrapolating the transverse field components to face centres. What one means with {\it divergence free} in discretized MHD methods is hence in part a matter of definition.

We note that a highly beneficial aspect of Cartesian CT is however the staggered storage of the vector potential at mesh corners (or equivalently the magnetic fluxes on mesh faces) while the conserved ordinary fluid quantities stay cell-centred. This staggering of the storage locations maximizes the spatial resolution of the method, and yields resolution benefits that in particular allow sharper jumps in the transverse magnetic field components compared to a purely cell-centred scheme such as ours. Because the faces and vertices of a dynamic Voronoi mesh are not persistent in time, a staggered storage of the vector potential is however not readily possible in our method.

By adopting a special gauge in terms of an auxiliary large-scale smooth velocity field, we have shown that a Galilean invariant and stable formulation of MHD can be realized. Because the gauge field can inject  irrational modes into the vector potential, we however needed to adopt divergence control in $\vec{A}$ in order to maintain long-term numerical stability. We have chosen a simple variant of Dedner cleaning for this purpose. Choosing appropriate cleaning coefficients is much more straightforward in this case than in cleaning methods for the magnetic field, and the problems due to local changes in timestep  pointed out by \citet{Tomida2026} are absent. The combination of using a smoothly differentiable velocity field as a local reference frame and of limiting the divergence in $\vec{A}$ is what overcomes the instability of the standard advective gauge in 3D that has frequently been reported in the literature.

In test problems we have been able to demonstrate the accuracy and robustness of our new formulation, including in demanding 3D problems. Importantly, our method works with local timesteps and allows seamless on-the-fly refinement and de-refinement of the mesh at the level of single cells, all without incurring stability problems in magnetohydrodynamical flows. This adds tremendous flexibility to the approach. The test results we found are competitive with constrained transport methods, and are superior in situations with strong advection, because in such cases our approach can drastically reduce or eliminate advection errors thanks to its Galilean invariance. 

In future work, it will also be interesting to add non-linear MHD extensions to our new solver. \citet{Zier2024a, Zier2024b} successfully implemented Ohmic dissipation, ambipolar diffusion, and the Hall effect in the Powell 8-wave formulation of MHD on a moving-mesh in {\small AREPO}. Modifying this implementation for use with our new vector potential approach appears very promising, and should be comparatively straightforward  as we have direct access to the electric current and if needed can compute it as a higher-order derivative with a double stencil. An important advantage of our new formulation compared to the Powell 8-wave approach is that it does not require  source terms in the momentum, energy and induction equations, which bear the risk of introducing spurious, hard-to-detect errors in certain situations.

We therefore expect that our method will also be helpful to inform the long-standing debate about whether results obtained with cleaning methods for the magnetic field (i.e.~either Dedner or Powell) are fully reliable and compatible with constrained transport and vector potential approaches in all situations of physical interest, or whether there are circumstances in which one of the methods can produce unphysical results. This in particular applies to dynamo problems. We note that here conservation of magnetic helicity, which is comparatively easily accessible in our formalism, could be a useful diagnostic that thus far has only rarely been considered in code validation.

\section*{Acknowledgements}

VS acknowledges insightful discussions with Kandaswamy Subramanian that helped to motivate this paper. The research ideas of this paper, its text, its figures, and the underlying simulation code have been created without the use of AI tools of any kind.

\section*{Data Availability}

The data underlying this article will be shared upon reasonable request to the corresponding authors.

\bibliographystyle{mnras}
\bibliography{main}

\appendix

\section{Cosmological equations} \label{SecCosmoEquations}

For reference and definiteness, let us here collect the set of MHD equations in the form we actually use in cosmological simulations, where we need to allow for expansion of space. The latter is described in the standard way with the cosmological scale factor $a(t) = 1/(1+z)$, where $z$ is the cosmological redshift, and  $H = \dot{a} /a$ is the Hubble rate. We use comoving coordinates $\vec{x} =\vec{r}/a$, `comoving densities $\rho_c = a^3\,\rho$, `comoving' magnetic fields  $\vec{B}_c = a^2 \vec{B}$ and vector potentials $\vec{A}_c = a \vec{A}$, peculiar velocities\footnote{Also useful is the conjugate velocity $\vec{v}_c = a \vec{v} = a^2 $, which is what we store for particles.} $\vec{v} = a\, \dot{\vec{x}}$, a `comoving' pressure $P_c \equiv (\gamma-1)\rho_c u = a^3 P$, and total pressure $P_{\rm tot} = P_c + \vec{B}_c^2/a$  as principle variables. We define a gradient operator $\nabla_c$ that acts on the comoving coordinates $\vec{x}_c$, i.e.~we have $\nabla = (1/a) \nabla_c$. For the magnetic field, we then still have $\vec{B}_c = \nabla_c \times \vec{A}_c$ for the comoving quantities, and we similarly  define $\vec{J}_c = \nabla_c \times \vec{B}_c$ to maintain this relation as well. Note that for this coordinate transformation, partial time derivatives at fixed physical location $\vec{r}$ and fixed comoving location $\vec{x}$ are related by $\left(\partial/\partial t\right)_{\vec{r}} = \left(\partial/\partial t\right)_{\vec{x}} - H \vec{x}\cdot \nabla_c$.

With these definitions, the cosmological MHD equations take on the form \citep[see][for more details on their derivation]{Springel2010, Pakmor2013}:
\begin{equation}
\frac{\partial \rho_c}{\partial t} + \frac{1}{a}\nabla_c \cdot \big[\rho_c
\vec{v}\big] = 0,
\end{equation}
\begin{equation}
\frac{\partial( a \rho_c \vec{v})}{\partial t} + \frac{1}{a}\nabla_c\cdot [
\rho_c \vec{v}\vec{v}^T + P_{\rm tot} - \vec{B}_c\vec{B}_c^T /a] = 0,
\end{equation}
\begin{equation}
\frac{\partial(a^2 \rho_c e)}{\partial t} + \frac{1}{a}\nabla_c \cdot \big[(\rho_c e + P_{\rm tot})\vec{v} -  (\vec{v}\cdot\vec{B}_c ) \vec{B}_c \big] =  H \frac{\vec{B}_c^2}{2} ,
\label{eqnEgycosmo}
\end{equation}
\begin{equation}
\frac{\partial( \rho_c s ) }{\partial t} + \frac{1}{a}\nabla_c \cdot \big[\rho_c s \vec{v}\big] = 0,
\end{equation}
\begin{equation}
\frac{\partial  \vec{B}_c  }{\partial t} + \frac{1}{a}\nabla_c \cdot \left[\vec{B}_c \vec{v}^T - \vec{v}\vec{B}_c^T\right] = 0,
\end{equation}
\begin{equation}
\frac{\partial  \vec{A}_c  }{\partial t} +  \frac{1}{a}\nabla_c \left[\vec{u} \cdot \vec{A}_c\right] = \frac{1}{a} \vec{v} \times \vec{B}_c  + \frac{1}{a}\nabla_c \phi_c,
\label{eqAcosmo}
\end{equation}
Here the partial time derivatives are at fixed comoving coordinate $\vec{x}$, and the specific energy per unit mass is defined in terms of the peculiar velocity as $e = u + \vec{v}^2/2 + \vec{B}_c^2 / (2 a\rho_c)$. In the last equation, we have included the large-scale velocity $\vec{u}$ in the gauge term, which is here a smoothed version of the peculiar velocity field $\vec{v}$, and on the right hand side we have also included the cleaning scalar $\phi_c = a\, \phi$ for controlling the divergence of the vector potential.

The form of these equations is very similar to their ordinary Newtonian form. Note in particular that the only places where cosmological terms appear explicitly is a source term in the energy equation that involves the magnetic field (which is easily dealt with in the time-integration in an operator split fashion),  through a $1/a$ prefactor in front of the spatial derivative operator (which is here defined in terms of comoving coordinates), and in the definitions of the conservative quantities stored in each cell for momentum and energy. As the cells are characterized by a comoving volume $V_c$, the conserved quantities are suitably taken as cell mass $m_i$, cell conjugate momentum $p_i = a m_i\vec{v}_i$, cell energy $E_i = a^2 m_i e_i$, and cell entropy $S_i = m_i P_i/\rho_i^\gamma$. The scale factors in these definitions ensure that source terms are largely absent from the above set of fluid equations, modulo the term on the right hand side of equation~(\ref{eqnEgycosmo}). Also of interest is that only thanks to our gauge choice,  the cosmological equation~(\ref{eqAcosmo}) for the vector potential looks again formally  similar to equation~(\ref{eqAnewton}) for Newtonian space. For other gauge choices, this is not the case. Finally, also note that for $a\equiv 1$, the Newtonian forms applicable for non-cosmological simulations are of course recovered.

The induction equation appears twice in the above system of equations, as the equations for the evolution of the magnetic field and the vector potential express the same physics, i.e.~one of these equations is in principle redundant. While the equation for the vector potential is our principal equation for dynamics of the magnetic field, we also use the magnetic field equation to predict the state of the magnetic field at the end of each timestep for our second flux computation, this is why we still list it here. While the equation for the vector potential looks deceptively like a conservation law with a source function, note that the second term on the left hand side is a gradient and not a divergence. It is originating in our gauge choice, whereas the source term on the right hand side expresses the key physics of the induction equation.

Also note that the equation for the entropy is in principle redundant, given the total energy equation. We nevertheless list it here as it may be optionally evolved in parallel with the total energy so that it could selectively be used to estimate the new temperature at the end of a timestep when the total energy formalism becomes inaccurate in highly supersonic or low-$\beta$ flows (see the discussion in Section~\ref{SecLowBetaFlow}).

\bsp
\label{lastpage}
 
\end{document}